 \documentclass[3p]{elsarticle}   

\usepackage{bbold}
\usepackage{hyperref,geometry,amsmath,appendix}
\usepackage{textgreek}

\def\openone{\mathbb I}
\def\R{\mbox{\boldmath$\displaystyle\mathbb{R}$}}

\usepackage{tikz-cd}  
\usepackage[font=itshape]{quoting}
\usepackage{geometry,appendix}

\usepackage[pagewise,displaymath]{lineno}
\usepackage{multirow}            
\usepackage[utf8]{inputenc}
\usepackage{hyperref,slashed,amsmath,amssymb,xcolor,mathrsfs}
\usepackage{graphicx,subfigure}
\usepackage{amsfonts,bbold,multicol,tikz,cancel,mathtools}
\usepackage{url}
\usepackage{tikz-feynman}

\biboptions{sort&compress}

\definecolor{nicered}{rgb}{0.7,0.1,0.1}
\definecolor{nicegreen}{rgb}{0.1,0.5,0.1}
\definecolor{red}{rgb}{1.0, 0, 0}
\definecolor{darkblue}{rgb}{.0,.0,.8}
\definecolor{niceblue}{rgb}{0,0,1}
\definecolor{niceviolet}{rgb}{0.5,0,1.0}
\definecolor{blue}{rgb}{0,0,1}
\hypersetup{colorlinks=true,citecolor=nicegreen,linkcolor=nicered,urlcolor=nicered}

\usepackage[normalem]{ulem}    
\modulolinenumbers[1]                

\journal{Physics Reports}

\def\I{\openone}
\def\openone{\mathbb I}

\newcommand{\gdualn}[1]{\overset{\:{}^{{}^{\boldsymbol{\neg}}}}{\smash[t]{#1}}} 

\def\df{\mbox{$\displaystyle\gdualn{\mathfrak{f}}$}}

\def\0{\mbox{\boldmath$\displaystyle\boldsymbol{0}$}}

\def\s{\mbox{\boldmath$\displaystyle\boldsymbol{\sigma}$}}

\def\x{\mbox{\boldmath$\displaystyle\boldsymbol{x}$}}
\def\p{\mbox{\boldmath$\displaystyle\boldsymbol{p}$}}

\begin{document}

\begin{frontmatter}

\title{\textsc{Mass dimension one fermions: Constructing darkness}}


\author{Dharam Vir Ahluwalia$^1$
\cortext[corrauth]{Corresponding author}}
\ead{d.v.ahluwalia@iitg.ac.in}

\author{Julio M. Hoff da Silva$^2$\, }
\author{Cheng-Yang Lee$^3$\,}
\author{Yu-Xiao Liu$^4$\,}
\author{Saulo H. Pereira$^2$\,}
\author{Masoumeh Moazzen Sorkhi$^5$\,}

\address{$^1$ Center for the Studies of the Glass Bead Game, \\ Normanby Road, Notting Hill, Victoria 3168, Australia}

\address{$^2$ Departamento de F\'isica, Faculdade de Engenharia de Guaratinguet\'a, Universidade
	Estadual Paulista, UNESP,\\ Av. Dr. Ariberto Pereira da Cunha, 333, Guaratinguet\'a, SP,
	Brazil.}

\address{$^3$ Center for Theoretical Physics, College of Physical Science and Technology,\\
Sichuan University, Chengdu, 610064, China}

\address{$^4$ Institute of Theoretical Physics and Lanzhou Center for Theoretical Physics, School of Physical Sciences and Technology, Lanzhou University, Lanzhou, 730000, China}

\address{$^5$  Department of Physics, Faculty of Basic Sciences, Kosar University of Bojnord, Iran}

\begin{abstract}
Let $\Theta$ be the Wigner time reversal operator for spin half and let $\phi$ be a Weyl spinor. Then, for a left-transforming $\phi$, the construct $\zeta_\lambda   \Theta \phi^\ast$  yields a right-transforming spinor. If instead,
$\phi$ is a right-transforming spinor, then the construct $\zeta _\rho \Theta \phi^\ast$  results in a left-transforming spinor
($\zeta_{\lambda,\rho}$ are phase factors). This allows us to introduce two sets of  four-component spinors. Setting $\zeta_\lambda$ and $\zeta_\rho$ to $\pm i$ render all eight spinors as  eigenspinor of the charge conjugation
operator~$\mathcal{C}$ (called ELKO). This allows us to introduce two quantum fields. A calculation of the vacuum expectation value of the time-ordered product of the fields and their adjoints reveals the mass dimension of the fields to be one. Both fields are local in the canonical sense of quantum field theory. Interestingly, one of the fields is fermionic and the other bosonic. 
 The mass dimension of the introduced fermionic fields and the matter fields of the Standard Model carry an intrinsic mismatch.  As such, they provide natural darkness for the new fields with respect to the Standard Model doublets. The statistics and locality are controlled by a set of phases. These are explicitly given. Then we observe that in $p_\mu p^\mu = m^2$,  Dirac  took the simplest square root of the $4\times 4$ identity matrix $\I$ (in $\I \times m^2 $, while introducing $\gamma_\mu p^\mu$ as the square root of the left hand side of the dispersion relation), and as such he implicitly ignored the remaining fifteen. When we examine the remaining roots, we obtain additional bosonic and fermionic dark matter candidates of spin half.
 We point out that by early nineteen seventies, Dirac had suspected the existence of spin half bosons, in the same space as his fermions. This is interweaved with a detailed discussion of duals and adjoints. We study the fermionic self-interaction and interactions with a real scalar field. We show that a consistent interacting theory can be formulated using the ELKO adjoint up to one-loop thus circumventing the earlier problem of unitarity violation. We then undertake  quantum field theoretic calculation that establishes the Newtonian gravitational interaction for a mass dimension one dark matter candidate.  The report ends: (a) by studying the partition function and main thermodynamic properties of the mass dimension one fermionic field in the context of the dark matter halo of galaxies. For the Milky Way, the observational data of rotation curve fits quite well for a fermionic mass of about 23 eV; and (b)
by introducing higher-dimensional ELKOs in braneworld scenario. After a brief introduction of some braneworld models, we review the localization of higher-dimensional ELKOs on flat and bent branes with appropriate  localization mechanisms. We discuss the massless and massive Kaluza-Klein modes of ELKO fields on branes and give a comparison with other fields.
\end{abstract}

\begin{keyword}
ELKO, mass dimension one fermions of spin half, mass dimension one boson of spin half, locality and causality phases, dark energy, dark matter,  brane

%
\end{keyword}


\end{frontmatter}
\newpage

\tableofcontents

\newpage

\section{Background and introduction}

The opening sentences of our 2005 PRD and JCAP papers that introduced the construct of mass dimension one fermions, read:

\begin{quote}
\begin{itemize}
\item[\empty]\textit{We report an unexpected theoretical discovery of a spin half matter field with mass dimension one.}
\item[\empty]
\hfill {\small D. V. Ahluwalia and D. Grumiller~\cite{Ahluwalia:2004sz}}
\end{itemize}
\begin{itemize}
\item[\empty]\textit{We provide the first details on the unexpected theoretical discovery of a spin-one-half matter field with mass dimension one.}
\item[\empty]
\hfill {\small D. V. Ahluwalia and D. Grumiller~\cite{Ahluwalia:2004ab}}
\end{itemize}
\end{quote}
\vspace{-5pt}while a referee report of the funding application based on these works read, in part
\begin{quote}
\begin{itemize}
\item[\empty]\textit{The problem has fueled intense debates in recent years and is generally considered fundamental for the advancement in the field. As for the proposed solution [by Ahluwalia], I find the approach advocated in the project a very solid one, and, remarkably, devoid of speculative excesses common in the field; the whole program is firmly rooted in quantum field theoretic fundamentals, and can potentially contribute to
them. If \textit{ELKO} and its siblings can be shown to account for
dark matter, it will be a major theoretical advancement that will
necessitate the rewriting of the first few chapters in any textbook
in quantum field theory.  If not, the enterprise will still have
served its purpose in elucidating the role of all representations of
the extended Poincar\'e group.}
\item[\empty]
\hfill{From a 2006 Referee report for Marsden application 07-UOC-055}
\\
\vspace{-11pt}
\hfill{Royal Society of New Zealand ``Dark Matter and its Darkness''.}
\end{itemize}
\end{quote}

\vspace{21pt}
In the intervening decade and a half, the formalism of mass dimension one fermions has evolved to the extent that: (a)  we now have a local mass dimension one fermionic quantum field of spin half, and (b)  we are able to construct additional new class of spin half fermions and bosons (surprisingly, also of spin half)~\cite{Vir_Ahluwalia_2020,Ahluwalia_2020}.  ELKO, being eigenspinors of charge-conjugation operator, rule out the usual $U(1)$ invariance and make them natural dark matter candidate. 
Furthermore, the mismatch of mass dimensionalities of the Standard Model (SM)  fermions with the new fermions forbids the latter to enter the SM doublets.
Specifically, a pair of mass dimension one fermions can form a doublet but it cannot form doublet with Dirac fermions of mass dimension three half.  The  mass dimensionality of one provides unsuppressed quartic self interaction for the new fermions. While the same is suppressed by two powers of the unification scale for the SM fermions.

 In parallel with the development of the formalism~\cite{Ahluwalia:2008xi,Ahluwalia:2018hfm,Lee:2019fni,Ahluwalia:2009rh,Ahluwalia:2010zn,Ahluwalia:2015vea,Speranca:2013hqa}, there are a significant number of authors who have explored viability of the new fermions for cosmology, including inflation,  dark matter and dark energy~\cite{Pereira:2020ogo,Boehmer:2007dh,Boehmer:2010ma,Boehmer:2008ah,Boehmer:2006qq,Pereira:2017efk,
Boehmer:2009aw,Boehmer:2007ut,Boehmer:2008rz,Pereira:2018hir,Fabbri:2009aj,Villalobos:2018shc,Sadjadi:2011uu,Basak:2011wp,Lee:2012qq,Basak:2012sn,Agarwal:2014oaa,daSilva:2014kfa,Basak:2014qea}. While others have given a serious thought to the implications of new spinors for five-dimensional branes and higher dimensions~\cite{Zhou:2020ucc,MoazzenSorkhi:2020fqp,Zhou:2018oib,Sorkhi:2018jhy,Dantas:2015mfi,Zhou:2017bbj,Liu:2011nb,Pereira:2014wta}. There are also studies devoted to the mass dimension one fermions in the context of Hawking radiation, their sensitivity to the topology of space-time and a dynamical system based analysis~\cite{daRocha:2014dla,
 Cavalcanti:2015nna,
 Bahamonde:2017ize,
 Bernardini:2012sc,
 CapelasdeOliveira:2013ams,
 daSilva:2016htz,Vaz:2017fac,
 daRocha:2011xb}.

 Thus this report extends the standard set of quantum fields of SM by constructing fundamentally new quantum fields. The darkness of many of these fields is not assumed but is intrinsic; that is constructed \emph{ab initio}. The guiding principles may be viewed in the tradition of Wigner's and Weinberg's foundational works on constructing quantum fields~\cite{Wigner:1962ep,
Wigner:1939cj,Weinberg:1995mt,Weinberg:1964cn,Basak_2013}: whatever dark matter is, it must furnish one of the representations of  the Lorentz algebra merged with parity, charge conjugation, and time reversal.

 \subsection{How is this possible?}

While Weinberg's formalism for spin half provides a no-go theorem for anything but, modulo the 1937 Majorana observation~\cite{Majorana:1937vz}, a quantum field with Dirac spinors as its expansion coefficients; Wigner's 1964 analysis offers three additional quantum field theoretic possibilities. Possibilities challenged by the 1966 work of Lee and Wick~\cite{PhysRev.148.1385}. To escape the no-go theorem of Weinberg, one must have an additional degeneracy and build unitarity and positivity of energy without insisting the Hamiltonian $H$ being Hermitian.\footnote{The demand of hermiticity and unitarity may be generalized to pseudo-hermiticity and pseudo-unitarity while ensuring the reality of the Hamiltonian~\cite{Mostafazadeh:2001jk}.} The latter route is provided by Carl Bender's observations on the subject contained in~~\cite{Bender:1998gh,Bender:1998ke,Bender:2005tb,Bender:2007nj}. Remarkably, the same work replaces the usual $\dagger$ in $H = H^\dagger$ in the Dirac adjoint by space-time reflection, that is $\mathcal{PT}$ symmetry. In the context of $\mathcal{CPT}$ preserving theories, this places the charge conjugation symmetry on a special footing. One of the mass dimension one quantum fields we construct here is built upon the eigenspinors of $\mathcal{C}$ as its expansion coefficients with locality controlled by phases attached to these~\cite{Ahluwalia:2016jwz,Ahluwalia_2020A}. This evades the Lee-Wick result.

An  accessible introduction to Wigner classes can be found in the referee reports published in the Acknowledgements section of~\cite{Ahluwalia:1995ur}. The Istanbul lectures that introduced Wigner classes is available in the proceedings~\cite{Wigner:1962ep}. In the said lecture itself, Wigner notes that the results were ``taken from a rather old but unpublished manuscript by V. Bargmann and A. S. Wightman.''

 In a recent monograph, we have argued that dark matter is a new aspect of reality in the same sense as spin and antiparticles.
The existence of spin resolved the doubling of the states of an electron in atoms.  The  insistence on the conservation of parity forced the right and left Weyl spinors to be treated on the same footing. This doubling of the degrees of freedom (from two to four), in our opinion, explains emergence of antiparticles in Dirac's 1928 paper. Here, we shall see that the notion of dark matter corresponds to yet another doubling of the degrees of freedom. Thus spin, antiparticles, and dark matter represent a trinity of duplexities -- or, so we suspect. For details of this argument, we refer the reader to~\cite[Sec. 2]{Ahluwalia:2019etz}.

When used for quarks and leptons, the Dirac formalism, in its quantum field theoretic incarnation, allows theorists to construct a beautiful edifice that fires the SM part of the universe. Without it, it is impossible to erect the local gauge symmetric fabric of unprecedented beauty. But only if we allow the structure to include general-relativistic description of gravity. Without gravity, there is very  little atomic varieties (no carbon, no oxygen). Without dark matter, structure formation is only a dream. So what we do here is an attempt to provide a theory of dark matter, just as Dirac's work, in the context of those years, was an attempt to provide a theory of electron. Modesty prevents us from making such remarks, but creating a larger context demands it.

\subsection{Notation}

Much of the notation we use is standard. To avoid such phrases as `$m$ is mass of the particle' we have decided to avoid most of  such phrases.  Instead, whenever an ambiguity arises, the reader may consult a recent monograph~\cite{Ahluwalia:2019etz}. We do make exceptions to this broad brush  guideline as and when we think the notation is too specialized. Apart from sec.~\ref{ElkoOnBrane}, we adopt the metric signature to be $(+,-,-,-)$.

\subsection{Errors and misconceptions in the textbook presentations of the Dirac quantum field}

As already noted, modulo the 1937 observation of Majorana~\cite{Majorana:1937vz}, all matter fields of the SM are described by the Dirac field. To the best of our knowledge, its full beauty and uniqueness is exposed best in the writings of Weinberg~\cite{PhysRev.133.B1318,Weinberg:1995mt}. While this report challenges the uniqueness, it is also necessary  to point out errors and misconceptions scattered in textbooks. One of the few exceptions to these flaws is the monograph of Srednicki~\cite{Srednicki:2007qs}, Weinberg's 1964 papers~\cite{Weinberg:1964cn,Weinberg:1964ev} and his classic \textit{The Theory of Quantum Fields}~\cite{Weinberg:1995mt}.

The following references document errors and misconceptions~\cite{Gaioli:1998ra,Ahluwalia:1998dv,Cahill:2005zb,Cahill:2021akg}. We collect them together below before proceeding to the task proper. This is particularly important because our earlier works were born in the shadows of these flaws and were to be corrected.

To every particle-antiparticle system of spin half, the SM associates an all-pervading quantum field $\Psi(x)$. The expansion coefficients of $\Psi(x)$ are the eigenspinors of the parity operator $\mathcal{P}$
(See, \cite{Speranca:2013hqa} and~\cite[Sec. 5]{Ahluwalia:2019etz})
 \begin{equation}
		\mathcal{P} = m^{-1} \gamma_\mu p^\mu. \label{eq:P}
\end{equation}
As shown in the indicated references, the above result relies entirely on Lorentz algebra without reference to a wave equation or a Lagrangian density. Each of the eigenspinors of $\mathcal{P}$, apart from a norm, has an important  phase freedom. If $u_\sigma(p^\mu)$ and  $v_\sigma(p^\mu)$ are eigenspinor of $\mathcal{P}$,  with eigenvalues $+1$ and $-1$ respectively, then so are $u^\prime_\sigma(p^\mu) \stackrel{\mathrm{def}}{=} e^{i \alpha_\sigma}  \times u_\sigma(p^\mu)$
and
$v^\prime_\sigma(p^\mu) \stackrel{\mathrm{def}}{=} e^{i \beta_\sigma}  \times u_\sigma(p^\mu)$; where $\alpha_\sigma~\mbox{and}~\beta_\sigma \in\R$.
Because the expansion coefficients multiplying the creation and annihilation operators appear as linear combinations in $\Psi(x)$, the phases  $e^{i \alpha_\sigma}~\mbox{and}~e^{i \beta_\sigma} $ must be fixed in order that the field be \textit{local} and \textit{covariant}~\cite{refId0}. This can be seen by explicit calculations or can be read off from the final results on $\Psi(x)$ in the Weinberg formalism.

Now a general $\mathcal{R}\oplus \mathcal{L}$ spinor $\psi(p^\mu)$, that enters the spin half quantum field $\Psi(x)$,
has the form
\begin{equation}
		\psi(p^\mu) = \left[ \begin{array}{c}
		\phi_\mathcal{R}(p^\mu)\\
		\phi_\mathcal{L}(p^\mu)
		\end{array}
		\right],\label{eq:gen-4-comp-psi}
\end{equation}
where the right($\mathcal{R}$)- and left($\mathcal{L}$)- handed Weyl spinors have certain constraints placed on them. The first is that they must transform under boosts and rotations in a way dictated by the Lorentz symmetry. But  each of the Weyl spinor cannot be assigned any helicity one `wants,' etc. As shown in~\cite[Chapter 5]{Ahluwalia:2019etz}, the demand that
$\psi(p^\mu)$  be an eigenspinor of $\mathcal{P}$ severely restricts this freedom, and restricts it so that at rest the
$\phi_{\mathcal{L}}({k^\mu})$ and $\phi_{\mathcal{R}}(k^\mu)$ are numerically identical for $u_\sigma(k^\mu)$, and differ by a minus sign for $v_\sigma(k^\mu)$:\footnote{Here we have defined $p^\mu$ at rest as $k^\mu \stackrel{\mathrm{def}}{=}  p^\mu {\big\vert}_{\lim_{\mathbf{p}\to 0}} $.}
\begin{equation}
\phi_\mathcal{R}(k^\mu) = \begin{cases} + \phi_\mathcal{L}(k^\mu), & \mbox{for}~u_\sigma(k^\mu)~ \mbox{spinors}
\\
- \phi_\mathcal{L}(k^\mu), &
\mbox{for}~v_\sigma(k^\mu)~\mbox{spinors}
\end{cases}.\label{eq:constraint12}
\end{equation}
This remains largely unknown in a significant fraction of literature, see, for example, references~\cite{Ryder:1985wq,Hladik:1999tt,Folland:2008zz} and the discussions in~\cite{Gaioli:1998ra,Ahluwalia:1998dv,Cahill:2005zb,Cahill:2021akg}.\footnote{A parenthetic remark: introduction of the four-component spinor $\psi(p^\mu)$ is necessitated by the covariance of the formalism under $\mathcal{P}$. This doubling in the degrees of freedom for a spin half particle from two to four, introduces an additional symmetry called charge conjugation. Thus antiparticles appear as a natural consequence of imposing $\mathcal{P}$ covariance. The existence of antiparticles is also required by causality. The latter argument is originally due to Feynman and can be found in~\cite{PhysRev.76.749,Feynman:1987gs} and in Section 13, Chapter 2, of~\cite{Weinberg:1972gc}.}

%
The $\psi(k^\mu)$ are constructed by taking
$\phi(k^\mu)$
 as eigenspinors of the helicity operator
\begin{equation}
\s\cdot\widehat \p\, \phi_\pm(k^\mu) = \pm \phi_\pm(k^\mu)\label{eq:hd}
\end{equation}
with an implicit choice of global phase factors
\begin{align}
\phi_+(k^\mu) & = \sqrt{m}  \left(
									\begin{array}{c}
									\cos(\theta/2)\exp(- i \phi/2)\\
									\sin(\theta/2)\exp(+i \phi/2)
											\end{array}
									\right),\label{eq:zimpok52-new-new2} \\
\phi_-(k^\mu) & = \sqrt{m}   \left(
									\begin{array}{c}
									- \sin(\theta/2)\exp(- i \phi/2)\\
									 \cos(\theta/2)\exp(+i \phi/2)
											\end{array}
									\right). \label{eq:zimpok52-new-new-new}
\end{align}
With this choice for the Weyl spinors at rest, a set of $\psi(k^\mu)$ follow upon incorporating the constraint (\ref{eq:constraint12}):
\begin{align}
& u_+(k^\mu) = e^{i \alpha_+}\left(
\begin{array}{c}
\phi_+(k^\mu) \\
\phi_+(k^\mu)
\end{array}
\right),
&u_-(k^\mu) &= e^{i \alpha_-}\left(
\begin{array}{l}
\phi_-(k^\mu) \\
\phi_-(k^\mu)
\end{array}
\right), &\empty &\empty\\
& v_+(k^\mu) = e^{i \beta_+}\left(
\begin{array}{c}
\phi_-(k^\mu) \\
- \phi_-(k^\mu)
\end{array}
\right),
&v_-(k^\mu)
&= e^{i \beta_-}  \left(\begin{array}{c}
 \phi_+(k^\mu) \\
-\phi_+(k^\mu)
\end{array}
\right), &\empty &\empty
\end{align}
where $\alpha_\sigma~ \mbox{and}~\beta_\sigma \in \R$. Two observations are now in order: (a) note that $v_+(k^\mu)$ is defined through $\phi_-(k^\mu)$, and not $\phi_+(k^\mu)$, with a similar comment applying to $v_-(k^\mu)$ and, (b) in order that we obtain a local quantum field satisfying the cluster decomposition principle and Lorentz symmetries, the set $\{\alpha_+,\alpha_-,\beta_+,\beta_-\}$ modulo an overall sign must be fixed to $\{0,0,0,\pi\}$. This leads to
\begin{align}
& u_+(k^\mu) =\left(
\begin{array}{c}
\phi_+(k^\mu) \\
\phi_+(k^\mu)
\end{array}
\right),
&u_-(k^\mu) &= \left(
\begin{array}{l}
\phi_-(k^\mu) \\
\phi_-(k^\mu)
\end{array}
\right), &\empty &\empty\\
& v_+(k^\mu) =\left(
\begin{array}{c}
\phi_-(k^\mu) \\
- \phi_-(k^\mu)
\end{array}
\right),
&v_-(k^\mu)
&= - \left(\begin{array}{c}
 \phi_+(k^\mu) \\
-\phi_+(k^\mu)
\end{array}
\right). &\empty &\empty
\end{align}
This is consistent with Weinberg's construction of spin half quantum field~\cite{PhysRev.133.B1318,Weinberg:1995mt}. We hope that our readers may find the foregoing discussion as a `correcting bridge' that fills the gap between where Weinberg leaves off and the more usual presentations of the subject. In fact, as already noted, the evolution of mass dimension one fermions from~\cite{Ahluwalia:2004sz,Ahluwalia:2004ab} to ~\cite{Ahluwalia:2016rwl,Ahluwalia:2019etz} is due, in part, to the insights collected in
the `correcting bridge.'

\subsection{Spinorial transformations, metric, and the defining spirit}

The Lorentz transformation for $\psi(p^{\mu})$ is formally given by
\begin{equation}
\psi'((\Lambda p)^{\mu})\equiv D(\Lambda)\psi(p^{\mu})
\end{equation}
where $D(\Lambda)$ is the spin half representation of the Lorentz transformation $\Lambda$ taking $p^{\mu}$ to $(\Lambda p)^{\mu}$. When $\Lambda=L(p)$ is a pure boost  that takes $k^{\mu}$ to momentum $p^{\mu}$, we have
\begin{eqnarray}
\psi'(p^{\mu})&=&D(L(p))\psi(k^{\mu})\nonumber\\
&=&\sqrt{\frac{E + m }{2 m}}
						\left[
						\begin{array}{cc}
						\I + \frac{\boldsymbol{\sigma}\cdot\mathbf{p}}{E +m} & \0 \\
						\0 & \I - \frac{\boldsymbol{\sigma}\cdot\mathbf{p}}{E +m}
						\end{array}
						\right]\psi(k^{\mu})
\end{eqnarray}
where $\I$ is a $4\times 4$ identity matrix in the 
$\mathcal{R}\oplus\mathcal{L}$ 
 representation space.
For convenience, from now onwards, we simply write the boost transformation as
$
\psi'(p^{\mu})\equiv\psi(p^{\mu})
$.
Under rotation $\Lambda=R(\hat{\boldsymbol{n}}\theta)$, we have
\begin{eqnarray}
\psi'(k^{\mu})&=&D(R(\hat{\boldsymbol{n}}))\psi(k^{\mu}) \nonumber\\
&=&\left[\begin{matrix}
\cos(\theta/2)\I+i\s\cdot\hat{\boldsymbol{n}}\sin(\theta/2) & \0\\
\0 & \cos(\theta/2)\I+i\s\cdot\hat{\boldsymbol{n}}\sin(\theta/2)
\end{matrix}\right]\psi(k^{\mu}).
\end{eqnarray}
Note that since the rotation leaves $k^{\mu}$ invariant, we must write $\psi'(k^{\mu})$ to denote the rotated spinor to avoid confusion. Since $\psi(k^{\mu})$ is chosen to be helicity eigenspinors, it follows that $\psi'(k^{\mu})$ is a linear combination of spinors of positive and negative helicity.

Once a spin half dual is defined for the expansion
coefficients $\psi(p^\mu)$
\begin{equation}
\overline{\psi}(p^\mu) \stackrel{\mathrm{def}}{=} \psi^\dagger(p^\mu)
\eta \label{eq:dn}
\end{equation}
the field adjoint follows. The spin half metric $\eta$ is determined from the constraint that bilinears constructed from spinors are Lorentz covariant. The constraint is that the metric $\eta$ commutes with each of the generators of rotation, $\zeta_i$, in the
$\mathcal{R}\oplus\mathcal{L}$ representation space, and anti-commutes with each of the generators of the boosts $\kappa_i$
\begin{equation}
\left[\zeta_i,\eta\right]=0,\quad
\left\{\kappa_i,\eta\right\} = 0,\qquad i = x,y,z.
\end{equation}
The resulting formalism supports the local gauge transformations of the SM and account for its `luminosity.'
For the mass dimension one fermions, how the metric $\eta$ enters the dual is  subtle and shall  be presented below.
For the moment, it suffices to note that
for the $\mathcal{R}\otimes\mathcal{L}$ representation space, up to a scale factor and a similarity transformation (see,~\cite[Ch. 17]{Ahluwalia:2019etz}), $\eta$ is found to be
\begin{equation}
\eta = e^{i\zeta}
\left(
\begin{array}{cccc}
1 & 0 & 0 &0\\
0 & -1 &0 &0\\
0&0&-1&0\\
0&0&0&-1
\end{array}
\right),\qquad \zeta \in \R \label{eq:ew}
\end{equation}
while for the $\mathcal{R}\oplus\mathcal{L}$ representation space the metric reads
\begin{equation}
 \eta = 
\left(
\begin{array}{cccc}
0 & 0 & a+i b &0\\
0 & 0 &0 & a + i b\\
c +i d&0&0&0\\
0&c + i d&0&0
\end{array}
\right), 
\qquad a,b,c,d \in \R.\label{eq:dd}
\end{equation}
The usual east and west coast metrics are special cases of (\ref{eq:ew}). The $\gamma_0$ of Dirac dual similarly follows from setting $a=1,\; b=0$ and $c=1,\;d=0$ in (\ref{eq:dd}). Such choices are not entirely arbitrary but may be dictated, for example, by symmetry considerations.

For dark matter, the matter fields that we seek must be one representation or the other of the Lorentz algebra with well-defined $\mathcal{P}$, $\mathcal{C}$, and $\mathcal{T}$ properties.
Yet its kinematic structure must be such that it does not support local gauge transformations of the SM. It turns out that these two requirements appear naturally and are not to be imposed separately or are to be seen as two independent requirements.


\section{Magic of Wigner time reversal operator}\label{Sec:Elko-b}


The argument for the above begins with the observation that  Wigner time reversal operator in its spin half incarnation
\begin{equation}
			\Theta= \left(\begin{array}{cc} 0 & -1 \\ 1 & 0\end{array}\right)
	\end{equation}
leads to the following theorem:

\begin{quote}
If $\phi(p^\mu)$ transforms as a left-handed Weyl spinor, then up to a multiplicative phase factor $\zeta_\lambda$, the $\Theta\phi^\ast(p^\mu)$ transforms as a right-handed Weyl spinor.
\end{quote}

As symmetry dictates
\begin{quote}

If $\phi(p^\mu)$ transforms as a right-handed Weyl spinor, then up to a multiplicative phase factor $\zeta_\rho$, the $\Theta\phi^\ast(p^\mu)$ transforms as a left-handed Weyl spinor.
\end{quote}
This logic is lost if, following Ramond, one mistakes $\Theta$ for  Pauli's $\sigma_y$ modulo a factor of $i$~\cite{Ramond:1981pw}. In that event, generalization to higher spins is not possible in a natural way but it can be naturally achieved by repeating the above arguments  with Wigner time reversal operator for any spin.

We are thus motivated to introduce a new set of four-component spinors. These Lorentz transform as the familiar eigenspinors of the parity operator (that is, Dirac spinors), but have intrinsically different properties under the discrete symmetries of $\mathcal{P}$, $\mathcal{C}$, and $\mathcal{T}$.  The resulting spinors from the very start do not allow the usual local $U(1)$ gauge transformation. Our task then is to use every insight from the `correcting bridge' and study properties of the quantum field built with the new spinors as expansion coefficients. Remarkably, the field thus constructed has mass dimension one, and not three half with numerous, previously unthought of, physical consequences. As the argument continues, we will see how the Lee-Wick no-go theorem
gets evaded~\cite{Lee:1966ik}.

\subsection{ELKO, the eigenspinors of charge conjugation operator}

With this briefest of brief prelude, we define the new spinors to have the form
\begin{equation}
\lambda_\alpha(p^\mu) \stackrel{\mathrm{def}}{=} \left(\begin{array}{c}
\zeta_\lambda \Theta\left[ \phi_\alpha(p^\mu)\right]^\ast \\
\phi_\alpha(p^\mu)
\end{array}
\right),\qquad
\rho_\alpha(p^\mu) \stackrel{\mathrm{def}}{=} \left(\begin{array}{c}
\phi_\alpha(p^\mu)\\
\zeta_\rho\Theta\left[ \phi_\alpha(p^\mu)\right]^\ast \\
\end{array}
\right).\label{eq:elko}
\end{equation}
Here, $\alpha$ is the helicity index of $\phi_{\alpha}(p^{\mu})$. While $\phi_{\alpha}(p^{\mu})$ have a definite helicity, the $\lambda_\alpha(p^\mu)$ and $\rho_\alpha(p^\mu)$ themselves do not. Their right-transforming and left-transforming components have opposite helicities
\begin{equation}
 \s\cdot\widehat{\p} \left[\Theta \phi_\pm^\ast(p^\mu)\right]  = \mp
  \left[\Theta \phi_\pm^\ast(p^\mu)\right]  .
 \end{equation}
As discussed at significant length in~\cite{Ahluwalia:2019etz}, the charge conjugation operator can be defined without reference to a Lagrangian density or a wave equation and it reads
\begin{equation}
	\mathcal{C} = \left(\begin{array}{cc}
					\0 & i\Theta \\
					-i \Theta & \0
					\end{array}
					\right) K . \label{eq:gamma2K}
	\end{equation}
The operator $K$  carries out the operation of complex conjugating to the right. The choice $\zeta_\lambda = +i$ yields a self-conjugate $\lambda^S(p^\mu) $, and $\zeta_\lambda = -i$ yields an anti-self conjugate $\lambda^A(p^\mu) $
\begin{equation}
\mathcal{C} \lambda^{S}_\pm(p^\mu) =  +  \lambda^{S}_\pm(p^\mu),\qquad
\mathcal{C} \lambda^{A}_\pm(p^\mu) =   - \lambda^{A}_\pm(p^\mu).
\end{equation}
In contrast, $\zeta_\rho = -i$ yields a self-conjugate $\rho^S(p^\mu) $, and $\zeta_\rho = +i$ yields an anti-self conjugate $\rho^A(p^\mu) $
\begin{equation}
\mathcal{C} \rho^{S}_\pm(p^\mu) =  +  \rho^{S}_\pm(p^\mu),\qquad
\mathcal{C} \rho^{A}_\pm(p^\mu) =   - \rho^{A}_\pm(p^\mu).\end{equation}
The new spinors are known in the literature as ELKO, an
acronym for the German
\textbf{E}igenspinoren des \textbf{L}adungs\textbf{k}onjugations \textbf{o}perators.\footnote{We encourage the community to refrain from using the phrase ELKO spinor(s) because ELKO already contains the word eigenspinors.}

 \subsection{ELKO at rest}

Introducing an appropriate set of locality phases (see chapter 8 of~\cite{Ahluwalia:2019etz}),
the rest spinors $\lambda_\alpha(k^\mu)$ are found to be
\begin{align}
 &\lambda^S_+(k^\mu)  =  \left[
					\begin{array}{c}
					+i \Theta\left[\phi_+(k^\mu)\right]^\ast\\
								\phi_+(k^\mu)
					\end{array}
					\right],\quad
\lambda^S_-(k^\mu)  =  \left[
				\begin{array}{c}
				+i \Theta\left[\phi_-(k^\mu)\right]^\ast\\
				\phi_-(k^\mu)
				\end{array}
				\right],  \label{eq:zimpok0new}\\
 &\lambda^A_+(k^\mu)  =  \left[
					\begin{array}{c}
					- i \Theta\left[\phi_-(k^\mu)\right]^\ast\\
								\phi_-(k^\mu)
					\end{array}
					\right],\quad
\lambda^A_-(k^\mu) =  - \left[
				\begin{array}{c}
				- i \Theta\left[\phi_+(k^\mu)\right]^\ast\\
					 \phi_+(k^\mu)
				\end{array}
				\right]
				\label{eq:zimpok91new}
				\end{align}
with similar expressions for $\rho^{S}(k^\mu)$ and $\rho^{A}(k^\mu)$:
\begin{align}
& \rho^S_+(k^\mu)   =  \left[
					\begin{array}{c}
					\phi_+(k^\mu)\\
					- i \Theta\left[\phi_+(k^\mu)\right]^\ast
					\end{array}
					\right],\quad 	  		
\rho^S_-(k^\mu)  =  \left[
					\begin{array}{c}
					\phi_-(k^\mu)\\
					- i \Theta\left[\phi_-(k^\mu)\right]^\ast
					\end{array}
					\right],	 \\
 &\rho^A_+(k^\mu)  =
 \left[
					\begin{array}{c}
					\phi_-(k^\mu)\\
					 +i \Theta\left[\phi_-(k^\mu)\right]^\ast
					\end{array}
					\right], \quad
	\rho^A_-(k^\mu) =
	- \left[
					\begin{array}{c}
					\phi_+(k^\mu)\\
					 +i \Theta\left[\phi_+(k^\mu)\right]^\ast
					\end{array}
					\right].
					\end{align}

\subsection{ELKO for arbitrary momentum}

These for arbitrary momentum are obtained from the rest spinors by the action of the
 boost transformation
as follows
\begin{equation}
\lambda^{S/A}_{\pm}(p^\mu) = D(L(p)) \; \lambda^{S/A}_{\pm}(k^\mu),\qquad
\rho^{S/A}_{\pm}(p^\mu) = D(L(p)) \; \rho^{S/A}_{\pm}(k^\mu).\label{eq:rho}
\end{equation}

In order that we are not lost in the story, we collect together some of the results and then develop the formalism.
We define two new spin half fields with ELKO as its expansion coefficients 
\begin{equation}
\mathfrak{f}(x) =\int
\frac{d^3 p}{(2\pi)^3}
\frac{1}{\sqrt{2 m E(\p)}}
\sum_\alpha
\left[
a_\alpha(\p)
\lambda^S_\alpha(\p) e^{-i p\cdot x}
+
b^\dagger_\alpha(\p)
\lambda^A_\alpha(\p) e^{i p\cdot x}
\right]                    \label{eq:f}
\end{equation}
 and
  \begin{equation}
\mathfrak{b}(x) =\int
\frac{d^3 p}{(2\pi)^3}
\frac{1}{\sqrt{2 m E(\p)}}
\sum_\alpha
\left[
c_\alpha(\p)
\rho^S_\alpha(\p) e^{-i p\cdot x}
+
d^\dagger_\alpha(\p)
\rho^A_\alpha(\p) e^{i p\cdot x}
\right].\label{eq:b}
\end{equation}
To construct the  propagator associated with each of the  fields we need to define appropriate adjoints. Doing this requires a  new dual as
the norm of ELKO  under the Dirac dual identically vanishes. Once the adjoint is defined, 
the Feynman-Dyson propagator is readily calculated. It
reveals two unexpected facts: 
\begin{itemize}
\item the \textit{mass dimension} of 
$\mathfrak{f}(x)$ is
\textit{one} and statistics fermionic 
\item 
 the \textit{mass dimension} of 
$\mathfrak{b}(x)$ is
\textit{one} and statistics bosonic.
\end{itemize}
 The mismatch of the mass dimensionality and statistics of the new fields with that of Dirac forbids these to enter the SM doublets. Thus, \textit{these naturally become first-principle dark matter candidates}.
Furthermore, self/anti-self conjugacy of the ELKO is lost under
$\lambda(p^\mu) \to \lambda^\prime(p^\mu) = e^{i\delta}  \times \lambda(p^\mu)$ with $\delta\in \Re$ and
$\rho(p^\mu) \to \rho^\prime(p^\mu) = e^{i\overline\delta}  \times \rho(p^\mu)$ with $\overline\delta\in \Re$.

Under the Dirac dual, the ELKO  norm identically vanishes
\begin{equation}
\overline{\lambda}^{S/A}_\alpha(p^\mu) {\lambda}^{S/A}_\alpha(p^\mu)
=0 = \overline{\rho}^{S/A}_\alpha(p^\mu) {\rho}^{S/A}_\alpha(p^\mu),\qquad \alpha = +,-
\end{equation}
with the Dirac dual defined as usual
\begin{equation}
\overline{\lambda}_\alpha(p^\mu) = \left[\lambda_\alpha(p^\mu)\right]^\dagger \gamma_0,\quad
\overline{\rho}_\alpha(p^\mu) = \left[\rho_\alpha(p^\mu)\right]^\dagger \gamma_0.
\end{equation}

For pedagogic reasons, we now bifurcate our discussion to 
$\lambda(p^\mu)$ and $\rho(p^\mu)$ sectors.

\subsection{The $\lambda(p^\mu)$ sector -- the new dual} \label{new_dual}

This circumstance necessitates the introduction of a new spinorial dual. The details can be found in a recent reference~\cite{Ahluwalia:2019etz}. Here we provide a self contained outline. 

The introduction of the new dual is implemented in two steps. First, we note that the introduction of (as definitions)
\begin{align}
& \widetilde{\lambda}_{+}^S(p^\mu) 
=  - i \left[ \lambda^S_{-} (p^\mu) \right]^\dagger
 \gamma_0,\quad
  \widetilde{\lambda}_{-}^S(p^\mu) 
=  + i \left[ \lambda^S_{+} (p^\mu) \right]^\dagger
 \gamma_0, \label{eq:33}\\
 & \widetilde{\lambda}_{+}^A(p^\mu) 
=  - i \left[ \lambda^A_{-} (p^\mu) \right]^\dagger
 \gamma_0,\quad
  \widetilde{\lambda}_{-}^A(p^\mu) 
=  + i \left[ \lambda^A_{+} (p^\mu) \right]^\dagger
 \gamma_0 \label{eq:34}
\end{align}
yields the following orthonormality relations
\begin{align}
& \widetilde\lambda^S_\alpha(p^\mu) \lambda^S_{\alpha\prime}
(p^\mu) =+ 2 m \delta_{\alpha\alpha\prime},\\
&\widetilde\lambda^A_\alpha(p^\mu) \lambda^A_{\alpha\prime}
(p^\mu) = -2 m \delta_{\alpha\alpha\prime},\\
&\widetilde\lambda^S_\alpha(p^\mu) \lambda^A_{\alpha\prime}(p^\mu) = 0 =
\widetilde\lambda^A_\alpha(p^\mu) \lambda^S_{\alpha\prime}(p^\mu)
\end{align}
along with the associated spin sums
\begin{align}
& \sum_\alpha \lambda^S_\alpha(p^\mu) \widetilde\lambda^S_\alpha(p^\mu) = +m \left[\I+\mathcal{G}(p^\mu)
\right], \\
& \sum_\alpha \lambda^A_\alpha(p^\mu) \widetilde\lambda^A_\alpha(p^\mu) = -m \left[\I-\mathcal{G}(p^\mu)
\right].
\end{align}
Where $\mathcal{G}(p^\mu)$ is defined as 
\begin{equation}
\mathcal{G}(p^\mu) =
\left(
\begin{array}{cccc}
0 & 0 &  0 &-i e^{-i\phi}\\
0 & 0 &i e^{i\phi} & 0\\
0 & -i e^{-i\phi} &0 &0 \\
i e^{i\phi} & 0 & 0 &0
\end{array}
\right).
\end{equation}
The obtained spin sums lead to the completeness relation
\begin{align}
\frac{1}{2m}\sum_{\alpha}\bigg[\lambda^S_\alpha(p^\mu)
\widetilde{\lambda}^S_\alpha(p^\mu) -
\lambda^A_\alpha(p^\mu)
\widetilde{\lambda}^A_\alpha(p^\mu)\bigg] = \I. \label{eq:sstilde}
\end{align}
The presence of $\mathcal{G}(p^\mu)$ introduces a preferred direction and manifests itself as a violation of locality~\cite{Ahluwalia:2010zn}. Both of these aspects can be attended to by a re-defintion of the ELKO duals
\begin{align}
&\widetilde{\lambda}^S_\alpha(p^\mu) 
\to \gdualn{\lambda}^S_\alpha(p^\mu) \stackrel{\mathrm{def}}{=}\widetilde{\lambda}^S_\alpha(p^\mu) \mathcal{A}
&\widetilde{\lambda}^A_\alpha(p^\mu) \to \gdualn{\lambda}^A_\alpha(p^\mu) \stackrel{\mathrm{def}}{=} \widetilde{\lambda}^A_\alpha(p^\mu) \mathcal{B}\label{eq:redef}
\end{align}
where $\mathcal{A}$ and $\mathcal{B}$ are constrained  to have the 
following properties:
\begin{itemize}
\item the $\lambda^S_\alpha(p^\mu)$ must be  eigenspinors of $\mathcal{A}$ with eigenvalue unity, 
$
\mathcal{A} \lambda^S_\alpha(p^\mu) = \lambda^S_\alpha(p^\mu),
$
\item the $\lambda^A_\alpha(p^\mu)$ must be  eigenspinors of $\mathcal{B}$ with eigenvalue unity, $
\mathcal{B} \lambda^A_\alpha(p^\mu) = \lambda^A_\alpha(p^\mu)$,
\item
and in addition, the following must hold
\begin{equation}
\widetilde{\lambda}^S_\alpha(p^\mu)\mathcal{A} \lambda^A_{\alpha^\prime}(p^\mu)=0,\quad \widetilde{\lambda}^A_\alpha(p^\mu)\mathcal{B} \lambda^S_{\alpha^\prime}(p^\mu) = 0.
\end{equation}
\end{itemize}
If $\mathcal{A}$ and $\mathcal{B}$ exist,  then with the new dual, the orthonormality relations  remain unaltered in 
 form
\begin{align}
& \gdualn\lambda^S_\alpha(p^\mu) \lambda^S_{\alpha^\prime}(p^\mu)
 = + 2 m \delta_{\alpha\alpha^\prime},\label{eq:zimpokJ9an}\\
&  \gdualn\lambda^A_\alpha(p^\mu) \lambda^A_{\alpha^\prime}(p^\mu)
 = - 2 m \delta_{\alpha\alpha^\prime}, \label{eq:zimpokJ9bn} \\
 &  \gdualn\lambda^S_\alpha(p^\mu) \lambda^A_{\alpha^\prime}(p^\mu) = 0 =
 \gdualn\lambda^A_\alpha(p^\mu) \lambda^S_{\alpha^\prime}(p^\mu),\label{eq:zimpokJ9cn}
\end{align}
while
the same very  re-definitions alter the spin sums to
\begin{align}
\sum_{\alpha} \lambda^S_\alpha(p^\mu) \gdualn{\lambda}^S_\alpha(p^\mu) & = +
m \big[\I + \mathcal{G}(p^\mu) \big] \mathcal{A}, \label{eq:sss-new}\\
\sum_{\alpha} \lambda^A_\alpha(p^\mu) \gdualn{\lambda}^A_\alpha(p^\mu) & =
- m \big[\I - \mathcal{G}(p^\mu)\big] \mathcal{B}. \label{eq:ssa-new}
\end{align}
Following the intricate arguments found in ~\cite[Section 14.7]{Ahluwalia:2019etz}, the $\mathcal{A}$ and $\mathcal{B}$ are found to be 
\begin{equation}
\mathcal{A} = 2 \left[\I + \tau\mathcal{G}(p^\mu)\right]^{-1},\quad
\mathcal{B} = 2 \left[\I - \tau\mathcal{G}(p^\mu)\right]^{-1}
\end{equation}
with $\tau\in \mathbb{R}$, and the limit $\tau\to 1$ taken as
 implicit.\footnote{We note that the introduction of $\mathcal{A}$ and $\mathcal{B}$ operators are also supported by the Clifford algebra construction. The $\tau$ deformation, as well as the limit 
 $\tau\rightarrow 1$, is further explored in~\ref{Appzero}. As discussed there, the inversion of $\I + \mathcal{G}$, performed by the introduced operator $\mathcal{A}$, are justified by the requirement that both operators are elements of a certain automorphism group (see the contextualisation in~\ref{Appzero}). Within this context, the $\tau$ deformation appears as a necessary procedure.}
This leads to spin sums
\begin{align}
& \sum_\alpha\lambda^S_\alpha(p^\mu) \gdualn{\lambda}^S_\alpha(p^\mu) = + 2 m \I,\\
&\sum_\alpha \lambda^A_\alpha(p^\mu) \gdualn{\lambda}^A_\alpha(p^\mu) = - 2 m \I
\end{align}
which are free from the problems that motivated us to introduce $\tau$
deformation. These spin sums lead to the completeness relation
\begin{equation}
\frac{1}{4 m}\left[ \lambda^S_\alpha(p^\mu) \gdualn{\lambda}^S_\alpha(p^\mu) - \lambda^A_\alpha(p^\mu) \gdualn{\lambda}^A_\alpha(p^\mu)\right]  = \I.
\end{equation}

\subsection{A hint for mass dimension one for $\mathfrak{f}(x)$}

For the $\lambda(p^\mu)$ sector $\left(\gamma_\mu p^\mu \mp m\I\right) \lambda(p^\mu) \ne 0$, instead we have (refer to~\cite{Dvoeglazov:1995eg,Dvoeglazov:1995kn} and~\cite[Chapter 9]{Ahluwalia:2019etz})
\begin{align}
&\gamma_\mu p^\mu \lambda^S_+(p^\mu) = im \lambda^S_-(p^\mu),\quad
\gamma_\mu p^\mu \lambda^S_-(p^\mu) = -  im \lambda^S_+(p^\mu),\\
& \gamma_\mu p^\mu \lambda^A_+(p^\mu) =-  im \lambda^A_-(p^\mu),\quad
\gamma_\mu p^\mu \lambda^A_-(p^\mu) = im \lambda^A_+(p^\mu).
\end{align}
Together, they imply
\begin{equation}
\left(p_\mu p^\mu \I - m^2 \I\right)\lambda^{S,A}_{\alpha}(p^\mu) =0,
\end{equation}
independent of self or anti-self conjugacy of $\lambda(p^\mu)$ under
$\mathcal{C}$, the charge conjugation operator. 
This hints that the field $\mathfrak{f}(x)$, with $\lambda(p^\mu)$ as its expansion coefficients, may be of mass dimension one.

\subsection{A hint for fermionic statistics for $\mathfrak{f}(x)$}  

To explore the statistics of  $\mathfrak{f}(x)$, we define its adjoint\footnote{Since we are dealing with a single field, the slight change in notation is entirely permissible, and is introduced for the ease of certain arguments.} 
\begin{equation}
\gdualn{\mathfrak{f}}(x) =
\int
\frac{d^3 p}{(2\pi)^3}
\frac{1}{\sqrt{2 m E(\p)}}
\sum_\alpha\left[
a^\dagger_\alpha(\p) \gdualn{\lambda}^S_\alpha(\p) e^{i p\cdot x}
+ b_\alpha(\p) \gdualn{\lambda}^A_\alpha(\p) e^{- i p\cdot x}
\right]\label{eq:df}
\end{equation}
without yet identifying the creation and annihilation operators with a specific statistics. Now, consider two space-like separated events, $x$ and $x^\prime$.  
For these, the time-ordering of events is not preserved.  To avoid causal paradoxes the very same process that was interpreted as a particle propagation for the set of observers with $x^\prime_0 > x_0$ must be re-interpreted as the propagation of an
antiparticle for observers for whom $x^\prime_0 < x_0$. We thus recognize the existence of two set of inertial observers, ones in which 
$x^\prime_0 > x_0$
and ones in which the opposite holds, $x_0 > x^\prime_0$. We call these sets of inertial observers as $\mathcal{O}$ and $\mathcal{O^\prime}$, respectively. 

In $\mathcal{O}$  we calculate the amplitude for a particle to propagate from $x$ to $x^\prime$, and in $\mathcal{O}^\prime$   
we calculate the amplitude for an antiparticle to propagate from $x^\prime$ to $x$. Causality requires that these two amplitudes may differ at most by a phase factor~\cite{Ahluwalia:2015vea}:
\begin{equation}
\mbox{Amp}(x\to x^\prime, \mbox{particle})\vert_\mathcal{O} =
e^{i \theta} \mbox{Amp}(x^\prime\to x, \mbox{antiparticle})\vert_{\mathcal{O}^\prime},\quad \theta \in \R.\label{eq:causality}
\end{equation}
We call $e^{i\theta}$ the  \textit{causality phase}.
Using the definitions of $\mathfrak{f}(x)$ and $\gdualn{\mathfrak{f}}(x)$, we readily obtain the needed amplitudes as follows
\begin{align}
\mbox{Amp}(x\to x^\prime, \mbox{particle}) \vert_\mathcal{O}
&= \langle~~\vert\mathfrak{f}(x^\prime)\gdualn{\mathfrak{f}}(x)\vert~~\rangle_{\mathcal{O}}  \nonumber
\\
&=
\int\frac{d^3 p}{(2\pi)^3 }
\left(\frac{1}{2 m E(\p)} \right)
e^{-i p\cdot(x^\prime -x)}
\sum_\alpha
 \lambda^S_\alpha(p^\mu) \gdualn{\lambda}^S_\alpha(p^\mu) 
 \end{align}
 and
 \begin{align}
  \mbox{Amp}( x^\prime\to x, \mbox{antiparticle})\vert_{\mathcal{O}^\prime} = \langle~~\vert\gdualn{\mathfrak{f}}(x)\mathfrak{f}(x^\prime)\vert~~\rangle\vert_{{\mathcal{O}^\prime}}
 &  =
  \left[\langle~~\vert\gdualn{\mathfrak{f}}(x)\mathfrak{f}(x^\prime)\vert~~\rangle\vert_{{\mathcal{O}}}\right]_{(x-x^\prime)\to(x^\prime-x)}\nonumber \\
 =
\int\frac{d^3 p}{(2\pi)^3 }
\left(\frac{1}{2 m E(\p)} \right)
e^{-i p\cdot(x^\prime -x)}
\sum_\alpha
 \lambda^A_\alpha(p^\mu) \gdualn{\lambda}^A_\alpha(p^\mu).
\end{align}
Use of ELKO spin sums in the above amplitudes leads to a remarkable results that $e^{- i \theta} = -1$. 
A direct calculation shows that
\begin{equation}
\mbox{Amp}(x^\prime\to x, \mbox{antiparticle})\vert_{\mathcal{O}^\prime} =\mbox{Amp}(x^\prime\to x, \mbox{antiparticle})\vert_{\mathcal{O}}.
\end{equation}
Substituting this result in (\ref{eq:causality}), and combing with 
$e^{- i \theta} = -1$, yields
\begin{equation}
\mbox{Amp}(x\to x^\prime, \mbox{particle})\vert_\mathcal{O} =
- \mbox{Amp}(x^\prime\to x, \mbox{antiparticle})\vert_{\mathcal{O}}.
\end{equation}
Translated in terms of the field and its adjoint, it leads to the following anticommutator
\begin{equation}\{
\gdualn{\mathfrak{f}}(x) , \mathfrak{f}(x^\prime)\}
=0.
\end{equation}
It strongly hints that $\mathfrak{f}(x)$ is fermionic. We thus tentatively assign fermionic statistics to the creation and annihilation operators used in the definition of the new field. Soon we will see that this assumption leads to an internally consistent local field theory.

\subsection{Departure from the ``second quantization''  formalism}

Now we refer the reader to monograph~\cite{Ahluwalia:2019etz} where the  events $x$ and $x^\prime$ are no longer confined to space-like separations. Calculations similar to those we have just performed yield the following additional results~\cite[Section 15.3-15.5]{Ahluwalia:2019etz}:
\begin{itemize}
\item Propagator is found to be
\begin{align}
S_{FD}(x^\prime-x) & = 
-\frac{i}{2} \langle~\vert
\mathfrak{T}[\mathfrak{f}(x^\prime)\mathfrak{f}(x)]\vert~\rangle \\  & =
\int\frac{d^4p}{(2\pi)^4}e^{-i p\cdot(x^\prime - x)}\frac{\I}{p^\mu p_\mu - m^2 + i\epsilon}.\label{eq:FDP}
\end{align}
\item
Mass dimension of the field is one
\begin{equation}
\mathcal{D}_\mathfrak{f} = 1.
\end{equation}
\item
Free field Lagrangian density is
\begin{equation}
\mathcal{L}_0(x) = \frac{1}{2}\left(\partial^\mu {\gdualn{\mathfrak{f}}} (x)
\partial_\mu\mathfrak{f}(x)
 - m^2 {\gdualn{\mathfrak{f}}}(x) \mathfrak{f}(x) \right). \label{eq:Lagrangian}
\end{equation}
\item
Dimension four quartic self interaction is natural
\begin{equation}
\left({\gdualn{\mathfrak{f}}}(x) \mathfrak{f}(x)\right)^2.
\end{equation}  
\item 
Locality structure of the field is canonical, and is given by
\begin{equation}
\left\{
\mathfrak{f}(t,\x),\mathfrak{p}(t,\x^\prime)\right\} = 
i \delta^3(\x-\x^\prime), \quad\left\{
\mathfrak{f}(t,\x),\mathfrak\mathfrak{f}(t,\x^\prime)\right\} = 0,\quad
\left\{
\mathfrak{p}(t,\x),\mathfrak{p}(t,\x^\prime)\right\} = 0 .
\end{equation} 
Where, with $\mathcal{L}_0(x)$ given by (\ref{eq:Lagrangian}), the momentum conjugate to $\mathfrak{f}(x)$ is defined as
\begin{equation}
\mathfrak{p}(x) =
\frac{\partial \mathcal{L}_0(x)}
{\partial       \dot{\mathfrak{f}}   (x)   } = \frac{1}{2} \frac{\partial}{\partial t}
\gdualn{\mathfrak{f}}(x) .
\end{equation}
\end{itemize}
These remarkable results follow from \emph{first} defining the field $\mathfrak{f}(x)$ with its expansion coefficients  (ELKO) chosen with appropriate locality phases, followed by calculating the Feynman-Dyson propagator (as in (\ref{eq:FDP})). It is only then, as argued in detail in~\cite{Ahluwalia:2019etz}, does the Lagrangian density appear on the scene. That allows us to define $\mathfrak{p}(x)$ and to calculate the locality anti-commutators. The mass dimensionality of the field is then an inevitable, and an unexpected consequence. In the formalism that we have developed, the Lagrangian density is not assumed but derived. The usual `second quantization method' could have never yielded the results we have obtained. 
 
 In all this the new dual and the associated adjoint is of inseparable significance.

\subsection{The $\rho(p^\mu)$ sector -- farewell to the spin statistics theorem.}

\begin{quote}
``\textit{There must be such boson variables connected with electrons. Their physical significance is a subject for further investigation.''
\hfill Dirac~\cite{Dirac:1974ke}}
\end{quote}

 We now begin with $\rho(p^\mu)$ defined in (\ref{eq:rho}) and 
 $\mathfrak{b}(x)$ introduced in (\ref{eq:b}). We choose the duals of $\rho(p^\mu)$ as
 \begin{align}
&{ \widetilde{\rho}}^S_{-}(p^\mu) = - i \,{\overline\rho}^S_+(p^\mu),\quad
{ \widetilde{\rho}}^S_{+}(p^\mu) = + i \,{\overline\rho}^S_-(p^\mu),\\
&{ \widetilde{\rho}}^A_{-}(p^\mu) = + i \,{\overline\rho}^A_+(p^\mu),\quad
{ \widetilde{\rho}}^A_{+}(p^\mu) = - i \,{\overline\rho}^A_-(p^\mu).
 \end{align}
We then implement a counterpart of $\tau$ deformation. After similar calculations as for the $\lambda$ sector, the result for the orthonormality relations and spin sums alter to
\begin{align}
& \gdualn\rho^S_\alpha(p^\mu) \rho^S_{\alpha^\prime}(p^\mu)
 = + 2 m \delta_{\alpha\alpha^\prime},\label{eq:zimpokJ9an2}\\
&  \gdualn\rho^A_\alpha(p^\mu) \rho^A_{\alpha^\prime}(p^\mu)
 = + 2 m \delta_{\alpha\alpha^\prime}, \label{eq:zimpokJ9bn2} \\
 &  \gdualn\rho^S_\alpha(p^\mu) \rho^A_{\alpha^\prime}(p^\mu) = 0 =
 \gdualn\rho^A_\alpha(p^\mu) \rho^S_{\alpha^\prime}(p^\mu),\label{eq:zimpokJ9cn2}
\end{align}
and
\begin{align}
&\sum_\alpha \rho^S_\alpha(p^\mu) \gdualn{\rho}^S_\alpha(p^\mu) = + 2 m \I,\\
&\sum_\alpha \rho^A_\alpha(p^\mu) \gdualn{\rho}^A_\alpha(p^\mu) = + 2 m \I.
\end{align}
As a result, the causality phase $e^{i \theta}$ in the counterpart of (\ref{eq:causality}) changes from   $-1 \to +1$. In consequence, the
$\mathfrak{b}(x)$ carries bosonic statistics. This is despite it being a spin half field. Its mass dimension is one in complete parallel to that of $\mathfrak{f}(x)$. The only other reference we know of spin half bosons ``appearing automatically'' along with spin half fermions is in Dirac's \emph{Spinors in Hilbert Space}~\cite{Dirac:1974ke}.

\subsection{The $\mathfrak{f}(x)$-$\mathfrak{b}(x)$ system and vanishing zero point energy}

The $\mathfrak{f}(x)$ and $\mathfrak{b}(x)$ make opposite contributions to the zero-point field energy
\begin{align}
&H_0^\mathfrak{f} = -\, 4 \,\times \;\frac{1}{(2\pi)^3}\int d^3 x\int d^3p\;\frac{1}{2} E(\p), \\
&H_0^\mathfrak{b} = +\, 4 \,\times \;\frac{1}{(2\pi)^3}\int d^3 x\int d^3p\;\frac{1}{2} E(\p) .
\end{align}
The net contribution thus identically vanishes. The fact that the fermion and boson mass is the same is not an assumption. Since $\lambda(p^\mu)$ and $\rho(p^\mu)$ are constructed from the same set of Weyl spinors,  the $\mathfrak{f}(x)$ and $\mathfrak{b}(x)$  must have the same mass.

\subsection{Lee-Wick theorem and the $\mathfrak{f}(x)$ and 
$\mathfrak{b}(x)$ fields}

The Lee-Wick no go theorem is manifestly evaded, as the theory presented here is local. The often quoted objection to the existence of unusual Wigner classes is thus challenged by the construct we are investigating. Using parity operator $\mathcal{P}= m^{-1} \gamma_\mu p^\mu$, and the time reversal operator $\mathcal{T}= i \gamma^{5}\mathcal{C}$, where
\begin{equation}
\gamma^{5} = \frac{i}{4!}\epsilon^{\mu\nu\lambda\sigma}\gamma_\mu\gamma_\nu\gamma_\lambda\gamma_\sigma =\left(
\begin{array}{cc}
\I &\0 \\
\0 & -\I
\end{array}
\right)
\end{equation}
with $\epsilon^{\mu\nu\lambda\sigma}$ defined as the completely antisymmetric $4th$ rank tensor with $\epsilon^{0123} = +1$, we find that for the $\lambda$ and $\rho$ sector spinors
\begin{align}
&\lambda\mbox{-sector}: \, (\mathcal{C}\mathcal{P}\mathcal{T})^2 = +\,\I, \quad  \left\{\mathcal{C},\mathcal{P}\right\}=0, \nonumber\\
&\rho\mbox{-sector}:  \, (\mathcal{C}\mathcal{P}\mathcal{T})^2 = -\,\I, \quad \left\{\mathcal{C},\mathcal{P}\right\}=0
\end{align}
hinting that the new fields belong to the unusual Wigner classes.

\section{Magic of square roots of $4\times 4$ identity matrix}

We take our reader back to 1928 when Dirac published  the spin half wave equation now known after his name. In the quantum field theoretic version its solutions, modulo certain phases, serve as expansion coefficients of a local fermionic field. This field serves as the kinematic foundation for all SM fermions. This we have already said. But it resets the stage. It is our contention here that Dirac's approach can still inspire us to construct a counterpart of his revolution, but now for dark matter. The starting point of Dirac's revolution was taking the square root of the dispersion relation:
\begin{equation}
p_\mu p^\mu\I = m^2\I.\label{eq:disperion}
\end{equation}
The `square root' obtained for the left-hand side of~(\ref{eq:disperion}) was $\gamma_\mu p^\mu$, while for the right-hand side it was $m \I$. The well-known $4\times 4$ matrices $\gamma_\mu$ satisfy $\{\gamma_\mu,\gamma_\nu \} = 2 \eta_{\mu\nu}\I$. The observations that opens the new possibility is two fold. 

Firstly, the $\gamma_\mu p^\mu$ as the `square root' of $p_\mu p^\mu\I$ is not unique~\cite{Vir_Ahluwalia_2020,Ahluwalia_2020}. There exists a class of matrices $a_\mu$ such that $\{a_\mu,a_\nu\}=2 \eta_{\mu\nu}\I$ but its concrete form differs -- in the \textit{same} Weyl realization --  from that of $\gamma_\mu$ by a global phase, and more importantly, by a relative sign between the off-diagonal $2\times 2$ blocks.\footnote{To avoid confusion while $a_{\mu}$ are formally related to $\gamma_{\mu}$ by a similarity transformation, this is not be taken as  getting out of the Weyl representation. That is, $a_\mu$ and $\gamma_\mu$ are in the same basis.}

Secondly, the `square root' that Dirac chose for the right-hand side of~(\ref{eq:disperion}), similarly, lacks uniqueness. Specifically, the choice of $\I$ as the square root of $\I$ is only one of the sixteen linearly independent possibilities $\Gamma_\ell$ (See, ~\cite[p. 71]{Schweber:1961zz}):
\begin{equation}
\begin{array}{lllllll}
\Gamma_{\ell}: & \I, &\empty &\empty &\empty &\empty &\empty \\
\empty & i\gamma_1, &  i \gamma_2, & i\gamma_3, & \gamma_0, &\empty&\empty \\
\empty &  i \gamma_2\gamma_3, & i\gamma_3\gamma_1, & i \gamma_1\gamma_2, &\gamma_0\gamma_1, &\gamma_0\gamma_2, & \gamma_0\gamma_3, \\
\empty &
i\gamma_0\gamma_2\gamma_3, &
i\gamma_0\gamma_1\gamma_3, &
i\gamma_0\gamma_1\gamma_2, &
\gamma_1\gamma_2\gamma_3, &\empty \\
\empty &i\gamma_{0}\gamma_{1}\gamma_{2}\gamma_{3} &\empty & \empty &\empty &\empty &\empty
\end{array}
\end{equation}
with $\Gamma_1$ being the first entry in the above array and $\Gamma_{16}$ being the last -- the $\ell$ assignment is in consecutive order.  

Two detailed constructions that study `square roots'  $\Gamma_7$ and $\Gamma_4$ have recently been published with the noteworthy result that the former supports a mass dimension one \textit{fermionic} quantum field of spin half, while the latter supports a mass dimension three halve \textit{bosonic} quantum field of spin half~\cite{Ahluwalia:2020miz,Ahluwalia:2020jkw}. Furthermore, both fields are local. None of these two fields can enter the doublets of the SM either because of the mismatch of mass dimensions, or differing statistics. Thus they become natural dark matter candidates.
\section{The S-Matrix for mass dimension one fermions}\label{dyn}


In quantum mechanics and field theory, transition probability is computed using Hermitian conjugation to ensure that the probability is positive-definite. But for mass dimension one fermions, the standard computational formalism is problematic. As demonstrated in~\cite{Lee:2015sqj}, the optical theorem is violated at one-loop. It should be noted that this result holds with and without the application of $\tau$-deformation because in both cases, the difficulties can be attributed to the fact that the theory of ELKO and mass dimension one fermion is non-Hermitian. Here, we will develop a new formalism by introducing a new adjoint $\ddag$ to compute transition probabilities and observables. Subsequently, we derive new consistency conditions and show that up to one-loop, these conditions are satisfied. At this moment, the formalism is developed for mass dimension one fermions but it can be easily extended to spin half bosons.

In analogy to the Dirac dual $\overline{\psi}(\p)=\psi^{\dag}(\p)\gamma_{0}$, we rewrite the ELKO dual given in~(\ref{eq:33}) and~(\ref{eq:34}) as
\begin{equation}
\widetilde{\lambda}^{I}_{\alpha}(\p)\equiv\left[\lambda^{I}_{\alpha}(\p)\right]^{\ddag}\gamma_{0},\quad I=S,A
\end{equation}
where
\begin{equation}
\left[\lambda^{I}_{\alpha}(\p)\right]^{\ddag}\equiv-i\alpha\left[\lambda^{I}_{-\alpha}(\p)\right]^{\dag}\label{eq:d0}
\end{equation}
so the action of $\ddag$ is to Hermitian conjugate the spinor, multiply it by a phase and flip the helicity label. Here, it is important to note that $\ddag$ only acts on the spinors and not on the creation and annihilation operators. The $\tau$ deformation on the spinorial dual will be performed when we compute the spin sums to obtain physical observables. In the subsequent section, we will explicitly demonstrate how to perform such calculations.

From~(\ref{eq:d0}), we obtain
\begin{equation}
\left[\lambda^{I}_{\alpha}(\p)\right]^{\dag}=-i\alpha\left[\lambda^{I}_{-\alpha}(\p)\right]^{\ddag}.\label{eq:d01}
\end{equation}
Apply $\ddag$ to~(\ref{eq:d01}) yields
\begin{align}
\left[\lambda^{I}_{\alpha}(\p)^{\dag}\right]^{\ddag}&=(-i\alpha)^{\ddag}\left[\lambda^{I}_{-\alpha}(\p)^{\ddag}\right]^{\ddag}\nonumber\\
&=i\alpha\left[\lambda^{I}_{-\alpha}(\p)^{\ddag}\right]^{\ddag} \label{eq:d02}
\end{align}
where on the second line of~(\ref{eq:d02}), we have used $(-i\alpha)^{\ddag}=i\alpha$. By demanding two successive applications of $\ddag$ to be the identity 
\begin{equation}
\left[\lambda^{I}_{\alpha}(\p)\right]^{\ddag\ddag}\equiv\lambda^{I}_{\alpha}(\p),\label{eq:d03}
\end{equation}
we obtain
\begin{equation}
\left[\lambda^{I}_{\alpha}(\p)^{\dag}\right]^{\ddag}=i\alpha\lambda^{I}_{-\alpha}(\p). \label{eq:d04}
\end{equation}
Next, we compute
\begin{align}
\left[\lambda^{I}_{\alpha}(\p)^{\ddag}\lambda^{J}_{\alpha'}(\p')\right]^{\ddag}
&=\left\{\sum^{4}_{\ell=1}\left[-i\alpha\lambda^{I}_{-\alpha}(\p)\right]^{\dag}_{\ell}\left[\lambda^{J}_{\alpha'}(\p')\right]_{\ell}\right\}^{\ddag}\nonumber\\
&=i\alpha\sum^{4}_{\ell=1}\left[\lambda^{I}_{-\alpha}(\p)^{\dag}\right]^{\ddag}_{\ell}\left[\lambda^{J}_{\alpha'}(\p')\right]^{\ddag}_{\ell}\nonumber\\
&=i\alpha\left[\lambda^{J}_{\alpha'}(\p')\right]^{\ddag}\left[\lambda^{I}_{-\alpha}(\p)^{\dag}\right]^{\ddag}.
\label{eq:d05}
\end{align}
Substituting~(\ref{eq:d04}) into~(\ref{eq:d05}), we obtain
\begin{equation}
[\lambda^{I}_{\alpha}(\p)^{\ddag}\lambda^{J}_{\alpha'}(\p')]^{\ddag}=\lambda^{J}_{\alpha'}(\p')^{\ddag}\lambda^{I}_{\alpha}(\p).
\label{eq:d06}
\end{equation}
Performing the same steps as~(\ref{eq:d05}), we also obtain the identity
\begin{equation}
\left[\widetilde{\lambda}^{I}_{\alpha}(\p)\lambda^{J}_{\alpha'}(\p')\right]^{\ddag}=\widetilde{\lambda}^{J}_{\alpha'}(\p')\lambda^{I}_{\alpha}(\p).
\end{equation}
We thus establish that the definition of the ELKO dual given in sec.~\ref{new_dual} is consistent with the definition of $\ddagger$.

\subsection{Computing observables using $\ddag$}

We now propose a new prescription for computing observables. Let $S_{ba}$ be the $S$-matrix for a generic process $a\rightarrow b$ involving mass dimension one fermions. We normalize the $S$-matrix as
\begin{align}
S_{ba}\equiv\delta(b-a)+ (2\pi)^{4}iM_{ba}\delta^{4}(p_{b}-p_{a}). \label{eq:s}
\end{align}

In a finite volume $V$ where the duration of interaction is $T$, we propose the new transition probability $P(a\rightarrow b)$ to be~\footnote{The use of $\ddag$ to compute observables is in analogy to its Dirac counterparts. In the Yukawa interaction $g\overline{\psi}\psi\phi$, the scattering amplitude for $\phi\rightarrow\overline{\psi}\psi$ is proportional to $\overline{u}v$ so the transition probability is of the form $(\overline{u}v)^{\dag}(\overline{u}v)$. In particular, the action of Hermitian conjugation takes $(\overline{u}v)^{\dag}$ to $\overline{v}u$. For mass dimension one fermions, given an amplitude of the form $\widetilde{\lambda}^{I}\lambda^{J}$, we want the transition probability to be proportional to $(\widetilde{\lambda}^{J}\lambda^{I})(\widetilde{\lambda}^{I}\lambda^{J})$. However, $(\widetilde{\lambda}^{J}\lambda^{I})^{\dag}\neq\widetilde{\lambda}^{I}\lambda^{J}$. For this reason, we have introduced $\ddag$.} 
\begin{equation}
P(a\rightarrow b)\equiv VT (2\pi)^{4}\delta^{4}(p_{b}-p_{a})(\wp_{ba} M^{\ddag}_{ba}M_{ba}).
\label{eq:pab}
\end{equation}
Comparing with the standard transition probability, there is one crucial difference, namely the Hermitian conjugation is replaced by $\ddag$. As a result, $M^{\ddag}_{ba}M_{ba}$ is not guaranteed to be positive-definite. Therefore, a phase $\wp_{ba}$ must be introduced to ensure that $P(a\rightarrow b)\geq0$. The value of the phase may depend on the interactions but it should not alter the absolute magnitude of $P(a\rightarrow b)$ so $|\wp_{ba}|=1$. Below, we will discuss how to determine $\wp_{ba}$. Apart from the use of $\ddag$ and the phase $\wp_{ba}$, equation~(\ref{eq:pab}) is identical to the standard transition probability. Therefore, the two-body cross-section for $12\rightarrow34$ is~\cite{Peskin:1995ev}
\begin{align}
\sigma(12\rightarrow34)
=&\frac{1}{4E_{1}E_{2}u_{12}}\int\frac{d^{3}p_{3}}{(2\pi)^{3}(2E_{3})}\int\frac{d^{3}p_{4}}{(2\pi)^{3}(2E_{4})}
\left[\wp_{(34)(12)}M^{\ddag}_{(34)(12)}M_{(34)(12)}\right]\nonumber\\
&\times(2\pi)^{4}\delta^{4}(p_{3}+p_{4}-p_{1}-p_{2}), \nonumber\\
u_{12}=&\frac{1}{E_{1}E_{2}}\sqrt{(p_{1}\cdot p_{2})^{2}-m^{2}_{1}m^{2}_{2}}.
\end{align}
In the center of mass frame, the differential cross-section is
\begin{equation}
\frac{d\sigma_{\mbox{\tiny{CM}}}}{d\Omega}=\frac{1}{64\pi^{2}}\frac{1}{E_{1}E_{2}u_{12}}\frac{|\p_{3}|}{E_{\mbox{\tiny{CM}}}}
\left[\wp_{(34)(12)}M^{\ddag}_{(34)(12)}M_{(34)(12)}\right].
\end{equation}
where $E_{\tiny{\mbox{CM}}}=E_{1}+E_{2}$ and $d\Omega=\sin^{2}\theta_{3}d\theta_{3}d\phi_{3}$ is the solid angle.

Let us now consider the interactions. Lorentz invariance demands the interacting potentials to be of the form $\gdualn{\mathfrak{f}}(x)\mathfrak{f}(x)$ and $\gdualn{\mathfrak{f}}(x)\mathcal{O}(x)\mathfrak{f}(x)$, where $\mathcal{O}(x)$ may be some differential operator or matrices. Here, we will focus on the former and leave the latter for future investigation. We consider the following interactions
\begin{eqnarray}
&& V_{\phi}(t)=\frac{g_{\phi}}{2}\int d^{3}x[\gdualn{\mathfrak{f}}(x)\mathfrak{f}(x)\phi^{2}(x)],\\
&& V_{\mathfrak{f}}(t)=\frac{g_{\mathfrak{f}}}{2}\int d^{3}x[\gdualn{\mathfrak{f}}(x)\mathfrak{f}(x)]^{2}.\label{eq:ip}
\end{eqnarray}
By the mass dimension of $\mathfrak{f}(x)$ and $\phi(x)$, these potentials are renormalizable. 
To distinguish the particle species, we take $m_{\mathfrak{f}}$ and $m_{\phi}$ to be the fermionic and bosonic mass. We note that for $V_{\mathfrak{f}}(t)$ and $V_{\phi}(t)$, whenever mass dimension one fermions appear as external states, the scattering amplitudes are products of $\gdualn{\lambda}^{I}_{\alpha}(\p)\lambda^{J}_{\alpha'}(\p')$. For the evaluation of the inner-products, we use the identities (in the limit $\tau\rightarrow1$)
\begin{align}
&\mathcal{A}\lambda^{S}_{\alpha}(\p)=\lambda^{S}_{\alpha}(\p),\quad \lambda^{S}_{\alpha}(\p)^{\dag}\mathcal{A}=\lambda^{S}_{\alpha}(\p)^{\dag}.\\
&\mathcal{B}\lambda^{A}_{\alpha}(\p)=\lambda^{A}_{\alpha}(\p),\quad \lambda^{A}_{\alpha}(\p)^{\dag}\mathcal{B}=\lambda^{A}_{\alpha}(\p)^{\dag}.
\end{align}
Additionally, using the fact that $\gamma_{0}$ commutes with $\mathcal{A}$ and $\mathcal{B}$, we obtain
\begin{equation}
\gdualn{\lambda}^{I}_{\alpha}(\p)\lambda^{J}_{\alpha'}(\p')=\widetilde{\lambda}^{I}_{\alpha}(\p)\lambda^{J}_{\alpha'}(\p').\label{eq:ELKO_IP}
\end{equation}
Note that this is in agreement with~(\ref{eq:zimpokJ9an}-\ref{eq:zimpokJ9cn}) when $\p'=\p$. The $\tau$ deformation is defined such that the inner-products of ELKO are unaltered, its purpose is to ensure the Lorentz invariance of the spin sums. Therefore, the transition probability is of the form
\begin{equation}
K_{IJ}\equiv \left[\widetilde{\lambda}^{I}_{\alpha}(\p)\lambda^{J}_{\alpha'}(\p')\right]\left[\widetilde{\lambda}^{J}_{\alpha'}(\p')\lambda^{I}_{\alpha}(\p)\right].
\end{equation}
The function $K_{IJ}$ has the following interpretation. When $I=J$, there are two external fermions or anti-fermions. When $I\neq J$, there is an external fermion-anti-fermion pair. Explicitly evaluating $K_{IJ}$, we find that it can be classified into the following two classes
\begin{eqnarray}
K_{SS}=K_{AA}\geq 0,\quad K_{SA}=K_{AS}\leq 0 \label{eq:ineq}
\end{eqnarray}
for all momenta and helicities. In other words, $M^{\ddag}_{ba}M_{ba}$ is positive- and negative-definite when there are an even and odd number of external anti-fermions in the interactions respectively. Therefore, for $V_{\mathfrak{f}}(t)$ and $V_{\phi}(t)$, we fix $\wp_{ba}$ to
\begin{equation}
\wp_{ba}=(-1)^{\bar{n}_{ba}}, \label{eq:wpsoln}
\end{equation}
where $\bar{n}_{ba}$ is the total number of incoming and outgoing anti-fermions. 

We now compute the cross-sections associated with $V_{\phi}$ and $V_{\mathfrak{f}}$, namely $\phi_{1}\phi_{2}\rightarrow\gdualn{\mathfrak{f}}_{3}\mathfrak{f}_{4}$ and $\mathfrak{f}_{1}\mathfrak{f}_{2}\rightarrow\mathfrak{f}_{3}\mathfrak{f}_{4}$. The scattering amplitude for $\phi_{1}\phi_{2}\rightarrow\gdualn{\mathfrak{f}}_{3}\mathfrak{f}_{4}$ at tree-level is given by
\begin{equation}
M_{(\bar{3}4)(12)}=\frac{g_{\phi}}{m_{\mathfrak{f}}}\widetilde{\lambda}^{A}_{3}\lambda^{S}_{4},\quad
M^{\ddag}_{(\bar{3}4)(12)}=\frac{g_{\phi}}{m_{\mathfrak{f}}}\widetilde{\lambda}^{S}_{4}\lambda^{A}_{3}.
\end{equation}
Summing over the helicity degrees of freedom, we have to perform the $\tau$ deformation. This gives us
\begin{align}
\sum_{\alpha_{3}\alpha_{4}}M^{\ddag}_{(\bar{3}4)(12)}M_{(\bar{3}4)(12)}&=\frac{g^{2}_{\phi}}{m^{2}_{\mathfrak{f}}}\sum_{\alpha_{3}\alpha_{4}}\mbox{tr}\left[(\lambda^{S}_{4}\widetilde{\lambda}^{S}_{4})(\lambda^{A}_{3}\widetilde{\lambda}^{A}_{3})\right]\nonumber\\
&\rightarrow\frac{g^{2}_{\phi}}{m^{2}_{\mathfrak{f}}}\sum_{\alpha_{3}\alpha_{4}}\mbox{tr}\left[(\lambda^{S}_{4}\gdualn{\lambda}^{S}_{4})(\lambda^{A}_{3}\gdualn{\lambda}^{A}_{3})\right]\nonumber\\
&=-16g^{2}_{\phi} \label{eq:ssff}
\end{align}
where $\lambda^{I}_{\alpha_{i}}(\p_{i})\equiv\lambda^{I}_{i}$. In the center of mass frame, the total differential cross-section is given by
\begin{equation}
\frac{d\sigma_{\tiny{\mbox{CM,tot}}}}{d\Omega}(\phi_{1}\phi_{2}\rightarrow\gdualn{\mathfrak{f}}_{3}\mathfrak{f}_{4})=
\frac{g^{2}_{\phi}}{4\pi^{2} E^{2}_{\tiny{\mbox{CM}}}}\left[\frac{1-(2m_{\mathfrak{f}}/E_{\tiny{\mbox{CM}}})^{2}}{1-(2m_{\phi}/E_{\tiny{\mbox{CM}}})^{2}}\right]^{1/2}.\label{eq:cssff}
\end{equation}
Since there is an anti-fermion in the external state, $\wp_{(34)(12)}=-1$. The scattering amplitude for $\mathfrak{f}_{1}\mathfrak{f}_{2}\rightarrow\mathfrak{f}_{3}\mathfrak{f}_{4}$ at tree-level is
\begin{equation}
M_{(34)(12)}=\frac{g_{\mathfrak{f}}}{m^{2}_{\mathfrak{f}}}
\widetilde{M}_{(34)(12)},\quad
M^{\ddag}_{(34)(12)}=\frac{g_{\mathfrak{f}}}{m^{2}_{\mathfrak{f}}}
\widetilde{M}^{\ddag}_{(34)(12)} \label{eq:wm}
\end{equation}
where
\begin{eqnarray}
&&\widetilde{M}_{(34)(12)}=\left[(\widetilde{\lambda}^{S}_{3}\lambda^{S}_{2})(\widetilde{\lambda}^{S}_{4}\lambda^{S}_{1})-(\widetilde{\lambda}^{S}_{4}\lambda^{S}_{2})(\widetilde{\lambda}^{S}_{3}\lambda^{S}_{1})\right],\\
&&\widetilde{M}^{\ddag}_{(34)(12)}=\left[(\widetilde{\lambda}^{S}_{2}\lambda^{S}_{3})(\widetilde{\lambda}^{S}_{1}\lambda^{S}_{4})-(\widetilde{\lambda}^{S}_{2}\lambda^{S}_{4})(\widetilde{\lambda}^{S}_{1}\lambda^{S}_{3})\right]
\end{eqnarray}
so that
\begin{eqnarray}
\widetilde{M}^{\ddag}_{(34)(12)}\widetilde{M}_{(34)(12)}&=&\Big[
\mbox{tr}(\lambda^{S}_{1}\widetilde{\lambda}^{S}_{1}\lambda^{S}_{4}\widetilde{\lambda}^{S}_{4})\mbox{tr}(\lambda^{S}_{2}\widetilde{\lambda}^{S}_{2}\lambda^{S}_{3}\widetilde{\lambda}^{S}_{3})
+\mbox{tr}(\lambda^{S}_{1}\widetilde{\lambda}^{S}_{1}\lambda^{S}_{3}\widetilde{\lambda}^{S}_{3})\mbox{tr}(\lambda^{S}_{2}\widetilde{\lambda}^{S}_{2}\lambda^{S}_{4}\widetilde{\lambda}^{S}_{4})\nonumber\\
&&-\mbox{tr}(\lambda^{S}_{1}\widetilde{\lambda}^{S}_{1}\lambda^{S}_{3}\widetilde{\lambda}^{S}_{3}\lambda^{S}_{2}\widetilde{\lambda}^{S}_{2}\lambda^{S}_{4}\widetilde{\lambda}^{S}_{4})-\mbox{tr}(\lambda^{S}_{1}\widetilde{\lambda}^{S}_{1}\lambda^{S}_{4}\widetilde{\lambda}^{S}_{4}\lambda^{S}_{2}\widetilde{\lambda}^{S}_{2}\lambda^{S}_{3}\widetilde{\lambda}^{S}_{3})\Big].
\end{eqnarray}
Performing $\tau$ deformation and summing over $\alpha_{1},\cdots,\alpha_{4}$, we obtain
\begin{equation}
\sum_{\alpha_{1}\cdots\alpha_{4}}\widetilde{M}^{\ddag}_{(34)(12)}\widetilde{M}_{(34)(12)}
=384m^{4}_{\mathfrak{f}}.\label{eq:wm2}
\end{equation}
The average differential cross-section is
\begin{eqnarray}
\frac{d\sigma_{\mbox{\tiny{CM,avg}}}}{d\Omega}(\mathfrak{f}_{1}\mathfrak{f}_{2}\rightarrow\mathfrak{f}_{3}\mathfrak{f}_{4})&=&\left(\frac{1}{16}\right)\frac{1}{64E^{2}_{\tiny{\mbox{CM}}}}\left(\frac{|\p_{3}|}{|\p_{1}|}\right)\left[\sum_{\alpha_{1}\cdots\alpha_{4}}\wp_{(34)(12)}M^{\ddag}_{(34)(12)}M_{(34)(12)}\right]\nonumber\\
&=&\frac{3g^{2}_{\mathfrak{f}}}{32\pi^{2}E^{2}_{\tiny{\mbox{CM}}}}.
\end{eqnarray}
Since there are no anti-fermions in the external states, $\wp_{(34)(12)}=1$.

\subsubsection{Consistency conditions} \label{con}

In Hermitian quantum field theories, the optical theorem is a consequence of unitarity. From our discussions above, we have shown that mass dimension one fermions is described by a non-Hermitian field theory. So in hindsight, one should not be surprised that the theory violates the optical theorem~\cite{Lee:2015sqj}. In other words, unitarity violation can be seen as a consequence of incorrectly applying Hermitian conjugation to compute physical observables. In light of this observation and the new formalism developed above, we propose the following consistency condition for the $S$-matrix involving mass dimension one fermions
\begin{equation}
\int db S^{\ddag}_{cb}S_{ba}=\delta(c-a).\label{eq:unit}
\end{equation}
Here, the integral $\int db$ includes momentum integration, summation over particle species and internal degrees of freedom. Now we expand the left-hand side of (\ref{eq:unit}). The elements of the $S$-matrix are $S^{\ddag}_{cb}=\langle c|S^{\ddag}|b\rangle$ and $S_{ba}=\langle b|S|a\rangle$ so we get
\begin{equation}
\int db S^{\ddag}_{cb}S_{ba}=\int db\langle c|S^{\ddag}|b\rangle\langle b|S|a\rangle.
\end{equation}
The integral involved is the completeness relation $\int db |b\rangle\langle b|=\openone$ which has the following expansion
\begin{eqnarray}
\int db |b\rangle\langle b|&=&\nonumber\int \frac{d^{3}p}{(2\pi)^{3}(2E)}\sum_{\alpha,n}|\p,\alpha,n\rangle\langle\p,\alpha,n|\nonumber\\
&&+\int \frac{d^{3}p_{1}}{(2\pi)^{3}(2E_{1})}\frac{d^{3}p_{2}}{(2\pi)^{3}(2E_{2})}  \nonumber\\
 &&\times \sum_{\alpha_{1},n_{1},\alpha_{2},n_{2}}
  \vartheta_{n_{1},n_{2}}|\p_{1},\alpha_{1},n_{1};\p_{2},\alpha_{2},n_{2}\rangle
\langle\p_{2},\alpha_{2},n_{2},\p_{1},\alpha_{1},n_{1}|+\cdots ~,\nonumber \label{eq:cr}\\
\end{eqnarray}
where we sum over the degrees of freedom $\alpha$ and particle species $n$. The phase $\vartheta_{n_{1},n_{2}}$ is defined as
\begin{equation}
\vartheta_{n_{1}n_{2}}=\begin{cases}
1 & n_{1}\neq n_{2} \\
\frac{1}{2} & n_{1}=n_{2}
\end{cases}.
\end{equation}
Normalizing the $S$-matrix as $S_{ba}\equiv \delta(b-a)+(2\pi)^{4}iM_{ba}\delta^{4}(p_{b}-p_{a})$, we obtain
\begin{equation}
-i\delta^{4}(p_{c}-p_{a})(M_{ca}-M^{\ddag}_{ca})=(2\pi)^{4}\int db
\left[\delta^{4}(p_{b}-p_{c})\delta^{4}(p_{b}-p_{a})M^{\ddag}_{cb}M_{ba}\right]. \label{eq:im}
\end{equation}
Setting $p_{c}=p_{a}$ and integrating over the Dirac $\delta$-function, we obtain the consistency condition
\begin{equation}
-i(M_{aa}-M^{\ddag}_{aa})=(2\pi)^{4}\int db\left[\delta^{4}(p_{b}-p_{a})
M^{\ddag}_{ab}M_{ba}\right].\label{eq:mm}
\end{equation}
Identifying the right-hand side of (\ref{eq:mm}) to cross-sections and decay rates requires care because  $\int db$ is not a pure momentum integral. To make such an identification, we must first expand the integral using (\ref{eq:cr}) and specify the particle states. Below, we show that $V_{\phi}(t)$ and $V_{\mathfrak{f}}(t)$ satisfy the consistency condition at the one-loop order.

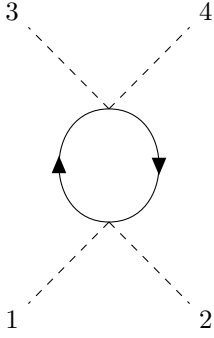
\begin{figure*}
\begin{center}
\begin{tikzpicture}
\begin{feynman}
\vertex (b);
\vertex [above left=of b] (a) {\(3\)};
\vertex [above right=of b] (f2) {\(4\)};
\vertex [below=of b] (c);
\vertex [below left=of c](f1) {\(1\)};
\vertex [below right=of c](f3){\(2\)};
\diagram* {
(b) --[scalar] (a),
(b) --[scalar] (f2),
(c) --[fermion, half left] (b) --[fermion, half left] (c),
(c) --[scalar] (f1),
(c) --[scalar] (f3),
};
\end{feynman}
\end{tikzpicture}
\caption{The $s$-channel contribution to $\phi_{1}\phi_{2}\rightarrow\phi_{3}\phi_{4}$ at one loop.}\label{one}
\end{center}
\end{figure*}

\subsubsection{Scalar interaction}

We consider the process $\phi_{1}\phi_{2}\rightarrow\gdualn{\mathfrak{f}}_{3}\mathfrak{f}_{4}$. The corresponding consistency condition is
\begin{eqnarray}
-i[M_{(12)(12)}-M^{\ddag}_{(12)(12)}]&=&\int \frac{d^{3}p_{3}}{(2\pi)^{3}(2E_{3})}
\frac{d^{3}p_{4}}{(2\pi)^{3}(2E_{4})}\sum_{\alpha_{3}\alpha_{4}}M^{\ddag}_{(12)(\bar{3}4)}M_{(\bar{3}4)(12)}\nonumber\\
&&\times(2\pi)^{4}\delta^{4}(p_{3}+p_{4}-p_{1}-p_{2}), \label{eq:22}
\end{eqnarray}
where $\vartheta_{34}=1$ since the final states are distinct. Here, $M_{(12)(12)}$ and $M_{(\bar{3}4)(12)}$ are the amplitudes for $\phi_{1}\phi_{2}\rightarrow\phi_{1}\phi_{2}$ and $\phi_{1}\phi_{2}\rightarrow\gdualn{\mathfrak{f}}_{3}\mathfrak{f}_{4}$ respectively. The amplitude for scalar bosons does not have explicit dependence on $\lambda^{I}_{\alpha}$ so $M^{\ddag}_{(12)(12)}=M^{\dag}_{(12)(12)}$. Therefore, the left-hand side of~(\ref{eq:22}) is
\begin{equation}
-i[M_{(12)(12)}-M^{\ddag}_{(12)(12)}]=2\mbox{Im}[M_{(12)(12)}]. \label{eq:im1212}
\end{equation}
The imaginary part of $M_{(12)(12)}$ is obtained from the one-loop $s$-channel diagram given in fig.~\ref{one} after setting $3\rightarrow2$ and $4\rightarrow1$. Because we are dealing with bosonic states, we can also set $3\rightarrow1$ and $4\rightarrow2$. If the final states are fermionic, one must adopt the former choice. Explicitly, the matrix elements are given by
\begin{equation}
M_{(12)(12)}=\langle2,1|M|1,2\rangle,\quad
M^{\ddag}_{(12)(12)}=\langle2,1|M^{\ddag}|1,2\rangle.
\end{equation}
The right-hand side of~(\ref{eq:22}) is related to the total cross-section of $\phi_{1}\phi_{2}\rightarrow\gdualn{\mathfrak{f}}_{3}\mathfrak{f}_{4}$ which can be obtained from~(\ref{eq:cssff})
\begin{eqnarray}
\sigma(\phi_{1}\phi_{2}\rightarrow\gdualn{\mathfrak{f}}_{3}\mathfrak{f}_{4})&=&-\frac{1}{4E_{1}E_{2}u_{12}}\int \frac{d^{3}p_{3}}{(2\pi)^{3}(2E_{3})}\frac{d^{3}p_{4}}{(2\pi)^{3}(2E_{4})}\sum_{\alpha_{3}\alpha_{4}}M^{\ddag}_{(12)(34)}M_{(34)(12)}\nonumber\\
&&\times(2\pi)^{4}\delta^{4}(p_{3}+p_{4}-p_{1}-p_{2})\nonumber\\
&=&\frac{g^{2}_{\phi}}{\pi E^{2}_{\tiny{\mbox{CM}}}}\left[\frac{1-(2m_{\mathfrak{f}}/E_{\tiny{\mbox{CM}}})^{2}}{1-(2m_{\phi}/E_{\tiny{\mbox{CM}}})^{2}}\right]^{1/2}.
\end{eqnarray}
Therefore, the consistency condition becomes
\begin{equation}
\mbox{Im}[M_{(12)(12)}]=-4E_{1}E_{2}u_{12}\sigma(\phi_{1}\phi_{2}\rightarrow\gdualn{\mathfrak{f}}_{3}\mathfrak{f}_{4}).\label{eq:im2}
\end{equation}
The left-hand side of (\ref{eq:im2}) is obtained by evaluating the one-loop $s$-channel $S$-matrix~\footnote{The fermionic propagator that enters the calculation is given by the time-oredered product $\langle\,\,|\mathcal{T}\mathfrak{f}(x)\gdualn{\mathfrak{f}}(x')|\,\,\rangle$.}
\begin{equation}
S_{(12)(12)}(s)=(-ig_{\phi})^{2}V(s)\delta^{4}(p_{3}+p_{4}-p_{1}-p_{2}),
\end{equation}
where $s=(p_{1}+p_{2})^{2}$ and
\begin{equation}
V(p^{2})=-16\int d^{4}k\left(\frac{i}{k^{2}-m^{2}_{\mathfrak{f}}+i\epsilon}\right)
\left[\frac{i}{(k+p)^{2}-m^{2}_{\mathfrak{f}}+i\epsilon}\right]. \label{eq:v}
\end{equation}
The minus sign comes from the fermionic statistics and the factor of 16 from taking the trace of $4\mathbb{I}$. In the center of mass frame $s=E^{2}_{\tiny{\mbox{CM}}}$. Evaluating the integral using dimensional regularization yields
\begin{equation}
V(s)=-16i\pi^{2}\int^{1}_{0}dx
\left\{\ln\left[m^{2}_{\mathfrak{f}}-E^{2}_{\tiny{\mbox{CM}}}(1-x)x\right]+\frac{2}{d-4}+\gamma\right\} , \label{eq:vs}
\end{equation}
where $d$ is the space-time dimension and $\gamma$ is the Euler constant. Therefore, the $s$-channel contribution to $M_{(12)(12)}$ at one-loop is
\begin{equation}
M_{(12)(12)}(s)=\frac{g^{2}_{\phi}}{\pi^{2}}
\int^{1}_{0}dx
\left\{\ln\left[m^{2}_{\mathfrak{f}}-E^{2}_{\tiny{\mbox{CM}}}(1-x)x\right]+\frac{2}{d-4}+\gamma\right\}
\end{equation}
and its imaginary part is given by
\begin{equation}
\mbox{Im}[M_{(12)(12)}]=-\frac{g^{2}_{\phi}}{\pi}\left(1-\frac{4m^{2}_{\mathfrak{f}}}{E^{2}_{\tiny{\mbox{CM}}}}\right)^{1/2}.
\label{eq:1212}\end{equation}
Explicit computation shows that~(\ref{eq:1212}) is equal to the right-hand side of~(\ref{eq:im2}) so the consistency condition is satisfied.

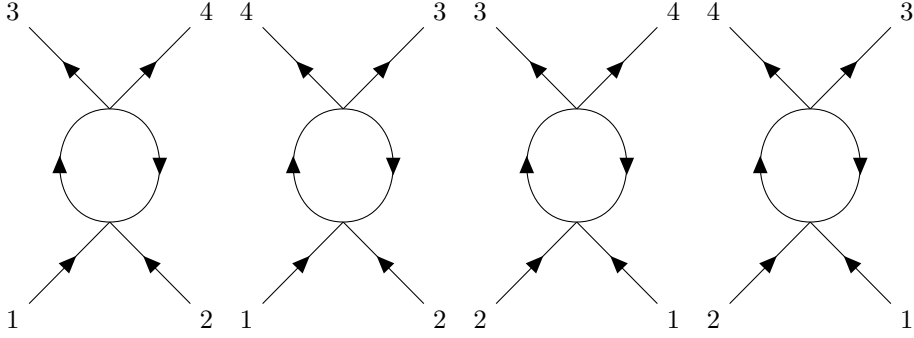
\begin{figure*}
\begin{center}
\begin{tikzpicture}
\begin{feynman}
\vertex (b);
\vertex [above left=of b] (a) {\(3\)};
\vertex [above right=of b] (f2) {\(4\)};
\vertex [below=of b] (c);
\vertex [below left=of c](f1) {\(1\)};
\vertex [below right=of c](f3){\(2\)};
\diagram* {
(b) --[fermion] (a),
(b) --[fermion] (f2),
(c) --[fermion, half left] (b) --[fermion, half left] (c),
(f1) --[fermion] (c),
(f3) --[fermion] (c),
};
\end{feynman}
\end{tikzpicture}
\begin{tikzpicture}
\begin{feynman}
\vertex (b);
\vertex [above left=of b] (a) {\(4\)};
\vertex [above right=of b] (f2) {\(3\)};
\vertex [below=of b] (c);
\vertex [below left=of c](f1) {\(1\)};
\vertex [below right=of c](f3){\(2\)};
\diagram* {
(b) --[fermion] (a),
(b) --[fermion] (f2),
(c) --[fermion, half left] (b) --[fermion, half left] (c),
(f1) --[fermion] (c),
(f3) --[fermion] (c),
};
\end{feynman}
\end{tikzpicture}
\begin{tikzpicture}
\begin{feynman}
\vertex (b);
\vertex [above left=of b] (a) {\(3\)};
\vertex [above right=of b] (f2) {\(4\)};
\vertex [below=of b] (c);
\vertex [below left=of c](f1) {\(2\)};
\vertex [below right=of c](f3){\(1\)};
\diagram* {
(b) --[fermion] (a),
(b) --[fermion] (f2),
(c) --[fermion, half left] (b) --[fermion, half left] (c),
(f1) --[fermion] (c),
(f3) --[fermion] (c),
};
\end{feynman}
\end{tikzpicture}
\begin{tikzpicture}
\begin{feynman}
\vertex (b);
\vertex [above left=of b] (a) {\(4\)};
\vertex [above right=of b] (f2) {\(3\)};
\vertex [below=of b] (c);
\vertex [below left=of c](f1) {\(2\)};
\vertex [below right=of c](f3){\(1\)};
\diagram* {
(b) --[fermion] (a),
(b) --[fermion] (f2),
(c) --[fermion, half left] (b) --[fermion, half left] (c),
(f1) --[fermion] (c),
(f3) --[fermion] (c),
};
\end{feynman}
\end{tikzpicture}
\end{center}
\caption{The $s$-channel contribution to $\mathfrak{f}_{1}\mathfrak{f}_{2}\rightarrow\mathfrak{f}_{3}\mathfrak{f}_{4}$ with a fermionic loop.}\label{two}
\end{figure*}

\subsubsection{Fermionic interaction}

We now consider the process $\mathfrak{f}_{1}\mathfrak{f}_{2}\rightarrow\mathfrak{f}_{3}\mathfrak{f}_{4}$. The corresponding consistency condition reads
\begin{eqnarray}
-i\left[M_{(12)(12)}-M^{\ddag}_{(12)(12)}\right]
&=&\frac{1}{2}\int \frac{d^{3}p_{3}}{(2\pi)^{3}(2E_{3})}\frac{d^{3}p_{4}}{(2\pi)^{3}(2E_{4})}\sum_{\alpha_{3}\alpha_{4}}M^{\ddag}_{(12)(34)}M_{(34)(12)}\nonumber\\
&&\times(2\pi)^{4}\delta^{4}(p_{3}+p_{4}-p_{1}-p_{2}) \label{eq:m12m12}
\end{eqnarray}
Here, because the final states are of the same species, we need to set $\vartheta_{34}=\frac{1}{2}$. The right-hand side of (\ref{eq:m12m12}) is identified with the total cross-section $\mathfrak{f}_{1}\mathfrak{f}_{2}\rightarrow\mathfrak{f}_{3}\mathfrak{f}_{4}$
\begin{align}
\sigma(\mathfrak{f}_{1}\mathfrak{f}_{2}\rightarrow\mathfrak{f}_{3}\mathfrak{f}_{4})&=\frac{1}{4E_{1}E_{2}u_{12}}\int \frac{d^{3}p_{3}}{(2\pi)^{3}(2E_{3})}\frac{d^{3}p_{4}}{(2\pi)^{3}(2E_{4})}\sum_{\alpha_{3}\alpha_{4}}M^{\ddag}_{(12)(34)}M_{(34)(12)}\nonumber\\
&\times(2\pi)^{4}\delta^{4}(p_{3}+p_{4}-p_{1}-p_{2})\nonumber\\
&=\frac{g^{2}_{\mathfrak{f}}}{16\pi E^{2}_{\tiny{\mbox{CM}}}}\left\{\left(\frac{8}{m^{2}_{\mathfrak{f}}}\right)
\left[(\widetilde{\lambda}^{S}_{1}\lambda^{S}_{1})(\widetilde{\lambda}^{S}_{2}\lambda^{S}_{2})-(\widetilde{\lambda}^{S}_{2}\lambda^{S}_{1})(\widetilde{\lambda}^{S}_{1}\lambda^{S}_{2})\right]\right\}.\label{eq:cs}
\end{align}
Therefore, the consistency condition becomes
\begin{equation}
-i[M_{(12)(12)}-M^{\ddag}_{(12)(12)}]=2E_{1}E_{2}u_{12}\sigma(\mathfrak{f}_{1}\mathfrak{f}_{2}\rightarrow\mathfrak{f}_{3}\mathfrak{f}_{4}).
\label{eq:m1212}
\end{equation}
The amplitude $M_{(12)(12)}$ of~(\ref{eq:m1212}) is obtained by summing diagrams in fig.~\ref{two} and their $t$- and $u$-channel counterparts with $3\rightarrow2$ and $4\rightarrow 1$. We will show that only the $s$-channel diagrams contribute to the left-hand side of (\ref{eq:m1212}). The first diagram is given by
\begin{eqnarray}
\begin{tikzpicture}
\begin{feynman}[small]
\vertex (b);
\vertex [above left=of b] (a) {\(4\)};
\vertex [above right=of b] (f2) {\(3\)};
\vertex [below=of b] (c);
\vertex [below left=of c](f1) {\(1\)};
\vertex [below right=of c](f3){\(2\)};
\diagram*{
(b) --[fermion] (a),
(b) --[fermion] (f2),
(c) --[fermion, half left] (b) --[fermion, half left] (c),
(f1) --[fermion] (c),
(f3) --[fermion] (c),
};
\end{feynman}
\end{tikzpicture}&=&\frac{(-ig_{\mathfrak{f}})^{2}}{4}
\left(\frac{1}{m^{2}_{\mathfrak{f}}}\right)
\left[(\widetilde{\lambda}^{S}_{3}\lambda^{S}_{1})
      (\widetilde{\lambda}^{S}_{4}\lambda^{S}_{2})
     -(\widetilde{\lambda}^{S}_{3}\lambda^{S}_{2})
     (\widetilde{\lambda}^{S}_{4}\lambda^{S}_{1})\right]V(s), \label{eq:mls}
\end{eqnarray}
where $V$ is given by~(\ref{eq:vs}). Summing over all the diagrams in fig.~\ref{two}, we obtain
\begin{equation}
M^{(1L)}_{(34)(12)}(s)=\frac{(-ig_{\mathfrak{f}})^{2}}{m^{2}_{\mathfrak{f}}}\left[(\widetilde{\lambda}^{S}_{3}\lambda^{S}_{1})(\widetilde{\lambda}^{S}_{4}\lambda^{S}_{2})-(\widetilde{\lambda}^{S}_{3}\lambda^{S}_{2})(\widetilde{\lambda}^{S}_{4}\lambda^{S}_{1})\right]V(s).
\end{equation}
Taking $3\rightarrow2$ and $4\rightarrow1$, we get
\begin{equation}
M^{(1L)}_{(12)(12)}(s)=M^{(1L)}_{\phi}(s)\left\{\left(\frac{8}{m^{2}_{\mathfrak{f}}}\right)
\left[(\widetilde{\lambda}^{S}_{1}\lambda^{S}_{1})(\widetilde{\lambda}^{S}_{2}\lambda^{S}_{2})
 -(\widetilde{\lambda}^{S}_{2}\lambda^{S}_{1})
  (\widetilde{\lambda}^{S}_{1}\lambda^{S}_{2})\right]\right\}, \label{eq:mls2}
\end{equation}
where $M^{(1L)}_{\phi}(s)$ is the one-loop $s$-channel contribution to $\phi_{1}\phi_{2}\rightarrow\phi_{1}\phi_{2}$ in the $g\phi^{4}/4!$ theory with $g_{\mathfrak{f}}$ replaced by $g$. Apply $\ddag$ to $M^{(1L)}_{(12)(12)}(s)$, we obtain
\begin{equation}
M^{(1L)\ddag}_{(12)(12)}(s)=M^{(1L)\dag}_{\phi}(s)\left\{\left(\frac{8}{m^{2}_{\mathfrak{f}}}\right)
\left[(\widetilde{\lambda}^{S}_{1}\lambda^{S}_{1})(\widetilde{\lambda}^{S}_{2}\lambda^{S}_{2})-(\widetilde{\lambda}^{S}_{2}\lambda^{S}_{1})(\widetilde{\lambda}^{S}_{1}\lambda^{S}_{2})\right]\right\}.\label{eq:3mls}
\end{equation}
In obtaining (\ref{eq:3mls}), we have used the fact that $M^{(1L)}_{\phi}$ is independent of $\lambda^{I}_{\alpha}$ so
$M^{(1L)\ddag}_{\phi}=M^{(1L)\dag}_{\phi}$ and observed that the term in the bracket on the right is invariant under $\ddag$. Therefore,
\begin{equation}
-i\left[M^{(1L)}_{(12)(12)}(s)-M^{(1L)\ddag}_{(12)(12)}(s)\right]=2\mbox{Im}\left[M^{(1L)}_{\phi}(s)\right]
\left\{\left(\frac{8}{m^{2}_{\mathfrak{f}}}\right)
\left[(\widetilde{\lambda}^{S}_{1}\lambda^{S}_{1})(\widetilde{\lambda}^{S}_{2}\lambda^{S}_{2})-(\widetilde{\lambda}^{S}_{2}\lambda^{S}_{1})(\widetilde{\lambda}^{S}_{1}\lambda^{S}_{2})\right]\right\}.\label{eq:mls3}
\end{equation}
Replacing $s$ with $t$ and $u$, we obtain the $t$- and $u$-channel contributions. Recall that in the $g\phi^{4}/4!$ theory, $\mbox{Im}[M^{(1L)}_{\phi}]=\mbox{Im}[M^{(1L)}_{\phi}(s)]$. Therefore,
\begin{equation}
\left[M^{(1L)}_{(12)(12)}-M^{(1L)\ddag}_{(12)(12)}\right]=\left[M^{(1L)}_{(12)(12)}(s)-M^{(1L)\ddag}_{(12)(12)}(s)\right].
\end{equation}
Substituting~(\ref{eq:mls3}) and (\ref{eq:cs}) into (\ref{eq:m1212}), it reduces to
\begin{equation}
\mbox{Im}\left[M^{(1L)}_{\phi}\right]=(2|\p_{1}|E_{\tiny{\mbox{CM}}})\left[\frac{g^{2}_{\mathfrak{f}}}{32\pi E^{2}_{\tiny{\mbox{CM}}}}\right].\label{eq:oscalar}
\end{equation}
Equation~(\ref{eq:oscalar}) is the optical theorem for the $g\phi^{4}(x)/4!$ theory at one-loop with $g_{\mathfrak{f}}$ replaced by $g$. Therefore, $\mathfrak{f}_{1}\mathfrak{f}_{2}\rightarrow\mathfrak{f}_{3}\mathfrak{f}_{4}$ is consistent at one-loop.

\section{Attractive gravitational potential}\label{gp}

In this section, we study the interaction between the mass dimension one fermion and the graviton in the weak field approximation. After deriving the tree-level fermion-graviton interaction vertex, we compute the scattering amplitude $\mathfrak{f}_{1}\mathfrak{f}_{2}\rightarrow\mathfrak{f}_{3}\mathfrak{f}_{4}$ mediated by a graviton. In the non-relativistic limit, we obtain  an attractive Newtonian potential as required for a dark matter candidate\footnote{We thank Laura Duarte for important discussions in this section.}.

We start with the action of mass dimension one fermionic fields in a curved space-time given by
\begin{eqnarray}
S = \int d^{4}x \, \sqrt{-g}\left(g^{\mu\nu}\nabla_{\mu}\gdualn{\mathfrak{f}}\nabla_{\nu}\mathfrak{f}- m^2\gdualn{\mathfrak{f}}\mathfrak{f}\right). \label{jm1}
\end{eqnarray}
In the action~(\ref{jm1}), the covariant derivatives are given by
\begin{eqnarray}\label{jm2}
&&\nabla_{\mu}\mathfrak{f} = \partial_{\mu}\mathfrak{f} - \omega_{\mu}\mathfrak{f},\\
&&\nabla_{\mu}\gdualn{\mathfrak{f}}= \partial_{\mu}\gdualn{\mathfrak{f}} + \gdualn{\mathfrak{f}}\omega_{\mu},
\end{eqnarray}
where the spin-connection is
\begin{eqnarray}
&&\omega_{\mu} =-\frac{i}{2}(e^{\nu}_{b}\Gamma^{\alpha}_{\mu\nu}e_{a\alpha}-e^{\nu}_{b}\partial_{\mu}e_{a\nu})S^{ab},\label{eq:gmu}\\
&&S^{ab}=\frac{i}{4}[\gamma^{a},\gamma^{b}].\label{eq:sab}
\end{eqnarray}
In~(\ref{eq:gmu}) and (\ref{eq:sab}), $\Gamma^{\alpha}_{\mu\nu}(x)$ is the Christoffel symbol, $a,b$ are non-holonomic indices and $e_{\nu}^{a}(x)$ is the tetrad field that admits the following decomposition of the metric
\begin{eqnarray}
e^{\mu}_{a}(x)e^{\nu}_{b}(x)g_{\mu\nu}(x) &= \eta_{ab},\\
e^{\mu}_{a}(x)e_{\nu}^{a}(x) &= \delta_{\nu}^{\mu},\\
 e^{a}_{\mu}(x)e_{b}^{\mu}(x) &= \delta_{b}^{a}.
\end{eqnarray}
By means of the weak field approximation, in orders of $\kappa^{2}=16\pi G$
\begin{align}
 g_{\mu\nu}&=\eta_{\mu\nu}+\kappa h_{\mu\nu}, \\
 g^{\mu\nu}&=\eta^{\mu\nu}-\kappa h^{\mu\nu}+\kappa^{2}h^{\mu\chi}h^{\nu}_{\chi}+\mathcal{O}(\kappa^{3}),\\
 e^{\alpha}_{a}&=\eta^{\alpha}_{a}-\frac{1}{2}\kappa h^{\alpha}_{a}+\frac{3}{8}\kappa^{2}h^{\chi}_{a}h^{\alpha}_{\chi}
+\mathcal{O}(\kappa^{3}),\\
 e^{a}_{\alpha}&=\eta^{a}_{\alpha}+\frac{1}{2}\kappa h^{a}_{\alpha}-\frac{1}{8}\kappa^{2}h_{\alpha\chi}h^{a\chi}+\mathcal{O}(\kappa^{3}),
\end{align}
one obtains~\cite{Rogerio:2019evl}
\begin{eqnarray}
\omega_{\mu}&=&\frac{iS^{\alpha\beta}}{8}\left[\kappa\partial_\beta h_{\mu\alpha}-\kappa\partial_{\alpha}h_{\mu\beta}+\frac{\kappa^2}{4}h^\rho_\beta\partial_\mu h_{\alpha\rho}-\frac{\kappa^2}{4}h^\rho_\alpha\partial_\mu h_{\beta\rho}\right. \\
	&&+\left.\frac{\kappa^2}{4}h^\rho_\beta\partial_\alpha h_{\mu\rho}-\frac{\kappa^2}{4}h^\rho_\alpha\partial_\beta h_{\mu\rho}+\frac{\kappa^2}{4}h^\rho_\alpha\partial_\rho h_{\mu\beta}-\frac{\kappa^2}{4}h^\rho_\beta\partial_\rho h_{\mu\alpha}\right]. \label{jm3}
\end{eqnarray}

Taking into account that $\sqrt{-g}\simeq 1+(\kappa/2)h+O(\kappa^2)$, the Lagrangian density in~(\ref{jm1}) may be recast as
\begin{eqnarray}
\mathfrak{L}&=&\partial_\mu\gdualn{\mathfrak{f}}\partial^\mu\mathfrak{f}-m^2\gdualn{\mathfrak{f}}\mathfrak{f}+\frac{\kappa}{4}h(\partial_\mu\gdualn{\mathfrak{f}}\partial^\mu\mathfrak{f}-m^2\gdualn{\mathfrak{f}}\mathfrak{f})-\frac{\kappa}{2}(\partial_\mu\gdualn{\mathfrak{f}})h^{\mu\nu}\partial_\nu\mathfrak{f}\nonumber\\
&&+\frac{\kappa}{32}\left[(\partial_\mu\gdualn{\mathfrak{f}})(\partial_\rho h_\sigma^\mu-\partial_\sigma h_\rho^\mu)\gamma^\sigma\gamma^\rho \mathfrak{f}+\gdualn{\mathfrak{f}}(\partial_\alpha h_{\beta\mu}-\partial_\beta h_{\alpha\mu})\gamma^\alpha\gamma^\beta \mathfrak{f}\right]. \label{extra1}
\end{eqnarray}
The above Lagrangian density may be decomposed as $\mathfrak{L}(x)=\mathfrak{L}_{0}(x) +\mathfrak{L}_{int}(x)$, where $\mathfrak{L}_{0}(x)$ stands for the free Lagrangian density and $\mathfrak{L}_{int}(x)$ represents the first order interaction between the mass dimension one fermion and the graviton. In the momentum space, the fermion-graviton interaction vertex can be obtained via a standard functional variation
\begin{equation}
V_{\alpha\beta}(p,q,r)=\frac{\delta}{\delta\df(p)}\frac{\delta}{\delta\mathfrak{f}(q)}\frac{\delta}{\delta h^{\alpha\beta}(r)} \mathfrak{L}_{int}, \label{extra2}
\end{equation}
where $p,q,r$ are the external momenta. Using the identity
\begin{equation}
\frac{\delta h^{\sigma\rho}}{\delta h^{\alpha\beta}}=\frac{1}{2}(\eta^{\alpha\rho}\eta^{\sigma\beta}-\eta^{\rho\beta}\eta^{\alpha\sigma}),
\end{equation}
the tree-level interaction vertex reads
\begin{eqnarray}\label{jm4}
V_{\alpha\beta}(p,q,r)&=&+\frac{i\kappa}{8}\delta^{4}(q-r-p)\Big\{4(p\cdot q - m^{2})
\eta_{\alpha\beta}\I-4(q_{\alpha}p_{\beta}+q_{\beta}p_{\alpha})\I \nonumber\\
&&+ \frac{1}{4}[\gamma_{\alpha},\gamma_{\mu}r^{\mu}](p+q)_{\beta}
+\frac{1}{4}[\gamma_{\beta},\gamma_{\nu}r^{\nu}](p+q)_{\alpha}\Big\}
\end{eqnarray}
where $p$ is the incoming fermionic momentum, $q$ is the outgoing fermionic momentum and $r$ is the incoming graviton momentum. The mass dimension one fermionic field has a unique feature where its dynamics is governed by the Klein-Gordon equation while the field furnishes spin half representation. This is reflected in the general form of the interaction vertex~(\ref{jm4}) where it can be decomposed into a sum of scalar-graviton (terms proportional to $\I$) and fermion-graviton contributions (terms proportional to $\gamma_{\alpha}$ and $\gamma_{\beta}$)\footnote{As a parenthetical remark, we stress that for the case at hand,f the gauge invariance is attained by the possibility of using~(\ref{jm4}) in order to satisfy the Ward-Takahashi relation \cite{Just:1965ola} \cite{Capper:1974vb}. For further details see \cite{Rogerio:2019evl}.} \cite{Holstein:2006bh}.

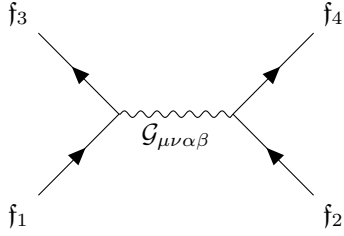
\begin{figure}
\begin{center}
\begin{tikzpicture}
\begin{feynman}
\vertex (a);
\vertex [above left=of a] (b) {\(\mathfrak{f}_{3}\)};
\vertex [below left=of a] (c) {\(\mathfrak{f}_{1}\)};
\vertex [right=of a] (d);
\vertex [below right=of d](f) {\(\mathfrak{f}_{2}\)};
\vertex [above right=of d](e){\(\mathfrak{f}_{4}\)};
\diagram* {
(a) --[fermion] (b),
(c) --[fermion] (a),
(a) --[boson, edge label'=\(\mathcal{G}_{\mu\nu\alpha\beta}\)] (d),
(d) --[fermion] (e),
(f) --[fermion] (d),
};
\end{feynman}
\end{tikzpicture}
\end{center}
\caption{Mass dimension one fermion scattering mediated by a graviton.}
	\label{fig3}
\end{figure}

Now we are able to compute the scattering process $\mathfrak{f}_{1}\mathfrak{f}_{2}\rightarrow\mathfrak{f}_{3}\mathfrak{f}_{4}$ as depicted in fig.~\ref{fig3}. Here, we are interested in the non-relativistic limit of the amplitude for which it can be associated with the interacting potential~\cite{Maggiore:2005qv}
\begin{eqnarray}\label{jjv}
V(\x)=\frac{1}{2E_{1}}\frac{1}{2E_{2}}\int \frac{d^3r}{(2\pi)^3}
e^{i\boldsymbol{r\cdot x}}[-iM_{(34)(12)}
(\boldsymbol{r})],
\end{eqnarray}
where $\boldsymbol{r}$ is the graviton three-momentum and the scattering amplitude reads \cite{Maggiore:2005qv}
\begin{eqnarray}\label{jja}
M_{(34)(12)}=-\frac{i}{m^2}\left[(\widetilde{\lambda}^{S}_{3} V^{\alpha\beta}\lambda^{S}_{1})\mathcal{G}_{\mu\nu\alpha\beta}(\widetilde{\lambda}^{S}_{4} V^{\mu\nu}\lambda^{S}_{2})\right]
\end{eqnarray}
and
\begin{equation}
{\mathcal{G}_{\alpha\beta\mu\nu}}=\frac{1}{2r^2}(\eta_{\alpha\mu}\eta_{\beta\nu}+\eta_{\beta\mu}\eta_{\alpha\nu}-\eta_{\alpha\beta}\eta_{\mu\nu})
\end{equation}
is the graviton propagator in the momentum space \cite{Capper:1973bk}. An important aspect associated with this process is that for the mass dimension one fermion to qualify as a dark matter candidate, the potential must be Newtonian. As we will show by explicit computation, the potential is indeed Newtonian.
The complete gauge fixing procedure and derivation of~(\ref{jja}) may be found in~\cite{Capper:1973pv}.

Adopting the vertex (\ref{jm4}) to the process at hand, we have
\begin{eqnarray}\label{j38}
V^{\mu\nu}&=&\frac{\kappa}{16}\Big{[}\underbrace{4(p_{2}\cdot p_{4} - m^{2})\eta^{\mu\nu}\I - 4(p^{\mu}_{4}p^{\nu}_{2}+p^{\nu}_{4}p^{\mu}_{2})\I}_{E^{\mu\nu}}
+ \underbrace{\frac{1}{4} [\gamma^{\mu},\slashed{r}](p_{2}+p_{4})^{\nu}+\frac{1}{4}[\gamma^{\nu},\slashed{r}](p_{2}+p_{4})^{\mu}}_{M^{\mu\nu}}\Big{]}\nonumber \\
&\equiv& \frac{\kappa}{16}(E^{\mu\nu}+M^{\mu\nu}),
\end{eqnarray}
where we separate out $E^{\mu\nu}$ and $M^{\mu\nu}$ to denote the scalar and fermionic sectors respectively. Substituting (\ref{j38}) into (\ref{jja}), we obtain
\begin{eqnarray}
M_{(34)(12)}&=&-\frac{i\kappa^2}{256m^2}\Big[\Big(E^{\alpha\beta}\widetilde{\lambda}^{S}_{3}\lambda^{S}_{1}+\widetilde{\lambda}^{S}_{3} M^{\alpha\beta}\lambda^{S}_{1}\Big)  \mathcal{G}_{\mu\nu\alpha\beta}
\Big(E^{\mu\nu}\widetilde{\lambda}^{S}_{4}\lambda^{S}_{2}+\widetilde{\lambda}^{S}_{4} M^{\mu\nu}\lambda^{S}_{2}\Big)\Big].\nonumber\\ \label{jjj}
\end{eqnarray}
Here, it is instructive to compute the scattering amplitude explicitly. For this purpose, we take the ELKO degrees of freedom to be $\alpha_{i}=+$ for all $i=1,\cdots,4$. The result is
\begin{eqnarray}\label{1111}
M_{(3^{+}4^{+})(1^{+}2^{+})}&=&-\frac{i\kappa^2}{256m^2 r^2}\Big{\{}
64m^2 \left[4m^2(p_{2}\cdot p_{4})-4m^4+2(p_{1}\cdot p_{2})(p_{3}\cdot p_{4})\right]\nonumber\\
&&+4m^3  \widetilde{\lambda}^{S}_{3^{+}}[\slashed{p}_{2}+\slashed{p}_{4}, \slashed{r}]\lambda^{S}_{2^{+}}
-4m p_{3}\cdot (p_{2}+p_{4})\widetilde{\lambda}^{S}_{4^{+}}[\slashed{p}_{1}, \slashed{r}]\lambda^{S}_{2^{+}}\nonumber\\
&&+4m^3\widetilde{\lambda}^{S}_{3^{+}}[\slashed{p}_{2}+\slashed{p}_{4}, \slashed{r}]\lambda^{S}_{1^{+}}
-4m p_{1}\cdot (p_{2}+p_{4})\widetilde{\lambda}^{S}_{4^{+}}[\slashed{p}_{3},\slashed{r}]\lambda^{S}_{2^{+}}\nonumber\\
&&-p_{2}\cdot (p_{1}+p_{3})\widetilde{\lambda}^{S}_{3^{+}}[\slashed{p}_{4}, \slashed{r}]\lambda^{S}_{1^{+}}-p_{4}\cdot (p_{1}+p_{3})\widetilde{\lambda}^{S}_{3^{+}}[\slashed{p'}, \slashed{r}]\lambda^{S}_{1^{+}}\nonumber\\
&&+\frac{1}{8}(p_{1}+p_{3})\cdot(p_{2}+p_{4})(\widetilde{\lambda}^{S}_{3^{+}}[\gamma^\alpha, \slashed{r}]\lambda^{S}_{1^{+}})(\widetilde{\lambda}^{S}_{4^{+}}[\gamma_\alpha, \slashed{r}]\lambda^{S}_{2^{+}})\Big{\}}.
\end{eqnarray}
In the center of mass frame $r^0=0$, we define the momentum for the fermions present in the interaction as
\begin{eqnarray}
&& p^{\mu}_{1} = (E,0,p,0), \\
&& p^{\mu}_{3} = (E, p\sin\theta, p\cos\theta, 0).
\end{eqnarray}
Therefore,
\begin{eqnarray}
&& r^{\mu} = p^{\mu}_{1} - p^{\mu}_{3} = (0,-p\sin\theta, p - p\cos\theta, 0),\\
&& p^{\mu}_{2} = (E, 0, -p, 0),\\
&& p^{\mu}_{4} = (E, -p\sin\theta, -p\cos\theta, 0).
\end{eqnarray}
These choices lead to a scattering amplitude whose sole momentum dependence is $p$. In the non-relativistic limit $p\rightarrow0$, the scattering amplitude simplifies to
\begin{equation}
\lim_{p\rightarrow0}M_{(3^{+}4^{+})(1^{+}2^{+})}=-\frac{8i\pi G m^4}{r^2}.\label{eq:mnr}
\end{equation}
Substituting (\ref{eq:mnr}) into (\ref{jjv}), we obtain
\begin{eqnarray}
V(\x)=-\frac{G m^2}{|\x|}
\end{eqnarray}
yielding an attractive Newtonian potential. This is indeed an important result since it reproduces the correct non-relativistic limit as expected for a dark matter candidate.


\section{Thermodynamic properties}{\label{Sec08}}

The thermodynamic properties {and the statistics of the new class of mass dimension one fields discussed in sec.~\ref{Sec:Elko-b} can be further ascertained from statistical thermodynamics using finite temperature field theory methods, in complete analogy to what is done for the SM fields~\cite{Bellac:2011kqa,Kapusta:2006pm,Das:1997gg,Bailin:1986wt}. 
In particular, the equilibrium thermodynamic properties and the statistics satisfied} by scalar fields, spin-half Dirac fields and spin one gauge fields can be extracted from the partition function $Z$ of the system, obtained through functional integration method and the imaginary time formalism.



In order to construct the equilibrium partition function $Z$ for a standard bosonic or fermionic field $\psi$, we must
introduce the temperature by means of the imaginary time formalism. For this purpose, we Wick rotate the real time axis $t$ to the imaginary one $\tau = it$. Once the system is in equilibrium, the equilibrium thermodynamic temperature ${T}$ is introduced by means of $\tau \equiv \beta = \frac{1}{k_B {T}}$. The temporal integration must be taken over the interval $0<\tau <\beta$ with
\begin{equation}
\mathfrak{\psi}(\tau=0,{\x})=\pm\mathfrak{\psi}(\tau=\beta,{\x})\label{eq:pe02}
\end{equation}
where the top and bottom signs are the periodic and anti-periodic boundary conditions for the bosonic or fermionic field respectively.


The starting point is the partition function for a generic bosonic or fermionic field $\psi(x)$, which can be written as \cite{Kapusta:2006pm,Bailin:1986wt}
\begin{equation}
Z=\int[D\mathfrak{\pi}]\int_{\pm}  [D\mathfrak{\psi}]\exp\Bigg[\int_0^\beta d\tau \int d^3x\,\bigg(i\pi\frac{\partial\mathfrak{\psi} }{\partial\tau}-\mathfrak{H}(\mathfrak{\psi},\mathfrak{\pi}) + \mu \mathfrak{N}(\mathfrak{\psi},\mathfrak{\pi}) \bigg)\Bigg]\label{eq:pe01}
\end{equation}
where the top and bottom sign denotes the periodic and anti-periodic functional integrations for bosonic or fermionic field respectively. The Hamiltonian density is given by $\mathfrak{H}(\mathfrak{\psi},\mathfrak{\pi}) = \pi \partial_{t}{\mathfrak{\psi}}-\mathfrak{L}$ with $\mathfrak{\pi}$ being the conjugate momentum and  $\mathfrak{N}$ is the conserved charge density to which it is associated with a chemical potential $\mu$.

Let us consider the {specific free field Lagrangian density for a mass dimension one field ${\mathfrak{a}}(x)$, where ${\mathfrak{a}}(x)={\mathfrak{b}}(x)$ for the bosonic field and ${\mathfrak{a}}(x)={\mathfrak{f}}(x)$ for the fermionic one:
\begin{equation}
\mathfrak{L}_{0}(x) =\frac{1}{2}\partial^\mu\gdualn{\mathfrak{a}}(x)\,\partial_\mu {\mathfrak{a}}(x) - \frac{m^2}{2} \gdualn{\mathfrak{a}}(x) \mathfrak{a}(x), \label{eq:pe03}
\end{equation}
where $\gdualn{\mathfrak{a}}(x)$} is the dual of the field. The momentum conjugate to the field and its dual are 
\begin{equation}
\mathfrak{\pi}(x)=\frac{1}{2}\partial_{t}{\gdualn{\mathfrak{a}}}(x)=\frac{i}{2}\partial_\tau\gdualn{\mathfrak{a}}(x),\quad
\gdualn{\mathfrak{\pi}}(x)=\frac{1}{2}\partial_{t}{\mathfrak{a}}(x)=\frac{i}{2}\partial_\tau\mathfrak{a}(x).
\end{equation}
Under a local dark-sector $U(1)$ symmetry {$\mathfrak{a} \to \mathfrak{a}'=\mathfrak{a}\,e^{-i\alpha(x)}$}, the conserved current is
{
\begin{equation}
    \mathfrak{j}_\mu = \frac{i}{2} \left[\gdualn{\mathfrak{a}}(\partial_\mu \mathfrak{a}) - (\partial_\mu\gdualn{\mathfrak{a}}) \mathfrak{a} \right]\label{eq:pe03u3}
\end{equation}
}
from which we obtain the total conserved charge {
\begin{equation}
    \mathfrak{Q} = \int d^3x\, \mathfrak{j}_0(x) =  \frac{i}{2}\int d^3x \left[\gdualn{\mathfrak{a}}(\partial_t \mathfrak{a}) - (\partial_t\gdualn{\mathfrak{a}}) \mathfrak{a} \right].\label{eq:pe03u4}
\end{equation}
}
Therefore, the partition function is\footnote{We should note that owing to the Klein-Gordon kinematics, apart from the trivial global U(1) symmetry, the Lagrangian for mass dimension one field is in fact also endowed with non-trivial finite-dimensional global symmetries. Associated with these global symmetries is a set of conserved charges $\mathfrak{Q}_{a}$ and chemical potentials $\mu_{a}$ which may yield a more complicated partition function. Here, we only consider the simplest possibility and leave the more complicated case for future investigations.}
{
\begin{eqnarray}
Z&=&\int[D\mathfrak{\pi}][D\gdualn{\mathfrak{\pi}}]\int_{\pm} [D\gdualn{\mathfrak{a}}] [D\mathfrak{a}]\exp\Bigg[\int_0^\beta d\tau \int d^3x\nonumber\\
&&\times\bigg(i\pi(\partial_\tau\mathfrak{a}) + i(\partial_\tau\gdualn{\mathfrak{a}})\gdualn{\mathfrak{\pi}}  -\mathfrak{H}(\mathfrak{a},\gdualn{\mathfrak{a}},\mathfrak{\pi},\gdualn{\mathfrak{\pi}}) + i\mu \Big(\gdualn{\mathfrak{a}} \gdualn{\mathfrak{\pi}} - \mathfrak{\pi} \,\mathfrak{a} \Big) \bigg)\Bigg]\,,\label{eq:pe01u21}
\end{eqnarray}
with
\begin{equation}
   \mathfrak{H}(\mathfrak{a},\gdualn{\mathfrak{a}},\mathfrak{\pi},\gdualn{\mathfrak{\pi}})= i\pi(\partial_\tau\mathfrak{a}) + i(\partial_\tau\gdualn{\mathfrak{a}})\gdualn{\mathfrak{\pi}} - \mathfrak{L} =  2\mathfrak{\pi}\gdualn{\mathfrak{\pi}} - \frac{1}{2}\partial^i\gdualn{\mathfrak{a}}\,\partial_i {\mathfrak{a}} + \frac{m^2}{2} \gdualn{\mathfrak{a}} \mathfrak{a}\,.
\end{equation}
}
To perform the functional integration in the conjugate momenta, we write the partition function as:
{
\begin{eqnarray}
Z&=&\int[D\mathfrak{\pi}][D\gdualn{\mathfrak{\pi}}]\int_{\pm} [D\gdualn{\mathfrak{a}}] [D\mathfrak{a}]\exp\Bigg[\int_0^\beta d\tau \int d^3x\nonumber\\
&&\times\Big(- 2\mathfrak{\pi}\gdualn{\mathfrak{\pi}} + i\mathfrak{\pi}( \partial_\tau\mathfrak{a} - \mu\mathfrak{a}) + i(\partial_\tau\gdualn{\mathfrak{a}} + \mu\gdualn{\mathfrak{a}})\gdualn{\mathfrak{\pi}}+ \frac{1}{2}\partial^i\gdualn{\mathfrak{a}}\,\partial_i {\mathfrak{a}} - \frac{m^2}{2} \gdualn{\mathfrak{a}} \mathfrak{a} \Big) \Bigg]\,,\label{eq:pe01u22}
\end{eqnarray}
}
so that the momentum integrals are of the form
{
\begin{eqnarray}
\int[D\mathfrak{\pi}][D\gdualn{\mathfrak{\pi}}]e^{-2\int_0^\beta d\tau \int d^3x\big( \mathfrak{\pi}\gdualn{\mathfrak{\pi}} - \frac{\mathfrak{\pi}A}{2} -\frac{B\gdualn{\mathfrak{\pi}}}{2}\big)} &=&\int[D\mathfrak{\pi}][D\gdualn{\mathfrak{\pi}}]e^{-\int_0^\beta d\tau \int d^3x \big[2\big( \mathfrak{\pi}-\frac{B}{2}\big)\big(\gdualn{\mathfrak{\pi}} - \frac{A}{2}\big)-\frac{BA}{2}\big]}\nonumber\\
&=& N e^{\big[\int_0^\beta d\tau \int d^3x\big(\frac{BA}{2}\big)\big]}\,,\label{eq:pe01u23}
\end{eqnarray}
with $A = i( \partial_\tau\mathfrak{a} - \mu\mathfrak{a})$ and $B=i(\partial_\tau\gdualn{\mathfrak{a}} + \mu\gdualn{\mathfrak{a}})$. The momentum integrations are performed by redefining $\mathfrak{\pi}\to \mathfrak{\pi}+\frac{B}{2}$ and $\gdualn{\mathfrak{\pi}}\to \gdualn{\mathfrak{\pi}}+\frac{A}{2}$. The result for momentum integration is just a normalization factor $N$. The $\mathfrak{a}$ and $\gdualn{\mathfrak{a}}$} integrations can be done by writing the partition function as
{
\begin{eqnarray}
Z=N\int_{\pm}[D\gdualn{\mathfrak{a}}][D\mathfrak{a}]\exp\Bigg[-\frac{1}{2}\int_0^\beta d\tau \int d^3x\int_0^\beta d\tau \int d^3x' \gdualn{\mathfrak{a}}(x)\mathbf{D}\,\mathfrak{a}(x')\Bigg]\,,\label{eq:pe04}
\end{eqnarray}
}
with
\begin{equation}
\mathbf{D}=\left[\overleftarrow{\partial}_\tau \overrightarrow{\partial}_\tau - \overleftarrow{\partial}^i \overrightarrow{\partial}_i + m^2  + \mu (\overrightarrow{\partial}_\tau - \overleftarrow{\partial}_\tau) -\mu^2 \right]\I\label{eq:pe05}
\end{equation}
where $\overleftarrow{\partial}$ and $\overrightarrow{\partial}$ indicate the action of the derivative operators to the left and right respectively. {The path integral over periodic $(+)$ or anti-periodic $(-)$ functions} is well-known\footnote{See, for instance, equation~(8.17) of~\cite{Bailin:1986wt}.} \cite{Kapusta:2006pm,Bailin:1986wt}. The result is
\begin{equation}
Z=N\exp\big(\mp\frac{1}{2}\mathrm{Tr}\ln \mathbf{D}\big)\,,\label{eq:pe06}
\end{equation}
where {the top and bottom signs refer to boson and fermion respectively}. The factor $N$ is just a $\beta$-dependent constant and it can be absorbed by renormalizing $Z$. The trace is evaluated by introducing the Fourier transform of the fields as
\begin{equation}
{\mathfrak{a}}(x)=\frac{1}{\sqrt{V}}\sum_n\int\frac{d^3p}{(2\pi)^3}{e}^{+i(\omega_n\tau+\boldsymbol{p\cdot x})}\,\tilde{\mathfrak{a}}(\omega_n,{\p}),\label{eq:pe07}
\end{equation}
\begin{equation}
\gdualn{\mathfrak{a}}(x)=\frac{1}{\sqrt{V}}\sum_n\int\frac{d^3p}{(2\pi)^3}{e}^{-i(\omega_n\tau+\boldsymbol{p\cdot x})}\,\stackrel{\neg}{\tilde{\mathfrak{a}}}(\omega_n,{\p})\label{eq:pe07b}
\end{equation}
where $\omega_n = \frac{2n\pi}{\beta}$ is the Matsubara frequency for bosons and $\omega_n = \frac{(2n+1)\pi}{\beta}$ is the Matsubara frequency for fermions, with integer $n$. We obtain
\begin{eqnarray}
\mp\frac{1}{2}\mathrm{Tr}\ln \mathbf{D}=\mp 2V\sum_n\int\frac{d^3p}{(2\pi)^3}\ln[\beta^2(\omega_n^2+|\p|^2+m^2+2i\mu\omega_n -\mu^2)]\label{eq:pe08}
\end{eqnarray}
where $V=\int d^3x$ is the total volume of the system. The summation must be carried over the Matsubara frequencies for bosons or fermions {(see \ref{AppA} for further details). The partition function~(\ref{eq:pe06}) for mass dimension one boson $\mathfrak{b}$ and mass dimension one fermion $\mathfrak{f}$ are
\begin{align}
\ln Z=\mp 2V\int\frac{d^3p}{(2\pi)^3}\Bigg[\beta \sqrt{ |\p|^2+m^2}&+ \ln\bigg(1\mp \exp(-\beta \sqrt{|\p|^2+m^2}+\mu)\bigg)\nonumber\\
 &+ \ln\bigg(1\mp\exp(-\beta \sqrt{|\p|^2+m^2}-\mu)\bigg)\Bigg].\label{eq:pe12}
\end{align}
}
{The fermionic partition function is the standard one. The bosonic partition function differs from its scalar field counterpart by a factor of four. This difference comes from the trace of the identity matrix. Each term in~(\ref{eq:pe12}) has a clear physical interpretation. The first term is the zero-point energy while the second and third terms are the contributions from particle and anti-particle respectively.}

The Helmholtz free-energy is
\begin{equation}
F=-\frac{1}{\beta}\ln Z.\label{eq:pe13}
\end{equation}
The main thermodynamic quantities namely, the total energy $E$, pressure $P$ and entropy $S$, are obtained from the Helmholtz free energy
\begin{equation}
E=F+{T}S,\quad P=-\Bigg(\frac{\partial F}{\partial V}\Bigg)_{{T}},\quad S=-\Bigg(\frac{\partial F}{\partial {T}}\Bigg)_V\,.\label{eq:pe14}
\end{equation}
Taking the temperature-dependent part of (\ref{eq:pe13}) and performing an integration by parts, we obtain{
\begin{equation}
\frac{F}{V}=-\frac{1}{3\pi^2}\int_0^\infty\frac{p^4}{\sqrt{p^2+m^2}}\bigg[\frac{1}{{e}^{\beta(\sqrt{p^2+m^2}-\mu)}\mp 1} + \frac{1}{{e}^{\beta(\sqrt{p^2+m^2}+\mu)}\mp 1}\bigg]dp.\label{eq:pe15}
\end{equation}
}
The term in the square bracket gives the standard particle distribution function. {The negative and positive signs are for the Bose-Einstein and Fermi-Dirac distributions respectively. These results further confirm that the statistics satisfied by $\mathfrak{b}(x)$ and $\mathfrak{f}(x)$ fields are bosonic and fermionic respectively.}


{As an application to the high-temperature limit $({T}\gg m)$, the integral in (\ref{eq:pe15}) is easily calculated~\cite{Bellac:2011kqa,Kapusta:2006pm,Das:1997gg,Bailin:1986wt}. The energy density $\epsilon$ from (\ref{eq:pe14}) are:
\begin{equation}
\mbox{boson: }\epsilon=\frac{2\pi^2{T}^4}{15}, \quad \mbox{fermion: }\epsilon=\frac{7\pi^2{T}^4}{60}.
\end{equation}
The result for the bosonic field differs from its scalar field counterpart by a factor of four. For the fermionic field, its energy density is identical to the Dirac field.\footnote{The above treatment for the fermionic field and main results were first obtained in \cite{Pereira:2018xyl}.} The high temperature energy density has important implications for the relic density of the universe and deserves further studies.
}

{The pressure integral (\ref{eq:pe14}-\ref{eq:pe15}) in the low-temperature limit is particularly interesting for fermions due to the Pauli exclusion principle. The fermionic degeneracy pressure has recently been proposed as a possible mechanism to prevent the gravitational collapse of gases, thereby leading to the formation of dark matter halos~\cite{Destri:2013pt, deVega:2013jfy, Domcke:2014kla, Randall:2016bqw, Pal:2019tqq, Barranco:2018gjg}. The dwarf spheroidal galaxies which are supposedly dominated by dark matter, are good laboratories for testing such models. Larger spheroidal and elliptic  galaxies can also be considered with some additional assumptions concerning their substructures.
}

{The equation for the hydrostatic equilibrium between gradient of pressure $P$ and gravitational attractive force is given by
\begin{equation}
    \frac{1}{r^2} \frac{d}{dr} \left(\frac{r^2}{\rho} \frac{dP}{dr} \right) = -4\pi G \rho\label{eq:pe21}
\end{equation}
where $\rho(r)$ is the mass density. Given the relation $P = P(\rho)$, equation~(\ref{eq:pe21}) can be solved for $\rho(r)$ satisfying the boundary conditions $\rho = \rho_0$ and $d\rho/dr=0$ at $r=0$. The total mass $M$ of the system is related to its characteristic radius $R$ by
\begin{equation}
M(R)=4\pi\int_0^R r^2 \rho(r) dr\,.\label{eq:pe23}
\end{equation}
For a degenerate gas ($T\rightarrow 0$), the Fermi-Dirac distribution function (\ref{eq:pe15}) becomes a step function
\begin{equation}
n_{FD}= \left\{ \begin{array}{rcl}
1 & \mbox{for}
& p\leq p_{F} \\ 
0 & \mbox{for} & p>p_{F} 
\end{array}\right.
\end{equation}
where $p_F = (3h^3\rho/8\pi m)^{1/3}$ is the Fermi momentum associated to a particle of mass $m$. The integral (\ref{eq:pe15}) for a non-relativistic particle satisfying $p_F \ll m$ gives the well known result for the equation of state relating pressure and mass density
\begin{equation}
P=\frac{h^2}{5m^{8/3}}\bigg(\frac{3}{8\pi}\bigg)^{2/3}\rho^{5/3}.\label{eq:pe25}
\end{equation}
Given a fermionic mass $m$ and a central density $\rho_0$, equation~(\ref{eq:pe21}) can be solved for $\rho(r)$.}

{For standard fermionic dark matter, Domcke and Urbano~\cite{Domcke:2014kla} used a completely degenerate model to constrain $m$ in dwarf galaxies in the range of $100\,\mbox{eV}\sim 200\,\mbox{eV}$. Randall et al.~\cite{Randall:2016bqw} treated the
dark matter as a quasi-degenerate Fermi gas surrounded by a thermal envelope which reduces these constraints to $m \gtrsim 50\,\mbox{eV}$.  For large galaxies such as the Milky Way, the main observational evidence for the existence of a dark matter halo around the galactic nuclei comes from the rotation curves. Contrary to dwarf galaxies, which do not rotate, it is well known that the presence of dark matter in large galaxies drastically alters the dynamic of rotation for stars around its centre, mainly at large distances. The rotation velocity of a star at a distance $r$ of the galactic centre is given by $V(R)=\sqrt{GM(R)/R}$, with $M(R)$ given by (\ref{eq:pe23}). Using data from a set of large
elliptical and spiral galaxies and from a small set of dwarf galaxies, Pal et al. \cite{Pal:2019tqq} found that their model can
explain most of the bulk galactic properties, as well as some of the features observed in the rotation
curves, provided that $m\sim 50\,\mbox{eV}$. By constraining light fermionic dark matter with Milky Way observations, J.~Barranco et al.~\cite{Barranco:2018gjg} found  $29\,\mbox{eV} <m<33\,\mbox{eV}$.
 }           

{
For a mass dimension one fermion, reference~\cite{Pereira_2021} presented a detailed study of the rotation curve for the Milky Way with a dark matter halo maintained by a degeneracy pressure of the form~(\ref{eq:pe25}) with a fermion mass $m \simeq 23$ eV. A study for the mass-ratio relation for dwarf galaxies was also presented in~\cite{Pereira_2021}.}



\section{ELKOs in braneworld models}\label{ElkoOnBrane}


{Recently, much consideration has been paid to higher-dimensional field
theory and the possibility that extra dimensions may be discovered in the
Large Hadron Collider or in the LIGO-Virgo detectors~\cite{Caldwell:2001ja,Yu:2016tar,Visinelli:2017bny,Pardo:2018ipy,LIGOScientific:2018dkp,Lin:2020wnp}. Therefore, the study of the higher-dimensional ELKO is a useful and important subject.}

{
In this section, we review the localization of higher-dimensional ELKO. For simplicity, we mostly focus on theories in five dimensions where the mass dimensionality of ELKO is three half. After introducing some braneworld models, we study the localization of free and interacting ELKO on flat and bent branes. These branes are four-dimensional so they correspond to the observable universe. In braneworld models, the branes are embedded in a higher (in this case five-dimensional) bulk space-time.}
 
 {The four-dimensional action is obtained by performing the Kaluza-Klein (KK) decomposition on the five dimensional action. The decomposition yields an infinite tower of discrete modes on the four-dimensional branes. Stability demands the zero mode to be massless whereas the higher modes are massive. We show that the former can be localized by introducing interactions (the Yukawa-like, non-minimal and derivative couplings) while the latter cannot be localized. Lastly, we compare the localization property of ELKO with the scalar, the U(1) vector, the Kalb-Ramond and the Dirac field. The application of ELKOs in cosmology can be found in~\cite{Fabbri:2010ws,Wei:2011yr}.
}

\subsection{Braneworld models}

Braneworld models are important subjects in the realm of theoretical
physics. They provide new insights to fundamental problems
such as the gauge hierarchy problem and the cosmological constant problem. Since the works of Arkani-Hamed-Dimopoulos-Dvali (ADD)~\cite{ArkaniHamed:1998rs} and Randall-Sundrum (RS)~\cite{Randall:1999ee,Randall:1999vf}, braneworld models have received increasing attention. In the ADD and RS models, our universe is a four-dimensional sub-manifold or brane embedded in a {higher-dimensional} space-time. The SM particles are localized on {the brane} while gravity can propagate in the bulk. {In the RS models, the branes are} assumed to be infinitely thin and {they do} not provide the smallest scale in a fundamental theory. When the inner structures of the branes are taken into account, one should consider realistic branes with finite thickness which can be dynamically generated by bulk matter fields~\cite{DeWolfe:1999cp,Kobayashi:2001jd,Bazeia:2002sd,Liu:2009ega}, or can be realized by pure gravity~\cite{BarbosaCendejas:2005kn,Zhong:2015pta}. Furthermore, various kinds of symmetry of {branes} lead to different braneworld models such as {Minkowski (also called the flat brane),} de Sitter (dS), anti-de Sitter (AdS), Friedmann-Robertson-Walker, and other branes. For more details on braneworld models, one can refer to~\cite{Dick:2000dt,Dick:2001sc,Dick:2001np,Hahn:2001ze,Boehmer:2007xh,Boehmer:2008zh,Boehmer:2009bmu,Lee:2009zzs,Cruz:2009ne,Cruz:2011kj,Costa:2013eua,Sousa:2014dpa,Cruz:2013uwa}.

In this section, we introduce several types of RS-like braneworld models. For reviews of more models, one can see~\cite{Dzhunushaliev:2009va,Liu:2017gcn}.  We use capital Latin letters {$M,N,\cdots$ and Greek letters $\mu,\nu,\cdots$} for the higher-dimensional and four-dimensional space-time indices. We mainly consider {the case of five-dimensional bulk. The extra dimension coordinate is denoted by $y$ or $z$}. Symbols with a sharp hat describe four-dimensional quantities on the brane.

\subsubsection{Randall-Sundrum-like thin branes}


The RS-I model~\cite{Randall:1999ee} provides a new solution to the hierarchy problem by using a non-factorizable warped geometry. This model contains a curved compact extra dimension in which only gravity propagates in the bulk while the SM fields are bounded to one of the two thin branes. The curvature of the bulk causes a scaling of the Planck scale $M_{\text{Pl}}$ down to the electroweak scale as expected. The ansatz for the metric is usually given by
\begin{equation}
ds^{2}=e^{2{A(y)}}ds^{2}_{\text{b}}+dy^{2}\label{5Dmetric_y}
\end{equation}
with the brane metric
\begin{equation} \label{BraneMetric1}
ds^{2}_{\text{b}}=\hat{g}_{\mu\nu}dx^{\mu}dx^{\nu}.
\end{equation}
Here, the warp factor is $e^{A(y)}$  where $y$ is the coordinate of the extra dimension. The factor $e^{2{A(y)}}$ means that this metric is non-factorizable. Specifically, the bulk manifold cannot be expressed as a product of the four-dimensional space-time and the extra dimension. The brane metric can be explicitly written as
\begin{equation}\label{BraneMetric2}
ds^{2}_{\text{b}}=\left\{\begin{array}{ll}
         \eta_{\mu\nu}dx^{\mu}dx^{\nu} & \text{for flat brane} \\
          -dt^{2}+e^{2Ht}\delta_{ij}dx^{i}dx^{j} & \text{for dS brane} \\
         e^{2{Hx_{3}}}(-dt^{2}+dx_{1}^{2}+dx_{2}^{2})+dx_{3}^{2} & \text{for AdS brane}
       \end{array}\right.
\end{equation}
with $H>0$. It will be convenient to work with a new extra dimensional variable $z$, which is related to the physical coordinate $y$ through
\begin{equation}\label{CoordinateTransformation1}
dz=e^{-{A(y)}}dy.
\end{equation}
With the above transformation, the metric~(\ref{5Dmetric_y}) can be rewritten as
\begin{equation}\label{5Dmetric_z}
ds^{2}=e^{2{A(z)}}(ds^{2}_{\text{b}}+dz^{2}).
\end{equation}

\begin{figure}
\begin{center}
  \centering
  \includegraphics[width=5.5cm,height=4.5cm]{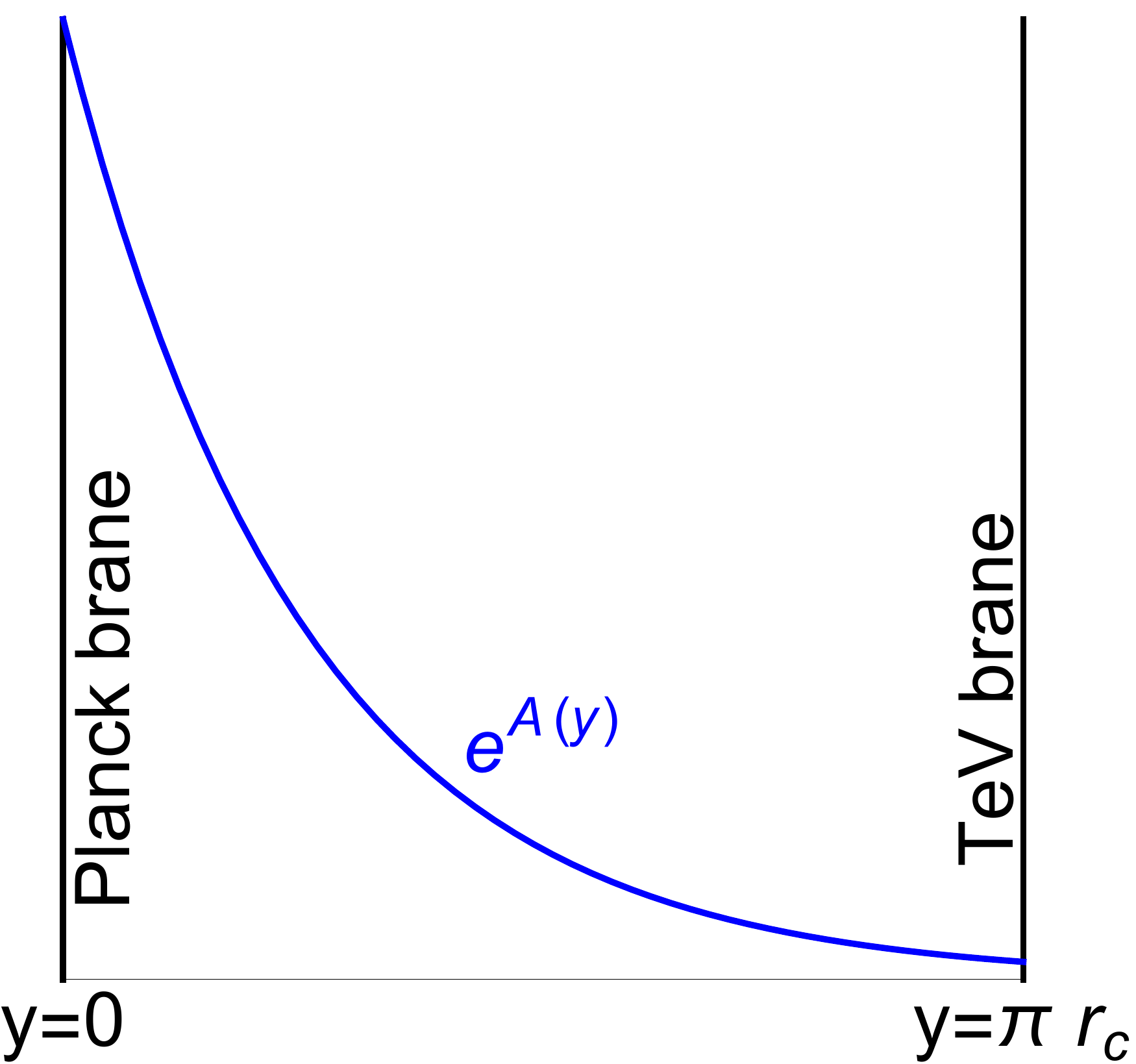}\\
  \end{center}
  \caption{The picture of the RS-I model.}\label{figRSIpicture}
\end{figure}

In the RS-I~\cite{Randall:1999ee} model, the extra dimension is compact on the orbifold ${S_{1}}/{Z_{2}}$ with a radius $r_{c}$. This model involves a `Planck brane' and a `TeV brane' located at the two fixed points of the orbifold, one at the origin $y = 0$
and another at $y=\pi r_{c}$. The space between the two branes is a slice of an AdS space. See fig.~\ref{figRSIpicture} for the picture of the RS-I model.

Using the induced metric on the flat brane, one can obtain the
following solution by solving the five-dimensional Einstein equations:
\begin{equation}\label{RSIsolution}
A(y)=-k|y|, \qquad k=\sqrt{\frac{-\Lambda}{24M^{3}}},
\end{equation}
where $k$ is a parameter of the order of the five-dimensional fundamental
scale $M$, and $\Lambda$ is the five-dimensional cosmological constant. Another result obtained from the Einstein equations is that the two-branes will have equivalent
but opposite tensions such that {$V_{\text{Pl}}=-V_{\text{TeV}}=24M^{3}k$}. Furthermore, the relation between the four-dimensional Planck scale $M_{\text{Pl}}$ and
the five-dimensional fundamental one $M$ is given by:
\begin{equation}
M^{2}_{\text{Pl}}=\frac{M^{3}}{k}(1-e^{-2kr_{c}\pi}).
\end{equation}
It is clear that $M^{2}_{\text{Pl}}$  weakly depends on the size of the extra dimension $r_{c}$. By assuming $k r_{c}\approx 10$, we have $M_{\text{Pl}}\approx M$, which means that the scale of the five-dimensional gravity itself will remain of the order of the Planck scale. While in this model, {the four-dimensional effective weak scale is exponentially suppressed (or redshifted) along the extra dimension from the fundamental Planck scale to the four-dimensional TeV one. Thus,
the RS-I model provides a completely new solution to the gauge Hierarchy Problem.}

In~\cite{Yang:2012dd,Guo:2018uxx}, a new mechanism with a warped extra dimension was proposed to solve the gauge hierarchy problem. In this scenario, the fundamental scale is assumed to be TeV while the four-dimensional scale of gravity is generated from the exponentially warped extra dimension, which can be ensured if our world is confined on the positive tension brane instead of the negative one. Such mechanism can be realized in the non-minimally coupled gravity theory, which is different from the minimally coupled one for the RS-like scenario. In this new scenario, there is a tower of spin-two excitations with very tiny mass gap $10^{-4}$eV. The experimental consequences are similar to the ADD model with large extra dimensions.

In contrast to the RS-I model, the RS-II model contains only one brane with positive brane tension and an infinitely large extra dimension. If we consider the decompactification limit $r_{c}\rightarrow\infty$ {in the RS-I model}, the effective Planck mass remains finite and the gravity is localized on the brane at $y=0$. For this model, the warp factor as well as cosmological constants in the bulk and on the flat brane can be achieved as
\begin{equation}\label{RSIIsolution}
A(y)=-k|y|, ~ \Lambda=-24M^{3}k^2, ~ V_{\text{Brane}}=24M^{3}k.
\end{equation}
The idea of infinite extra dimension provides us with more interesting results for studying the localization mechanisms of matter fields in the braneworld models. So in the following discussions, we mainly consider the RS-II model for the localization of ELKOs.

In the above, we have reviewed the models with flat thin branes. It was shown in~\cite{Kaloper:1999sm,Nihei:1999mt,DeWolfe:1999cp,Karch:2000ct} that the solutions of the warp factor $A(y)$ for dS and AdS thin branes are respectively given by
\begin{eqnarray}
A(y)\!\!\! &=& \!\!\! \ln\left[\frac{H}{b}\sinh(b|y|-\sigma)\right], \label{WarpFactorRSIIdS}\\
A(y)\!\!\! &=&\!\!\! \ln\left[\frac{H}{b}\cosh(b|y|-\sigma)\right], \label{WarpFactorRSIIAdS}
\end{eqnarray}
where $b$ and $\sigma$ are two parameters.

\subsubsection{Flat thick branes}


In the RS-II model, the thickness and inner structure of the brane are neglected. In more realistic models, where the most fundamental theory would have a minimal length scale, the thickness of the brane should be taken into account and the brane can be dynamically generated by matter fields.
Thick braneworld models that have been proposed as a smooth generalization of
the RS-II model give us new possibilities and problems. These branes are usually generated by one or more bulk scalar fields with kink configurations which can be minimally~\cite{Kehagias:2000au} or non-minimally~\cite{Guo:2011wr} coupled to gravity. Various sorts of bulk scalar fields lead to different kinds of braneworld models, see~\cite{Wang:2002pka,Caldwell:1999ew,Minamitsuji:2005gt,Olasagasti:2003fw,German:2012rv} or the reviews~\cite{Dzhunushaliev:2009va,Liu:2017gcn}. We will review some thick brane models constructed by a scalar field where the branes are embedded in five- or six-dimensional AdS space-times.

First, we consider the flat thick brane model with a single scalar field $\phi$ minimally coupled to gravity. The action of {this brane model} is given by
\begin{equation}\label{flatThickBrane1}
S=\int d^{5}x\sqrt{-g}\left[{ \frac{1}{2\kappa_{5}^{2}} R}-\frac{1}{2}g^{MN} \partial_M\phi \partial_N \phi -V(\phi)\right ]
\end{equation}
where $R$ is the five-dimensional scalar curvature, $g=\det(g_{MN})$, {and $\kappa^{2}_{5}=8 \pi G_{N}^{(5)}=1/(4M^3)$ with $G_{N}^{(5)}$ the five-dimensional Newton constant and $M$ is the fundamental mass scale. In order to be consistent with the symmetry of the Minkowski brane, the scalar field $\phi(y)$ and the warp factor $e^{A(y)}$ depend only on
the extra dimension coordinate $y$.} By using the flat brane metric~(\ref{BraneMetric2}) and the following sine-Gordon potential
\begin{equation}\label{flatThickBrane1V}
V(\phi)=\frac{3}{2}c^{2}\left[3b^{2} \cos^{2}(b\phi)-4\sin^{2}(b\phi)\right],
\end{equation}
one can get the following solutions of the warp factor and scalar field~\cite{Gremm:1999pj,Afonso:2006gi,Dzhunushaliev:2009va}
\begin{equation}\label{flatThickBrane1Solution1}
e^{A(y)}=\text{sech}^{\frac{1}{3b^{2}}}(ky),\qquad \phi(y)=\frac{2}{b}\text{arctan}\left[\tanh\Big(\frac{3}{2}ky\Big)\right].
\end{equation}
Here, the parameters $b$ and $c$ are positive. They are also related to the thickness of
the brane and the AdS curvature with  $k=cb^{2}$. In this solution, we have set $\kappa^{2}_{5}=1/2$.

{In addition, similar thick brane solutions have also been obtained for other scalar potentials or in other gravity theories. Most of these solutions have the following form or similar shape~\cite{Dzhunushaliev:2009va,Liu:2012gv,Bogdanos:2006qw}:
\begin{equation}\label{flatThickBrane1Solution2}
e^{A(y)}=\text{sech}^{\gamma}(ky),\qquad \phi(y)=\phi_{0}\tanh(ky),
\end{equation}
where $\gamma$, $\phi_{0}$, and $k$ constants.

For the string-like defect in six-dimensional space-time, one can refer to the Gherghetta-Shaposhnikov model~\cite{Gherghetta:2000qi} and the Hamilton string-cigar model~\cite{Silva:2012yj}.}

\subsubsection{de Sitter/anti de Sitter thick branes}


In previous introduction, we have reviewed branes with flat geometry~\cite{Randall:1999vf,Liu:2017gcn,Guo:2018uxx,Kaloper:1999sm,Afonso:2006gi,Liu:2012gv,Bogdanos:2006qw,Gherghetta:2000qi}. Now, we focus on warped dS/AdS thick branes with the brane metric~(\ref{BraneMetric2}). Solutions of single scalar field generated dS/AdS thick branes {can be found} in~\cite{Afonso:2006gi,Zhang:2016ksq}
\begin{eqnarray}
A(z)\!\!\! &=& \!\!\!-\frac{1}{2}\ln \left[a^{2}s(1+\Lambda_{4})\cosh^{2}(hz)\right] , \label{dS_AdS_ThickBrane1_Az}\\
\phi(z)\!\!\! &=& \!\!\!\frac{1}{b}\text{arcsinh}[\sinh(hz)] , \label{dS_AdS_ThickBrane1_phiz}
\end{eqnarray}
where $h=\sqrt{\frac{1+\Lambda_{4}}{s}}$. The parameters $a$, $b$, $s$ are real with $s\in (0,1]$ and $b=\sqrt{\frac{2+(1+\Lambda_{4})}{(3+(1+s)\Lambda_{4})}}$. The four-dimensional cosmological constant of the dS/AdS brane $\Lambda_{4}$, is related to the parameter $H$ by $\Lambda_{4}=\pm3H^{2}$. Note that the coordinate $z$ is related to the extra dimensional coordinate $y$ with~(\ref{CoordinateTransformation1}). {Other dS and AdS brane solutions were considered in~\cite{Wang:2002pka,Kamali:2015yya,Bazeia:2003cv,German:2012rv,Herrera-Aguilar:2014oua,Chen:2015ana,Liu:2009uca}.}

\subsection{Localization of ELKOs}

In the context of the braneworld physics, the localization of various matter fields by a natural mechanism is an important and interesting subject. This investigation guides us as to which kind of brane structure is more acceptable phenomenologically. In order to recover the SM of particle physics in our four-dimensional universe (the brane), the zero modes of matter fields should be localized on the brane. In~\cite{Gherghetta:2000qi,Randall:1999vf,Bajc:1999mh,Liang:2009zzc,Tofighi:2014jla,Guo:2010az}, it was found that the massless {free scalar fields} and {gravitons} can be localized on the RS thin brane, its generalizations and thick branes.
However, {free gauge fields} can be localized only in some non-RS-II models with
non-flat branes~\cite{Liu:2008wd,Liu:2009dwa,Guo:2011qt,Gogberashvili:2011ua,Herrera-Aguilar:2014oua}. In order to localize gauge vector and tensor fields on thick RS-like branes in five-dimensional AdS space-time, one or more extra terms in the standard action are needed~\cite{Moazzen:2017jlf,Guerrero:2009ac,Tofighi:2015lpn,Chumbes:2011zt,Cruz:2010zz}. {By introducing} the Yukawa coupling between
a background scalar field and fermions, the issue of fermion localization and fermion resonances can be achieved~\cite{Cruz:2011ru,Barbosa-Cendejas:2015qaa,Correa:2010zg,Xie:2013rka,Almeida:2009jc}.
{The idea of fermion localization in some extra-dimensional models was also used
for solving the doubling problem on the lattice~\cite{Kaplan:1992bt} and the fermion mass hierarchy problem~\cite{Gherghetta:2000qt,Huber:2000ie,Gherghetta:2010cj}. Also, the existence of a new phase, the so-called Layer phase, on the lattice has been found by using
higher-dimensional gauge field localization~\cite{Dimopoulos:2006qz,Farakos:2020inf}.}


The theory of ELKO in higher-dimensional space-time was investigated in~\cite{Liu:2011nb,Zhou:2017bbj,Sorkhi:2018jhy,MoazzenSorkhi:2020fqp}. Here, we give a brief review of the localization of five-dimensional ELKOs on branes. A general action of ELKO coupled with the geometry and background scalar field in five-dimensional curved space-time is usually given by
\begin{eqnarray}
S_{\text{ELKO}}\!\!\! &=& \!\!\!\int d^5 x\sqrt{-g}\left[-\frac{1}{4}fg^{MN}\left(\mathfrak{D}_{M}\mathop \lambda\limits^\neg\mathfrak{D}_{N}\lambda+\mathfrak{D}_{N}\mathop \lambda\limits^\neg\mathfrak{D}_{M}\lambda\right)
 -\eta F \mathop \lambda\limits^\neg\lambda
 \right]\label{n-dim_action for Elko}
\end{eqnarray}
where $f$ and $F$ are two general functions of the Ricci scalar $R$ and the background scalar field $\phi$ generating the brane while $\eta$ is the `Yukawa coupling constant'.
The covariant derivative in the above action~(\ref{n-dim_action for Elko}) is defined as
\begin{eqnarray}
\mathfrak{D}_{M}\lambda=(\partial_{M}+\omega_{M})\lambda,~~\mathfrak{D}_{M}\mathop \lambda\limits^\neg=\partial_{M}\mathop \lambda\limits^\neg-\mathop \lambda\limits^\neg\omega_{M}, \label{covariant derivatives}
\end{eqnarray}
with the spin-connection $\omega_{M}$ given by
\begin{eqnarray}
   \omega_{M}=-\frac{i}{2}\left(e_{P\bar{A}}{e^{N}}_{\bar{B}} \Gamma^{P}_{MN}
                             -{e^{N}}_{\bar{B}}\partial_{M}e_{N\bar{A}}\right)
                  S^{\bar{A}\bar{B}} \label{SpinConnection1}
\end{eqnarray}
where $\Gamma^{P}_{MN}$ is the connection, $e_{M\bar{A}}$ is the vierbein satisfying the relation $e_{M\bar{A}} e_{N\bar{B}} \eta^{\bar{A}\bar{B}} = g_{MN}$, and  $S^{\bar{A}\bar{B}}$ is determined by the $\Gamma^{\bar{A}}$ in five-dimensional flat space-time:
\begin{eqnarray}
   S^{\bar{A}\bar{B}} = \frac{i}{4}[\Gamma^{\bar{A}},\Gamma^{\bar{B}}].
    \label{tangent space connection}
\end{eqnarray}
Here, the capital letters with bar $\bar{A},~\bar{B},\cdots=0,1,2,3, 5$ denote the five-dimensional local Lorentz indices and $\Gamma^{\bar{A}}$ satisfy $\big\{ \Gamma^{\bar{A}} , \Gamma^{\bar{B}} \big \} = 2\eta^{\bar{A}\bar{B}}$. The spin-connection~(\ref{SpinConnection1}) can also be written as
\begin{equation}
    \omega_M=\frac{1}{4}\omega_M^{\,\,\,\,\bar{A}\bar{B}}\Gamma_{\bar{A}}\Gamma_{\bar{B}}
    \label{SpinConnection2}
\end{equation}
with
\begin{eqnarray}
     \omega_M^{\,\,\,\,\bar{A}\bar{B}}
     = \frac{1}{2}e^{N\bar{A}}\Big(\partial_Me^{\,\,\,\,\bar{B}}_N-\partial_Ne^{\,\,\,\,\bar{B}}_M \Big)
     -\frac{1}{2}e^{N\bar{B}}\Big(\partial_Me^{\,\,\,\,\bar{A}}_N-\partial_Ne^{\,\,\,\,\bar{A}}_M \Big)
     -\frac{1}{2}e^{P\bar{A}}e^{Q\bar{B}}e^{\,\,\,\,\bar{C}}_M \Big(\partial_P e_{Q\bar{C}}-\partial_Q e_{P\bar{C}}\Big).
\end{eqnarray}
{
The gamma matrices are $2^{D/2} \times 2^{D/2}$ or $2^{(D-1)/2} \times 2^{(D-1)/2}$ matrices for even and odd $D$ dimensions respectively where $D>2$~\cite{Kugo:1982bn}.
Therefore, the five-dimensional ELKO is a four-component spinor}
From~(\ref{n-dim_action for Elko}), the equation of motion for the non-minimally coupled ELKO is given by
\begin{eqnarray}
\frac{1}{\sqrt{-g}f}\mathfrak{D}_{M}(\sqrt{-g}f g^{MN}\mathfrak{D}_{N}\lambda) -2\eta F \lambda=0.
\label{5DElkoEoM1}
\end{eqnarray}

We will focus on the localization of ELKOs on branes in the five-dimensional space-time with the metric ansatz~(\ref{5Dmetric_y}) or~(\ref{5Dmetric_z}).
The non-vanishing components of the spin-connection~(\ref{SpinConnection1}) for the metric~(\ref{5Dmetric_z}) in five-dimensional space-time are given by
\begin{eqnarray}
\omega_\mu=\frac{1}{2}(\partial_z A) \gamma_\mu\gamma_5+\hat{\omega}_{\mu} \label{NonZeroSpinConnection1}
\end{eqnarray}
where $\hat{\omega}_\mu$ is the spin-connection with respect to the four-dimensional induced metric $\hat{g}_{\mu\nu}$ on the brane and vanishes for a flat brane.

From the metric~(\ref{5Dmetric_z}) and the non-vanishing spin-connection~(\ref{NonZeroSpinConnection1}), the equation of motion~(\ref{5DElkoEoM1}) for ELKO can be rewritten as
\begin{align}
  \frac{1}{\sqrt{-\hat{g}}}\hat{\mathfrak{D}}_{\mu}(\sqrt{-\hat{g}}\hat{g}^{\mu\nu}\hat{\mathfrak{D}}_{\nu}\lambda)
  &+\frac{1}{2}(\partial_z A)\left[\hat{\mathfrak{D}}_{\mu}(\hat{g}^{\mu\nu}\gamma_{\nu}\gamma_{5}\lambda)
 +\hat{g}^{\mu\nu}\gamma_{\mu}\gamma_{5}\hat{\mathfrak{D}}_{\nu}\lambda\right]
 \nonumber\\
 &-\frac{1}{4}(\partial_z A)^{2}\hat{g}^{\mu\nu}\gamma_{\mu}\gamma_{\nu}\lambda
 +e^{-3A}f^{-1}\partial_{z}(e^{3A}f\partial_{z}\lambda)
 -2\eta\,e^{2A} F\lambda = 0
  \label{5DElkoEoM2}
\end{align}
with $\hat{\mathfrak{D}}_{\mu} = \partial_{\mu} + \hat{\omega}_{\mu}$.
Noting that  $\hat{\mathfrak{D}}_{\mu}\hat{g}^{\lambda\rho}=\hat{\mathfrak{D}}_{\mu}(\hat{e}_{a}^{~\lambda}\hat{e}^{a\rho})=0$ and $\hat{\mathfrak{D}}_{\mu}\hat{e}^{a}_{~\nu}=0$, we can simplify the above equation to
\begin{eqnarray}
 \frac{1}{\sqrt{-\hat{g}}}\hat{\mathfrak{D}}_{\mu}(\sqrt{-\hat{g}}\hat{g}^{\mu\nu}\hat{\mathfrak{D}}_{\nu}\lambda)
  -\! (\partial_z A)\gamma^{5}\gamma^{\mu} \hat{\mathfrak{D}}_{\mu}\lambda
   \!-\! (\partial_z A)^{2}\lambda
   +e^{-3A}f^{-1}\partial_{z}(e^{3A}f\partial_{z}\lambda)-2\eta\,e^{2A} F\lambda=0.~~
  \label{5DElkoEoM3}
\end{eqnarray}

Next, we perform a KK decomposition for ELKO to solve~(\ref{5DElkoEoM3}) and reduce the higher-dimensional fundamental action~(\ref{n-dim_action for Elko}) to a four-dimensional effective one.
{We start with the following relations in curved space-time:
\begin{align}
\gamma^{\mu}\hat{\mathfrak{D}}_{\mu}\varsigma_{\pm}(x) &=\mp {i}m \varsigma_{\mp}(x),
&\gamma^{\mu}\hat{\mathfrak{D}}_{\mu}\tau_{\pm}(x) &= \pm {i}m \tau_{\mp}(x), &\empty &\empty \label{Elko4D1}\\
\gamma^5\varsigma_{\pm}(x) &= \pm\tau_{\mp}(x),
&\gamma^5\tau_{\pm}(x) &= \mp\varsigma_{\mp}(x)&\empty&\empty \label{Elko4D2}
\end{align}
where $\varsigma_\pm(x)$ and $\tau_{\pm}(x)$ are four types of ELKOs in four-dimensional space-time. Similarly, their duals satisfy
\begin{align}
\hat{\mathfrak{D}}_{\mu}{\mathop \varsigma\limits^\neg}_{\pm}\gamma^{\mu}
 &= \pm {i}m{\mathop \varsigma\limits^\neg}_{\mp}, &\hat{\mathfrak{D}}_{\mu}{\mathop \tau\limits^\neg}_{\pm}\gamma^{\mu}&=\mp {i}m{\mathop \tau\limits^\neg}_{\mp}, \label{eq:dElko4D1} &\empty&\empty\\
{\mathop \varsigma\limits^\neg}_{\pm}\gamma^5
 &=\mp{\mathop \tau\limits^\neg}_{\mp},
&{\mathop \tau\limits^\neg}_{n\pm}\gamma^5&=\pm{\mathop \varsigma\limits^\neg}_{\mp}. &\empty&\empty\label{eq:dElko4D2}
\end{align}
 These relations are generalizations of their flat space-time counterparts obtained by replacing $\partial_{\mu}$ with $\hat{\mathfrak{D}}_{\mu}$. From~(\ref{Elko4D1}-\ref{Elko4D2}) and the term $-(\partial_z A)\gamma^{5}\gamma^{\mu} \hat{\mathfrak{D}}_{\mu}\lambda$ that appears in~(\ref{5DElkoEoM3}), we observe that the general solution is a linear combination of two different types of ELKO
\begin{align}
\lambda&=\lambda_{+} + \lambda_{-},\\
\gdualn{\lambda}&=\gdualn{\lambda}_{+} + \gdualn{\lambda}_{-} \label{KK decomposition0}
\end{align}
where $\lambda_{\pm}$ and $\gdualn{\lambda}_{\pm}$ are given by the following general KK decomposition
\begin{align}
\lambda_{\pm}&\equiv \sum_n \lambda_{n\pm}
 = \sum_n  e^{-3A/2} f^{-\frac{1}{2}} \left[\alpha_{n\pm}(z)\varsigma_{n\pm}(x)+\beta_{n\pm}(z)\tau_{n\pm}(x)\right], \label{KK decomposition}\\
\gdualn{\lambda}_{\pm}&\equiv\sum_{n}\gdualn{\lambda}_{n\pm}=
 \sum_{n} e^{-3A/2} f^{-\frac{1}{2}}
          \left[\alpha_{n}^{*}(z){\mathop \varsigma\limits^\neg}_{n\pm}(x)
               +\beta_{n}^{*}(z){\mathop \tau\limits^\neg}_{n\pm}(x)
          \right]. \label{KK decomposition2}
\end{align}
Here, we impose the following orthonormality conditions for $\alpha_{n}$ and $\beta_{n}$
\begin{eqnarray}
\int \alpha^{*}_{n}\alpha_{m}dz\!\!\! &=& \!\!\!\delta_{nm},\label{orthonormality relation 1}\\
\int \beta_{n}^{*}\beta_{m}dz\!\!\! &=& \!\!\!\delta_{nm},\\
\int \alpha^{*}_{n}\beta_{m}dz\!\!\! &=& \!\!\!\int \alpha_{n}\beta^{*}_{m}dz=\delta_{nm}.
\end{eqnarray}
Note that the factor $e^{-3A/2} f^{-\frac{1}{2}}$ is not necessary in the extra-dimensional part but its introduction removes the first derivative terms of $\alpha_{n\pm}(z)$ and $\beta_{n\pm}(z)$ in their equations of motion. Here, the linearly-independent four-dimensional ELKO modes $\varsigma_{n\pm}(x)$ and $\tau_{n\pm}(x)$ satisfy~(\ref{Elko4D1}-\ref{eq:dElko4D2}) with $m$ replaced by $m_{n}$.}

The corresponding equations for the ELKO KK modes $\alpha_{n\pm}$ and $\beta_{n\pm}$ are given by
\begin{eqnarray}
\left[\alpha_{n\pm}''
        -V(z) \alpha_{n\pm}
        +m_n^2\alpha_{n}
        - {i} m_{n} A' \beta_{n\pm}
  \right]
  \varsigma_{n\pm}
  +\left[\beta_{n\pm}''\,
        -V(z) \beta_{n\pm}
        +m_n^2\beta_{n}\,
        - {i}m_{n}A'\alpha_{n\pm}
    \right]
    \tau_{n\pm}
    = 0 \label{KKequation3}
\end{eqnarray}
where
\begin{eqnarray}
 V(z) =  \frac{3}{2}A'' +\frac{13}{4}A'^2
               -\frac{1}{4} \frac{f'^2}{f^2} + \frac{3}{2} \frac{f'}{f}A' +\frac{1}{2}\frac{f''}{f}
               +2\eta\,e^{2A}F.   
                \label{effective potential Vz} 
\end{eqnarray}
The primes denote differentiation with respect to $z$ from now on.
Note that $\lambda_+$ and $\lambda_-$ are linearly-independent and the equations for ($\alpha_{n+},~\beta_{n+}$) and ($\alpha_{n-},~\beta_{n-}$) are the same. Since the operators in~(\ref{5DElkoEoM3}) do not change the subscripts ``$+$" and ``$-$", we only consider one of them and omit the $\pm$ subscript in $\alpha_{n\pm}$ and $\beta_{n\pm}$ for simplicity.

By considering the linear independence of $\varsigma_{n\pm}$ and $\tau_{n\pm}$, we obtain the following coupled equations for $\alpha_{n}$ and $\beta_{n}$:
\begin{eqnarray}
\alpha_{n}''-V(z) \alpha_{n}+m_n^2\alpha_{n}- {i}m_{n}A'\beta_{n}\!\!\! &=& \!\!\!0, \label{EoMalpha}\\
\beta_{n}''-V(z) \beta_{n} +m_n^2 \beta_{n}- {i}m_{n}A'\alpha_{n}\!\!\! &=& \!\!\!0. \label{EoMbeta}
\end{eqnarray}
Subsequently, by introducing the new functions $a_n$ and $b_{n}$
\begin{eqnarray}
a_{n}\!\!\! &=& \!\!\!\frac{1}{2}(\alpha_n+\beta_n), \\
b_{n}\!\!\! &=& \!\!\!\frac{1}{2}(\alpha_n-\beta_n),
\end{eqnarray}
we can transform the coupled equations of motion~(\ref{EoMalpha}) and~(\ref{EoMbeta}) to the decoupled ones
\begin{eqnarray}
a_{n}''-\left[V(z)-m_n^2+ {i}m_{n}A'\right]a_{n}\!\!\! &=& \!\!\!0, \label{EoM_a_n}\\
b_{n}''-\left[V(z)-m_n^2- {i}m_{n}A'\right]b_{n}\!\!\! &=& \!\!\!0.  \label{EoM_b_n}
\end{eqnarray}
Note that there is a factor of imaginary unit $i$ in the above equations for the mass terms. This feature is very different from the cases of scalar, Dirac spinor, and vector fields.

Next, we consider the reduction of the fundamental action by using the above KK decomposition and equation of motion for ELKO. The procedure is the same as the ones of scalar, vector, and Dirac fields.
By substituting the KK decompositions~(\ref{KK decomposition}) and~(\ref{KK decomposition2}) into the action~(\ref{n-dim_action for Elko}), we obtain the effective action of a four-dimensional massless ELKO and a series of four-dimensional massive ELKO KK modes
\begin{eqnarray}
S_{\text{ELKO}}
              \!\!\! &=& \!\!\!-\frac{1}{2}\sum_{n}\int d^4x\left[\frac{1}{2}\hat{g}^{\mu\nu}(\hat{\mathfrak{D}}_{\mu}\hat{\mathop \lambda\limits^\neg}_{n}\hat{\mathfrak{D}}_{\nu}\hat{\lambda}_{n}
              +\hat{\mathfrak{D}}_{\nu}\hat{\mathop \lambda\limits^\neg}_{n}\hat{\mathfrak{D}_{\mu}}\hat{\lambda}_{n})
              +m_{n}^2\hat{{\mathop \lambda\limits^\neg}}_{n}\hat{\lambda}_{n}\right].
\end{eqnarray}
The general four-dimensional ELKOs on the brane {$\hat{\lambda}_{n\pm}$} satisfy the four-dimensional massive Klein-Gordon equation:
\begin{eqnarray}
\frac{1}{\sqrt{-\hat{g}}}\hat{\mathfrak{D}}_{\mu}
 (\sqrt{-\hat{g}}\hat{g}^{\mu\nu}\hat{\mathfrak{D}}_{\nu}\hat{\lambda}_{n\pm})
 =m^2_n\hat{\lambda}_{n\pm}.
\end{eqnarray}

It is easy to see that the corresponding conditions for $a_{n}$ and $b_{n}$ are given by
\begin{eqnarray}
\int a^{*}_{n}a_{m}dz\!\!\! &=& \!\!\!\delta_{nm},\\
\int b^{*}_{n}b_{m}dz\!\!\! &=& \!\!\!0, \label{orthonormality relation 2}
\end{eqnarray}
which shows that $b_{n}=0$ and $\alpha_{n}=\beta_{n}$. Therefore, the KK modes for different types of ELKOs are the same so one can simplify the KK decomposition as
\begin{eqnarray}
\lambda_{\pm} = \sum_{n} \text{e}^{-3A/2}f^{-\frac{1}{2}} \alpha_{n}(z) \hat{\lambda}_{n\pm}(x)\label{Elkodecomposition}
\end{eqnarray}
where
\begin{eqnarray}
\hat{\lambda}_{n\pm}(x) =\varsigma_{n\pm}(x) + \tau_{n\pm}(x).
\end{eqnarray}
The equation of motion of the KK mode reads
\begin{eqnarray}
 \left[ -\partial_{z}^{2}+V(z) + {i}m_{n}A' \right] \alpha_{n}(z) = m_n^2 \alpha_{n}(z), \label{KKequationforElko}
\end{eqnarray}
which is just~(\ref{EoM_a_n}) with the effective potential $V(z)$ given by~(\ref{effective potential Vz}). 

{The zero mode $\alpha_0$ describes the ground state with minimal energy so it is  massless $m_{0}=0$. Therefore, equation~(\ref{KKequationforElko}) simplifies to
\begin{eqnarray}
 \left[ -\partial_{z}^{2}+V(z) \right] \alpha_{0}(z)=0 \label{KKequationforzeromode}
\end{eqnarray}
with the normalization condition
\begin{equation}
\int dz \alpha^{*}_{0}\alpha_{0}=1. \label{NormalizationConditionForZeroMode}
\end{equation}
For $n\geq1$, the excited states are massive $m_{n}>0$. In what follows, we will mainly focus on the localization of the massless zero mode in various models. The difficulties in localizing the massive modes is discussed in sec.~\ref{m_modes}.}



\subsubsection{Free ELKOs}


{Firstly, we consider the localization of the free massless zero mode. This corresponds to the choice $f=1$ and $\eta=0$ in~(\ref{n-dim_action for Elko}) for which the effective potential~(\ref{KKequationforzeromode}) is given by
 \begin{equation}\label{5D_EffectivePotential_Vz}
V(z)=\frac{3}{2}A^{\prime\prime}+\frac{13}{4}A^{\prime2}.
\end{equation}}

{In the RS-II flat brane scenario, by using the solution $A(z)=-\ln (k|z|+1)$ derived from the coordinate transformation~(\ref{CoordinateTransformation1}),  one obtains the potential~\cite{Liu:2011nb}
  \begin{equation}\label{5D_FreeElko_flatBrane_Vz}
V(z)=\frac{19k^{2}}{4(1+k|z|)^{2}}-\frac{3k\delta(z)}{1+k|z|}.
\end{equation}
The general solution of the zero mode is
  \begin{equation}
\alpha_{0}(z)=\mathcal{C}_{1}(1+k|z|)^{\frac{1}{2}+\sqrt{5}}
 +\mathcal{C}_{2}(1+k|z|)^{\frac{1}{2}-\sqrt{5}} \label{zeroMode_RS_II}
\end{equation}
where $\mathcal{C}_{1}$ and $\mathcal{C}_{2}$ are free parameters. It can be shown that the normalization condition~(\ref{NormalizationConditionForZeroMode}) cannot be satisfied. Therefore, the free zero mode cannot be localized on the RS-II brane~\cite{Jardim:2014cya}.}

{For a flat thick brane generated by a real scalar field in five-dimensional asymptotic AdS space-time, the typical warp factor can have the following form
 \begin{equation}\label{WarpFactorEAyUnifiedForm}
e^{A(y)}=\text{sech}^{b}(ky)
\end{equation}
where $k$ is a parameter related to the fundamental scale of the five-dimensional theory and $b$ is a positive constant. Such a solution has a similar asymptotic behavior to the RS-II one at the boundary of the extra dimension. This indicates that the zero mode has a similar asymptotic behavior as~(\ref{zeroMode_RS_II}) in the limit $|z|\rightarrow\infty$. Therefore, the zero mode of a five-dimensional
free massless ELKO cannot be localized on flat thick branes in asymptotic AdS space-time either.}

{Furthermore, the results are the same for dS/AdS thin and thick branes/strings~\cite{Zh1,Sorkhi:2018jhy,MoazzenSorkhi:2020fqp}. Therefore, new mechanisms are needed to localize the zero mode in braneworld models. Next, we consider these mechanisms.}

\subsubsection{ELKOs with Yukawa-like coupling}


We now review the localization of ELKOs with a coupling term in~(\ref{n-dim_action for Elko}). In this case, we show that the zero mode can be bounded to the branes by introducing an interaction between ELKO and the background scalar. This class of interaction is known as the Yukawa-like coupling and it can take various forms. Here, we consider the simplest choice with $f=1$ and $F=F(\phi)$ or $F=F(R)$. {With these assumptions, the effective potential~(\ref{KKequationforzeromode}) reads
 \begin{equation}\label{5D_YukawaCoupling_Vz}
V_{\text{Y}}(z)=\frac{3}{2}A^{\prime\prime}+\frac{13}{4}A^{\prime2}+2\eta e^{2A}F.
\end{equation}
Below, we show that for appropriate choices of $F$, the zero mode can be localized.}

{We first consider the RS thin brane model. Since there is no background scalar field in this model, we choose the following interacting term~\cite{Jardim:2014xla}
 \begin{equation}\label{YukawaFlatBrane_etaF1}
\eta F=\frac{1}{2}(M_{\text{ELKO}}^{2}+c\delta(z)),
\end{equation}
 where $M_{\text{ELKO}}$ is the five-dimensional ELKO mass. The effective potential~(\ref{5D_YukawaCoupling_Vz}) reads
  \begin{equation}\label{5D_YukawaCoupling_FlatBrane_Vz1}
V_{\text{Y}}(z)=-\left(\frac{19}{4}k^{2}+M_{\text{ELKO}}^{2}\right)(k|z|+1)^{-2}+(3k-c)\delta(z).
\end{equation}
Using~(\ref{5D_YukawaCoupling_FlatBrane_Vz1}), we obtain the localized zero mode~\cite{Zhou:2020ucc}
  \begin{equation}\label{SorkhiEq:54}
\alpha_{0}(z)=\frac{(k|z|+1)^{\frac{1}{2}-\nu}}{\sqrt{k(\nu-1)}}
\end{equation}
that satisfies the normalization condition~(\ref{NormalizationConditionForZeroMode}). In~\cite{Jardim:2014xla}, a geometrical coupling with Ricci scalar was considered where 
\begin{equation}
F=R,\quad R=-4(2A^{\prime\prime}+3A^{\prime2})e^{-2A}.
\end{equation}
 The corresponding normalized zero mode is
 \begin{equation}
\alpha_{0}(z)=\sqrt{\frac{2}{3k}}(k|z|+1)^{-2}.
\end{equation}
For a flat thick brane solution~(\ref{flatThickBrane1Solution2}), by considering the coupling $F = F(\phi)=\phi^{2}$, we can also obtain a localized zero mode for some given parameters~\cite{Liu:2011nb}.}

{The localization of ELKO with a Yukawa-like coupling in string-like
scenarios in six dimensions was investigated in~\cite{Dantas:2015mfi}. It was shown that the constant coupling $\eta {F}=-\frac{5}{32}c^2$ leads to a localized zero mode which is similar to the zero modes of gravity and the scalar field in six-dimensional space-time. The authors of~\cite{Dantas:2015mfi} also proposed a way to prevent the appearance of imaginary mass term in the ELKO KK modes by adding a new term to the covariant derivative. As a result, it was found that the zero mode can be obtained for the Gherghetta-Shaposhnikov model~\cite{Gherghetta:2000qi} and the Hamilton string-cigar model~\cite{Silva:2012yj}.}

{
Lastly, we turn to the localization of ELKO on dS/AdS thick branes. With the warp factor solution~(\ref{dS_AdS_ThickBrane1_Az}) and the following choice
\begin{equation}\label{SorkhiEq:74}
F(\phi)=-\frac{h^2a^2s}{16\eta}(1+\Lambda_4)\left[25-4p(2+p)+(4p^2-13)\cosh(2b\phi)\right],
\end{equation}
the effective potential~(\ref{5D_YukawaCoupling_Vz}) and zero mode are~\cite{Zhou:2018oib}:
\begin{eqnarray}
V_{\text{Y}} \!\!\! &=& \!\!\!
     -ph^2\text{sech}^2(hz)+p^2h^2\tanh^2(hz), \label{VYdSAdSZhou}\\
\alpha_0\!\!\! &\propto&\!\!\! e^{pA(z)} \propto  {\text{sech}^p(hz)}. \label{ZeroModedSAdSZhou}
\end{eqnarray}
Note that the zero mode solution~(\ref{ZeroModedSAdSZhou}) has a Gaussian profile so the normalization condition is satisfied for $p>0$. Therefore, the zero mode can be localized on the dS/AdS thick brane for any $p>0$.}

\subsubsection{ELKOs with other couplings}

{

Apart from the Yukawa-like coupling, the non-minimal coupling and derivative coupling can also be used to localize ELKO on the brane. The non-minimal coupling corresponds to $f=f(\phi)$ and $F=0$ in~(\ref{n-dim_action for Elko}). With some choices of the coupling function $f=f(\phi)$, the zero mode can be trapped on various kinds of branes~\cite{Zhou:2018oib,Sorkhi:2018jhy}. For derivative coupling, the five-dimensional action reads
\begin{equation}\label{SorkhiEq:action of ndc}
S=\int d^{5}x\sqrt{-g}\left[-\frac{1}{4}g^{MN} \Big(\mathfrak{D}_M\mathop \lambda\limits^\neg\mathfrak{D}_{N}\lambda+\mathfrak{D}_N\mathop \lambda\limits^\neg\mathfrak{D}_{M}\lambda\Big)
 -\tilde{\eta}\mathop\lambda\limits^\neg {\Gamma^{M}\partial_{M}\tilde{F}(\phi)\Gamma^{5}}\lambda\right]
\end{equation}
where $\tilde{F}$ is a function of the background scalar field $\phi$ and $\tilde{\eta}$ is a coupling constant. This mechanism was first applied to solve the problem of the fermion localization on thick branes generated by an even background scalar field~\cite{Liu:2013kxz}. For a suitable choice of $\tilde{F}(\phi)$, the localized zero mode is given by $\alpha_{0}(z)\varpropto e^{pA(z)}$~\cite{MoazzenSorkhi:2020fqp} where $p$ is a real parameter.}

\subsection{Massive ELKO modes} \label{m_modes}

In the previous section, we have only {investigated the localization of} the zero mode in various braneworld models with $m_{0}=0$ in~(\ref{KKequationforElko}). We now study the massive modes corresponding to the general solution of~(\ref{KKequationforElko}). It should be noted that because of the imaginary part in~(\ref{KKequationforElko}), the massive KK mode $\alpha_n$ is a complex function. Therefore, we can take
\begin{eqnarray}
\alpha_{n}(z)=R_{n}(z)+iI_{n}(z)  \label{alpha_R_I}
\end{eqnarray}
with $R_n$ and $I_n$ being real functions. The orthonormality condition~(\ref{orthonormality relation 1}) requires
\begin{eqnarray}
\int dz\left(R_{n}^2+I_{n}^2\right)=1.  \label{orthonormalityRelationMassiveModes}
\end{eqnarray}

We first consider a five-dimensional free spinor field. Substituting~(\ref{alpha_R_I}) into~(\ref{KKequationforElko}) yields
\begin{eqnarray}
R_n''-\left(\frac{3}{2}A''+\frac{13}{4}(A')^2-m_n^2\right)R_n\!+\!m_n A'I_n
+i\left[I_n''-\left(\frac{3}{2}A''\!+\!\frac{13}{4}{A'}^2\!-\!m_n^2\right)I_n\!-\!m_n A'R_n\right]=0,
\end{eqnarray}
which results in the following coupled equations
\begin{eqnarray}
-R_n''+V_{\text{F}}R_n-m_n A'I_n\!\!\! &=& \!\!\!m_n^2R_n,\label{KKequationforR}\\
-I_n''~+V_{\text{F}}I_n+m_n A'R_n\!\!\! &=& \!\!\!m_n^2I_n\label{KKequationforI}
\end{eqnarray}
where the effective potential is given by
\begin{eqnarray}
V_{\text{F}}=\frac{3}{2}A''+\frac{13}{4}A'^2. \label{EffectivePotentialVF}
\end{eqnarray}
Here, we would like to consider branes embedded in an asymptotically AdS space-time. In this case, the mass terms $m_n A'I_n$ and $m_n A'R_n$ in~(\ref{KKequationforR}) and~(\ref{KKequationforI}) can be neglected at the boundary of the extra dimension since $A'\rightarrow0$ as $z\rightarrow\infty$. Therefore, we obtain the following Schr\"{o}dinger-like equations
\begin{eqnarray}
-R_n''+V_{\text{F}}R_n\!\!\! &=& \!\!\!m_n^2R_n,\\
-I_n''~+V_{\text{F}}I_n~\!\!\! &=& \!\!\!m_n^2I_n.
\end{eqnarray}
Although the above equations are approximations, we can use them to analyse the localization properties of the massive KK modes. For the thick branes embedded in an AdS or asymptotically AdS space-time, the effective potential $V_{\text{F}}$ is usually a volcano potential that vanishes at the boundary of the extra dimension. Therefore, the massive KK modes cannot be localized on the branes~\cite{Liu:2011nb}.

{For ELKOs with Yukawa-like interaction considered in the last subsection, it can be shown that the above result also applies to any thick branes embedded in an asymptotically AdS space-time with a kink scalar field~\cite{Liu:2011nb,Jardim:2014xla}.
It was also found that the massive modes are oscillating and divergent in the Gherghetta-Shaposhnikov string-like model~\cite{Dantas:2015mfi}. Therefore, the massive KK modes cannot be bounded in this model. A few studies have been done for massive modes on bent branes~\cite{Zh1}. It was shown that the result is the same for massive KK modes in the cases of dS and AdS thin branes. Therefore, new mechanisms are needed to localize massive modes on branes.}

\subsection{Comparison with other fields}\label{comparison}

Comparing with other fields, we consider the equations of motion for KK modes of gravity (general relativity), massless free scalar field, U(1) vector field, Kalb-Ramond field, and Dirac spinor field in five-dimensional RS-like flat braneworld models with $A(z\rightarrow\pm\infty)\rightarrow 1/|kz|$ (see~\ref{kkd} for the explicit KK decompositions)
\begin{eqnarray}
 \left[-\partial_{z}^{2}+V(z) \right] f_{n}(z) = m_n^2 f_{n}(z) \label{KKequationforElko3}
\end{eqnarray}
where the effective potentials are given by~\cite{Randall:1999vf,DeWolfe:1999cp,Csaki:2000fc,Liu:2017gcn}
\begin{eqnarray}
 V(z) =  \left\{
 \begin{array}{ll}
   +\frac{3}{2}A'' +\frac{9}{4}A'^2  & \text{for tensor perturbation of gravity} \\
   +\frac{3}{2}A'' +\frac{9}{4}A'^2  & \text{for a scalar field} \\
   +\frac{1}{2}A'' +\frac{1}{4}A'^2 & \text{for a U(1) vector field} \\
   -\frac{1}{2}A'' +\frac{1}{4}A'^2 & \text{for a Kalb-Ramond field} \\
   0  & \text{for a Dirac spinor field}
 \end{array}
 \right. \label{potentials}
\end{eqnarray}
The normalization condition is
\begin{eqnarray}
\int_{-\infty}^{\infty} |f_n|^2 dz < \infty. \label{normalization}
\end{eqnarray}

For gravity and the scalar field, the Schr\"{o}dinger-like equation~(\ref{KKequationforElko3}) can be rewritten in the form of supersymmetric quantum mechanics
\begin{eqnarray}
   \mathcal{P} \mathcal{P}^{\dag} f_{n}(z) = m_n^2 f_{n}(z) \label{PPf=m2f}
\end{eqnarray}
where the operators are given by
$\mathcal{P} =  \partial_z +\frac{3}{2}A' $ and $\mathcal{P}^{\dag} =  -\partial_z +\frac{3}{2}A' $.
The form of~(\ref{PPf=m2f}) excludes the presence of tachyon modes with $m^{2}_{n}<0$.
It is easy to show that the graviton and scalar zero modes $f_0(z)\propto \exp\left[\frac{3}{2}A(z)\right]$ can be localized on the flat branes. Thus, the four-dimensional Newtonian potential on the brane can be recovered.

The Schr\"{o}dinger-like equations for the vector and Kalb-Ramond fields are supersymmetric dual since they have the following relation:
\begin{eqnarray}
  \mathcal{K} \mathcal{K}^{\dag} f_{n}(z) \!\!\! &=& \!\!\! m_n^2 f_{n}(z) ~~~ \text{for a vector field}, \label{supersymmetricFormForScalar}\\
  \mathcal{K}^{\dag} \mathcal{K} f_{n}(z) \!\!\! &=& \!\!\! m_n^2 f_{n}(z)~~~\text{for a Kalb-Ramond field}, \label{supersymmetricFormForKR}
\end{eqnarray}
where $m^{2}_{n}\geq0$ and $\mathcal{K}=\partial_z +A'/2$ is the same for both fields. However, the zero modes of the vector and Kalb-Ramond fields are of the respective form $f_0(z)\propto \exp\left[\frac{1}{2}A(z)\right]$ and $\exp\left[-\frac{1}{2}A(z)\right]$ so they cannot be localized on the flat brane since the normalization condition is not satisfied. Note that the free U(1) vector field can be localized on the dS brane~\cite{Guo:2011qt}. In order to localize these fields on the flat brane, some localization mechanisms have been proposed~\cite{Chumbes:2011zt,Cruz:2012kd,Zhao:2014gka,Vaquera-Araujo:2014tia,Alencar:2014moa,Sui:2020fty}.

As for the free Dirac field, both the left- and right-chiral KK modes cannot be localized since their effective potentials vanish. To localize the Dirac field on flat branes, one has to introduce general couplings between the Dirac spinor and background scalar field. 
{The action is given by~(see app.~\ref{kkd} for details)
\begin{eqnarray}
    S_{\Psi}=\int d^5x\sqrt{-g}\big[F_1\bar{\Psi}\Gamma^{M}D_M\Psi
    +\lambda F_2\bar{\Psi}\Psi+\eta \bar{\Psi}\Gamma^{M}(\partial_M F_3)\gamma^5\Psi\big]. \label{nonMinimumalFermionAction}
\end{eqnarray}
The left- and right-chiral KK modes of the Dirac
fermionic field also satisfy the Schr\"{o}dinger-like equation~(\ref{KKequationforElko3}) and the corresponding effective potentials are given by
    \begin{align}
    &V_{L}(z)= \mathcal{F}^2(z)+\partial_z\mathcal{F}(z),\\
    &V_{R}(z)=\mathcal{F}^{2}(z)-\partial_{z}\mathcal{F}(z)     \label{potentialnew}
    \end{align}
where 
\begin{equation}
\mathcal{F}(z)= \lambda e^{A(z)}{F_2}/{F_1}+ \eta {\partial_zF_3}/{F_1}.
\end{equation}}
The two Schr\"{o}dinger-like equations can be rewritten as
    \begin{eqnarray}
     \begin{array}{c}
       \mathcal{K}^{\dag}\mathcal{K}\, f_{Ln}=m^2_nf_{Ln}, \\
       \mathcal{K} \mathcal{K}^{\dag} \, f_{Rn}=m^2_nf_{Rn}
     \end{array}
    \label{operator2}
    \end{eqnarray}
with $\mathcal{K}=\partial_z-\mathcal{F}(z)$ which ensures that $m_n^2 \ge 0$. The orthonormality conditions for $f_{Ln,Rn}$ are
    \begin{align}
    &\int {F_1f_{Lm}f_{Ln}dz}
      =\int {F_1f_{Rm}f_{Rn}dz}=\delta_{mn},\\
    &\int {F_1f_{Ln}f_{Rn}dz}=0.\label{orthonormalityForFermion}
    \end{align}
The solutions of the corresponding chiral zero modes are
{    
    \begin{align}
    &f_{L0} \propto
    e^{+ \int {dz} \mathcal{F}(z)},
    \label{newzero mode}\\
    &f_{R0} \propto
    e^{-\int {dz} \mathcal{F}(z)}
    \end{align}
which shows that the left- and right-chiral fermion zero modes cannot simultaneously satisfy~(\ref{normalization}). Therefore, only one of the chiral fermion zero modes can be localized on the branes.}

{The fermion localization mechanism described above can be realized in two cases.}
The first one is the simplest Yukawa coupling between fermions and the background scalar fields~\cite{Bajc:1999mh,Ringeval:2001cq,Melfo:2006hh,Slatyer:2006un,Liu:2008pi,Liu:2009mga,Liang:2009zzg,Liang:2009zzc,Cruz:2009ne,Liu:2009ve,Chumbes:2010xg,Liu:2011zy,Cruz:2011ru,Correa:2010zg,Castro:2010au,CastilloFelisola:2012ez,Andrianov:2013moa,Barbosa-Cendejas:2015qaa,Agashe:2014jca,Bazeia:2017nzd,Paul:2017dav,Mitra:2017run,Hundi:2011dc,Koley:2008tn,Koley:2008dh,Xie:2019jkq}, which corresponds to $F_1=1$ and $F_3=0$. Recently, the fermion localization on a brane array generated by multi-scalar fields was studied in~\cite{Xie:2019jkq}. It was found that the left-chiral fermion zero mode is the only bound state that can be localized between the two outermost sub-branes. This mechanism works when the background scalar field is an odd function of the extra dimension.
If the background scalar field is an even function of the extra dimension, the Yukawa coupling mechanism fails since the $Z_2$ reflection symmetry of the effective potentials for the fermion KK modes is not guaranteed~\cite{Liu:2013kxz}. The second mechanism was presented in~\cite{Liu:2013kxz} with $F_1=1$ and $F_2=0$. Such a choice is called the derivative coupling.  In particle physics, it describes the interactions between $\pi$-meson and nucleons in quantum field theory.

To implement the two mechanisms, we need one or more background scalar fields as evident from the choice of $F_{1},F_{2}$, and $F_{3}$. As for thick brane models without background scalar fields, we can consider geometric coupling
{where the bulk fermionic field couples to the scalar curvature $R(z)$ of the background space-time via $\mathcal{L}_{\text{int}}=\eta\bar{\Psi}\Gamma^M\partial_M{F_3(R)}\gamma^5\Psi$. This takes the same form as the derivative coupling since $R(z)$ is an even function of the extra dimension.}
It has been shown that, for some suitable {$F_3(R)$}, one of the left- and right-chiral fermion zero modes is localized on the brane~\cite{Liu:2013kxz,Li:2017dkw,Guo:2014nja,Xie:2015dva}.

Now, we come back to ELKO. For the free massless zero mode,
the effective potential~(\ref{5D_EffectivePotential_Vz}) has an additional term $A'^2(z)$ compared with gravity and the scalar field. As a result, it cannot be localized on the flat, dS, and AdS branes. {The $A'^2(z)$ term also prevents us from writing the equation of motion of $\alpha_{0}(z)$ in the supersymmetric form.} This is different to the non-localized free massless vector and Kalb-Ramond fields where the equations of motion can be written as~(\ref{supersymmetricFormForScalar}) and~(\ref{supersymmetricFormForKR}) respectively.

{One can localize ELKOs in braneworld models by introducing mass terms or interactions. Specifically, the zero mode can be localized on the flat, dS, and AdS branes via Yukawa-like coupling, non-minimal coupling and derivative coupling. When the effective potential takes the form $ V_{\text{ELKO}}(z) = p A'' + p^2 A'^2$, the equation of motion of the zero modes can be put in the supersymmetric form. For the Dirac field, only one of the left- and right-chiral fermion zero mode can be localized on the branes. For ELKO, both the left- and right- zero modes satisfy the same equations of motion so they can be localized simultaneously. On the other hand, we have not been able to localize the massive ELKO KK modes. In braneworld models, dark matter must be localizable. Therefore, if one takes ELKO as a dark matter candidate, its zero mode must acquire a mass via particle interactions such as the Higgs mechanism.
}



\section{Concluding remarks}

On the theoretical side this report announces several unexpected results. The first is announced in the title itself, the mass dimension one fermions. At the heart of it is a careful formulation of the eigenspinors of the charge conjugation operator, and certain specific implications of Wigner's time reversal operator. The formulation exposes 
a fundamental incompleteness in our understanding of Majorana spinors. Under Dirac dual, these spinors have a null norm. Here, the Majorana spinors coincide with $\lambda^S(p^\mu)$. By incorporating 
$\lambda^A(p^\mu)$ in the formalism we bring Majorana spinors at par with Dirac spinors, now in their incarnation as ELKO. Both the
$\lambda^S(p^\mu)$, and $\lambda^A(p^\mu)$, are found to have null norms, under the usual spinorial dual. A new dual is thus introduced. Under the new dual, ELKO have the exactly similar orthonormality and completeness relations as  the  Dirac spinors. The spin sums too are formally identical.

We point out that  $(\gamma_\mu p^\mu \pm m )$ does not annihilate the 
$\lambda^{S,A}(p^\mu) $. This fact, under quantum field theoretic formulation, invites mass dimensionality of one for 
a quantum field with the $\lambda^{S,A}(p^\mu)$ as its expansion coefficients.
The new field, under detailed examination, turns out to be local. Similar results follow for $\rho(p^\mu)$. 

In the process, we come to appreciate that the spin-statistics result for spin half crucially depends on the spin sums associated with the expansion coefficients of the field. When this exercise is repeated for the
$\rho(p^\mu)$ spinors -- the siblings of $\lambda(p^\mu)$ in disguise -- a second surprise sprouts. The resulting spin sums provide a local bosonic field, of spin half! This new  `understanding' not only fulfils Dirac's expectation of a spin half boson associated with `electron' but it also provides the sought after deeper relation between spin and statistics.
Specification of spin does not necessarily determine its statistics.

At the core of the unexpected element about spin and statistics relation is an argument recently published by one of us on spin-statistics relation~\cite{Ahluwalia:2021dva}:

\begin{quote}

 Consider two events, $x$ and $y$, with a space-like separation.  The fact that time ordering for such events cannot be preserved  allows us to introduce two set of observers, $\mathcal{O}$ and $\mathcal{O}^\prime$.
As part of the definition of these observers for the former $y_{0} > x_{0}$, while for the latter $x_{0} > y_{0}$.
The observer in $\mathcal{O}$ calculates the amplitude for a particle to propagate from $x$ to $y$, while  the  observer in $\mathcal{O}^\prime$
calculates the amplitude for an associated antiparticle to propagate from $y$ to $x$. Causality, dictates that these two amplitudes can only differ, at most, by a phase factor
\begin{equation}
\mbox{Amp}(x\to y, \mbox{particle})\vert_{\mathcal{O}}
= e^{i \theta} \mbox{Amp}(y\to x, \mbox{antiparticle})\vert_{\mathcal{O}^\prime},\qquad \theta \in \Re \nonumber
\end{equation}
Knowledge of the spin sums~\cite{Vir_Ahluwalia_2020, Ahluwalia:2020jkw}, allows us to arrive at the following remarkable result:
\begin{equation}
e^{i\theta} ={\bigg\{}
\begin{array}{ll}
  -1 & \mbox{spin half fermions} \\
                            +1 & \text{spin half bosons}
\end{array}\nonumber
\end{equation}
It is now a straight forward argument to see how the fermionic and bosonic statistics follows and the new bosons subvert the spin-statistics theorem.
\end{quote}
Therefore, the causality phase $e^{i\theta}$ determines the statistics. For a given spin, as seen here for spin half, it can acquire both of the indicated values.

\vspace{11pt}
ELKO based quantum fields, and those arising from considering the complete set of square roots of the identity matrix in the $\mathcal{R}\oplus\mathcal{L}$ representation space, because of their different mass dimensionality or statistics, carry mismatch with respect to the SM fields. Consequently, they are first principle dark matter candidates. Their gravitational interactions in the Newtonian limit coincide with the usual scenario.

\appendix

\section{A mathematical substrate for the new dual}{\label{Appzero}}

In this appendix we shall further explore, on formal grounds, the re-definition of the ELKO dual, namely
\begin{equation}\label{matjm1}
\widetilde{\lambda}_\alpha^S(p^{\mu})\rightarrow\gdualn{\lambda}^{S}_{\alpha}(p^{\mu})\equiv \widetilde{\lambda}_\alpha^S(p^{\mu})\mathcal{A}, \quad
\widetilde{\lambda}_\alpha^A(p^{\mu})\rightarrow \gdualn{\lambda}^{A}_{\alpha}(p^{\mu})\equiv\widetilde{\lambda}_\alpha^A(p^{\mu})\mathcal{B}
\end{equation}
from the Clifford algebra perspective~\cite{Vaz:2016qyw,daSilvaa:2019kkt}. For the sake of exposition, we start with a mathematical review from the very basic concept of pseudo-Euclidean space-time endowed with a quadratic form $g$ (the metric). Then we move to contextualize the $\tau$ deformation and investigate the $\tau\rightarrow 1$ limit.

\subsection{Freedoms in Clifford algebra}

The associated Clifford algebra,\footnote{We took the liberty to expose here the topic directly related to our purposes and, as such, we already particularize the space-time dimensions, the divisional ring, and the Clifford product is denoted by juxtaposition.} $\mathcal{C}l_{1,3}$, is the unital associative algebra such that the linear application $\gamma:\R^{1,3}\rightarrow \mathcal{C}l_{1,3}$ and satisfies $\{\gamma(v),\gamma(w)\}=2g(v,w)$ for all $u,v\in\Re^{1,3}$, where $\{\cdot,\cdot\}$ denotes the anti-commutator. Additionally, if $(\mathcal{K}l_{1,3},\gamma')$ is another associative unital algebra such that $\gamma':\R^{1,3}\rightarrow \mathcal{K}l_{1,3}$ and $\{\gamma'(v),\gamma'(w)\}=2g(v,w)$, then there exists a unique homomorphism $\phi:\mathcal{C}l_{1,3}\rightarrow \mathcal{K}l_{1,3}$ such that $\gamma'=\phi\circ \gamma$.

It is quite useful to recall a less known definition of spinors, the so called algebraic definition~\cite{Che:1997oi}. According to this point of view, spinors are the left minimal ideals originated from primitive idempotents, $f$, of the Clifford algebra. These ideals are denoted by $\mathcal{C}l_{1,3}f$. Along with a simple application formalizing the multiplication of elements of $\mathcal{C}l_{1,3}f$ by complex numbers, the resulting structure $\mathcal{C}l_{1,3}f\equiv \mathcal{S}_{1,3}$ is then called algebraic spinor space, whose elements are algebraic spinors.
Analogously, right minimal ideals (also built up from primitive ideals) lead to $f\mathcal{C}l_{1,3}$ and the resulting algebraic spinor space is denoted by $\mathcal{S}_{1,3}^{\star}$. A linear action of $\psi^{\star}\in\mathcal{S}_{1,3}^{\star}$ from the left on $\psi\in\mathcal{S}_{1,3}$ yields a complex number.\footnote{It should be noted that the $\star$ operation on the spinor defines a general adjoint and not simply its complex-conjugation.} Therefore, an isomorphism between $\mathcal{S}^{\star}_{1,3}$ and the space of linear action on $\mathcal{S}_{1,3}$ into the space of complex numbers $\mathbb{C}$ appears. This leads to an inner-product $\beta:\mathcal{S}_{1,3}\times \mathcal{S}_{1,3}\rightarrow \mathbb{C}$. In terms of the adjoint $\psi^{\star}$, we have $\beta(\psi,\xi)=\psi^{\star}\xi\in\mathbb{C}$.

A few additional remarks are still in order as they are in the very heart of the dual freedoms explored. The left and right ideals are mapped to each other by the so called algebraic involutions. In such a procedure, the idempotents are not always preserved~\cite{Che:1997oi}. Explicitly, let $\kappa$ be an algebraic involution. Its action upon an $\mathcal{C}l_{1,3}f$ element is given by
\begin{equation}
\kappa(\mathcal{C}l_{1,3}f)=\kappa(f)\mathcal{C}l_{1,3} \label{matj1}
\end{equation}
with $\kappa(f)\neq f$. Nevertheless, there always exists a Clifford algebra element $h$ such that $\kappa(f)=h^{-1}fh$ with $\kappa(h)=h$, where the ideals are interchanged~\cite{Benn:1987fw}
\begin{eqnarray}
\kappa(\mathcal{C}l_{1,3}hf)=\kappa(hf)\mathcal{C}l_{1,3}=\kappa(h)\kappa(f)\mathcal{C}l_{1,3}=fh\mathcal{C}l_{1,3},\label{matj2}
\end{eqnarray}
so that $\kappa: \mathcal{S}_{1,3}\rightarrow  \mathcal{S}^{\star}_{1,3}$ and vice-versa. The adjoint is then defined as
\begin{eqnarray}
\psi^{\star}=\kappa(h\psi)=h\kappa(\psi) \label{matj3}
\end{eqnarray}
which in abstract, taking into account the very nature of the algebraic idempotent ideal, reads
\begin{eqnarray}
\psi^{\star}=h\kappa(\psi f)=h\kappa(f)\kappa(\psi)=fh\kappa(\psi). \label{matj4}
\end{eqnarray}
The inner product is defined as
\begin{eqnarray}
\beta(\psi,\xi)=h\kappa(\psi)\xi\label{matj5}
\end{eqnarray}
whose abstract version reads
\begin{eqnarray}
\beta(\psi,\xi)=h\kappa(\psi f)\xi f=fh\kappa(\psi)\xi f, \label{matj6}
\end{eqnarray}
so that $\beta$ is an element of $f\mathcal{C}l_{1,3}f\simeq \mathbb{C}$.

Finally, for complexified Clifford algebras $\mathbb{C}\otimes\mathcal{C}l_{1,3}$ there is also the hermitian conjugation as a relevant complement to be taken in the algebraic involution representation. Within this case, it is sufficient that \cite{Benn:1987fw}
\begin{eqnarray}
\kappa^\star(\psi)=h^{-1}\psi^\dagger h,\quad h=h^{\dag}\label{matj7}
\end{eqnarray}
for all $\psi\in \mathbb{C}\otimes\mathcal{S}_{1,3}$ and $h\in \mathbb{C}\otimes\mathcal{C}l_{1,3}$. In (\ref{matj7}), the operator $\dag$ denotes Hermitian conjugation. It is now possible to foresee the form of $\mathbb{C}\otimes\mathcal{S}^{\star}_{1,3}\ni \psi^{\star}$ elements as
\begin{equation}
\psi^{\star}=h\kappa^{\star}(\psi)\label{matj8}
\end{equation} which, via~(\ref{matj7}), leads to the widespread form
\begin{equation}
\psi^{\star}=hh^{-1}\psi^\dagger h=[h\psi]^\dagger.\label{matj9}
\end{equation}
Notice that from the formal algebraic point of view, some freedom to be explored in the spinorial duals is encoded in $h$. Whatever the possible dual generalization is, it must preserve the inner-product structure of the Clifford algebra. Here, the particular form of $h\equiv\eta\Delta$ is suitable. We depict some possibilities for duals in a non-exhaustive list in table~\ref{tabua}.\footnote{See \cite{Cavalcanti:2020obq} for an exploration of many other possibilities and respective analysis.} In table~\ref{tabua}, the first choice leads to the Dirac dual $\overline{\psi}\equiv\psi^{\dag}\gamma^{0}$ while the second choice leads to the ELKO dual $\widetilde{\psi}\equiv\psi^{\dag}\gamma^{0}\Xi$, where we have used the identity $\Xi^{\dag}=\gamma^{0}\Xi\gamma^{0}$.\footnote{For the definition of $\Xi$ the reader is referred to reference~\cite[Section 14.4]{Ahluwalia:2019etz}.}
\begin{table}[h!]
	\begin{center}
		\caption{{Possible adjoints for $\psi$ coming from $h=\eta\Delta$.}}
		\label{tabua}
		\begin{tabular}{l|c|c}
			$\eta$ & $\Delta$ & $\psi^{\star}$ \\ 
			\hline
			$\gamma^0$ & $\I$ & $\psi^{\dag}\gamma^{0}$ \\ 
			$\gamma^0$ & $\Xi$ & $\psi^{\dag}\gamma^{0}\Xi$ \\ 
		\end{tabular}
	\end{center}
\end{table} A simple but important aspect shall be noticed here. The operation $\Delta\psi$ leads, of course, to another spinor. This crevice is behind the exploration of the so-called symmetric dual between $\lambda^{S/A}$ and $\rho^{S/A}$ spinors. 

Having framed the structure of the Clifford algebra and a possible generalization thus far, we call attention to the last piece of the dual re-definition, allowed by exploiting the eigenspinor relations
\begin{equation}
\mathcal{A}\lambda^S=\lambda^S,\quad\mathcal{B}\lambda^A=\lambda^A.
\end{equation} This is the additional freedom exploited in the dual formulation and renders~(\ref{matjm1}) possible.

\subsection{The $\tau$ deformation}

Now we proceed to argue for the $\tau$ deformation and the $\tau\rightarrow1$ limit. Let $\mathcal{M}$ be a space comprising of all relativistic (self and anti-self conjugated) ELKOs. Then $\mathcal{M}=[\mathbb{C}\otimes\mathcal{C}l_{1,3}f]^S\oplus[\mathbb{C}\otimes\mathcal{C}l_{1,3}f]^A$  is endowed with the existence of the null spinor\footnote{This point, although important, is not problematic. In fact, under quite general assumptions, $\mathcal{M}$ may be treated as a vector spinor space \cite{Cavalcanti:2015adn}.}.
We focus on the operator $\mathcal{A}$ but same results, with a parallel analysis, also apply to $\mathcal{B}$. As already discussed in the main text, there are two constraints imposed on $\mathcal{A}$ in order to keep the physical properties already obtained by the initial dual without $\tau$ deformation.\footnote{The index denoting the ELKO degrees of freedom $\alpha$ is suppressed here for the sake of exposition.} Firstly
\begin{eqnarray}
\mathcal{A}\lambda^S=\lambda^S \label{mjm1}
\end{eqnarray} and also
\begin{eqnarray}
\mathcal{A}\lambda^A=\sigma\lambda^A. \label{mjm2}
\end{eqnarray}
Explicit calculation shows that $\sigma=1/(1-\tau)$. The fact that $\sigma$ is divergent as $\tau\rightarrow 1$ is not an issue, for the limit shall be taken at the end of calculations (if this limit is taken before the ending of operations, the operator itself is not well-defined). From these properties it is clear that $\mathcal{A}:\mathcal{M}\rightarrow\mathcal{M}$ and $\ker\mathcal{A}=\{0\}$. Moreover, from a general spinor $\lambda=\alpha_1\lambda^S+\alpha_2\lambda^A \in\mathcal{M}$, $\alpha_i\in \mathbb{C}$ $(i=1,2)$, there is always another one $\lambda_0=\alpha_1\lambda^S+(\alpha_2/\sigma)\lambda^A\in \mathcal{M}$ with $\sigma\neq 0$ such that $\mathcal{A}\lambda_0=\lambda$. Notice that this $\lambda$ spinor is nothing but a linear combination in $\mathcal{M}$. Collecting all these simple properties, we find that $\mathcal{A}$ is a linear bijection from $\mathcal{M}$ in itself, an element of the automorphism set.

As discussed, the inclusion of the operator $\mathcal{A}$ comes from a demand within the theoretical scope. Essentially $\mathcal{A}$ acts as the inverse of $[\I+\mathcal{G}(\p)]$. From the properties of $\mathcal{G}(\p)$ acting upon elements of $\mathcal{M}$, notably from $\mathcal{G}\lambda^{S}=\lambda^{S}$, it is simple to see that $[\I+\mathcal{G}(\p)]$ is not an element of the automorphism set. In fact $\ker[\I+\mathcal{G}(\p)]=\mathbb{C}\otimes\mathcal{C}l_{1,3}f^A$. The implementation of the $\tau-$deformation procedure regularizes the problematic kernel in such a way that $\ker[\I+\tau\mathcal{G}(\p)]=\{0\}$, $\tau\in\Re$, rendering an isomorphic operator. The final step is to require both operators, $[\I+\tau\mathcal{G}(\p)]$ and $\mathcal{A}$, be members of $\mbox{Aut}(\mathcal{M})$, the automorphism group. In doing so, as elements of the same group, $\mathcal{A}$ may be taken as the inverse of $[\I+\tau\mathcal{G}(\p)]$.

The additional issue to consider is the $\tau\rightarrow 1$ limit undertaken at the end of calculations~\cite{Rogerio:2016mxi}. Notice that in $[\I+\tau\mathcal{G}(\p)]$, both matrices $\I$ and $\mathcal{G}(\p)$ are non-singular. Therefore, by writing
\begin{equation}
[\I+\tau\mathcal{G}(\p)]=[\I\tau+\mathcal{G}^{-1}(\p)]\mathcal{G}(\p), \label{mjm3}
\end{equation}
we see that $[\I+\tau\mathcal{G}(\p)]$ is invertible as long as $[\I\tau+\mathcal{G}^{-1}(\p)]$ is non-singular. Take $Z=-\mathcal{G}^{-1}(\p)$ and let $P_Z(\delta)=\det|Z-\delta\I|$ be its associated polynomial. Since~(\ref{mjm3}) may be recast as $-(Z-\I\tau)\mathcal{G}(\p)$ we see that the non-singular condition for $[\I+\tau\mathcal{G}(\p)]$ reads $\tau\notin \{\delta_k | P_Z(\delta_k)=0\}$. Now let
\begin{eqnarray}
&&\delta_{min}=\min\{|\delta_k| , \delta_k\neq 0\},\nonumber\\
&&\delta_{max}=\max\{|\delta_k|\}.\label{mjm4}
\end{eqnarray}
It is fairly simple to see that there exists two dense regions for $\tau \in \Re$ in which the inverse of $[\I+\tau\mathcal{G}(\p)]$ is well-defined. They are $0<|\tau|<\delta_{min}$ and $|\tau|>\delta_{max}$. Finally, as
\begin{equation}
P_Z=\delta^4-2\delta^2+1,
\end{equation}
we have a multiplicity two for each root given by $\delta_{k}=\pm 1$. Therefore the inverse of $[\I+\tau\mathcal{G}(\p)]$ in the neighborhood of $\tau=1$ may be made precise as the allowed region for $\tau$ is given by $\Re^+-\{1\}$.
In this vein, $\tau$ may be taken to be sufficiently close to $1$ as required (taken here as an adherent point of the domain in real line) from both sides. As previously shown, at the end of calculations, the only function that requires us to take the $\tau\rightarrow 1$ limit is $1/(1+\tau)$. Therefore, the limit is well-defined.

We end our discussions on the ELKO dual with a few comments. Firstly, by performing the $\tau$ deformation, we obtain a local theory upon quantization. Secondly, the results presented in this appendix show that a generalization to the Dirac dual is possible from the Clifford algebra perspective and that the inverse of $[I+\tau\mathcal{G}(\p)]$ is well-defined. Moreover, it also proposed a hypothesis towards justifying the $\tau$ deformation. The hypothesis states that for the theory to be physical, the spin sums and $\mathcal{A}$ must belong to the same group of automorphism $\mbox{Aut}(\mathcal{M})$, where $\mathcal{M}=[\mathbb{C}\otimes\mathcal{C}l_{1,3}f]^{S}\oplus[\mathbb{C}\otimes\mathcal{C}l_{1,3}f]^{A}$. This amounts to the requirement that their respective kernels must be null.

\section{Matsubara frequency sums for bosons and fermions}{\label{AppA}}

{
The Matsubara frequency sum in~(\ref{eq:pe08}) must be taken over $\omega_n = \frac{2n\pi}{\beta}$ for bosons and $\omega_n = \frac{(2n+1)\pi}{\beta}$ for fermions, where $n$ is an integer.\footnote{See \cite{Kapusta:2006pm,Bailin:1986wt} for further details.} We have:
\begin{equation}
    2\sum_n\ln[\beta^2(\omega_n+i\mu)^2+\omega^2]=\sum_n\Big\{\ln\left[\beta^{2}\left(\omega_{n}^{2}+(\omega-\mu)^2\right)\right]+\ln\left[
    \beta^{2}\left(\omega^{2}_{n}+(\omega+\mu)^2\right)\right]\Big\},\label{eq:pe11}
\end{equation}
where $\omega = \sqrt{|\p|^2+m^2}$.
}

{
For bosons, we use the following identities
\begin{align}
    \ln[(2n\pi)^2 +\beta^2(\omega\pm\mu)^2]&=\int_1^{\beta^2(\omega\pm \mu)^2} \frac{d\theta^2}{(2n\pi)^2 +\theta^2}+\ln[1+(2n\pi)^2],\\
        \sum_{n=-\infty}^{\infty}\frac{1}{(2n\pi)^2 +\theta^2}&=\frac{1}{\theta}\Bigg(1+\frac{2}{e^\theta -1}\Bigg),
\end{align}
from which the $\theta$ integration can be performed. For fermions, we use
\begin{align}
    \ln[(2n+1)^2\pi^2 +\beta^2(\omega\pm\mu)^2]&=\int_1^{\beta^2(\omega\pm \mu)^2} \frac{d\theta^2}{(2n+1)^2\pi^2 +\theta^2}+\ln[1+(2n+1)^2\pi^2],\\
    \sum_{n=-\infty}^{\infty}\frac{1}{(2n+1)^2\pi^2 +\theta^2}&=\frac{1}{\theta}\Bigg(\frac{1}{2}-\frac{1}{e^\theta +1}\Bigg).
\end{align}
}

\section{Kaluza-Klein decompositions}\label{kkd}

{
We present the actions for the real scalar field $\Phi$, the U(1) vector field $A_{M}$, the Kalb-Ramond field $B_{NL}$ and the Dirac spinor field $\Psi$ in five dimensions. Subsequently we will write down their KK decompositions to four dimensions. Their actions are~\cite{Bajc:1999mh,Ringeval:2001cq,Melfo:2006hh,Slatyer:2006un,Liu:2017gcn,ZhangLiu2021}
\begin{align}
S_{\Phi}&=\int d^{5}x\sqrt{-g}\left[-\frac{1}{2}g^{MN}\partial_{M}\Phi\partial_{N}\Phi\right], \\
S_{A}&=-\frac{1}{4}\int d^{5}x\sqrt{-g}F^{MN}F_{MN},\quad F_{MN}=\partial_{M}A_{N}-\partial_{N}A_{M}, \\
S_{B}&=-\int d^{5}x\sqrt{-g}H_{MNL}H^{MNL},\quad H_{MNL}=\partial_{[M}B_{NL]},\\
S_{\Psi}&=\int d^{5}x\sqrt{-g}
\left[F_{1}\overline{\Psi}\Gamma^{M}D_{M}\Psi+\lambda F_{2}\overline{\Psi}\Psi+\eta\overline{\Psi}\Gamma^{M}(\partial_{M}F_{3})\gamma^{5}\Psi\right],\label{eq:5dpsi}
\end{align}
where $F_{1}$, $F_{2}$, and $F_{3}$ are functions of the background scalar field with $\lambda$ and $\eta$ being the coupling constants. For the U(1) vector field and the Kalb-Ramond field, we choose the gauge $A^{5}=0$ and $B^{\mu4}=0$. Their KK decompositions to four dimensions are
\begin{align}
\Phi(x,z)&=e^{-3A(z)/2}\sum_{n}\varphi_{n}(x)\chi_{n}(z), \\
A^{\mu}(x,z)&=e^{-A(z)/2}\sum_{n}a_{n}^{\mu}(x)\rho_{n}(z),\\
B^{\mu\nu}(x,z)&=e^{-7A(z)/2}\sum_{n}b^{\mu\nu}_{n}(x)U_{n}(z),\\
\Psi(x,z)&=e^{-A(z)}\sum_{n}\left[\psi_{Ln}(x)f_{Ln}(z)+\psi_{Rn}(x)f_{Rn}(z)\right].
\end{align}
For convenience, in~sec.~\ref{comparison}, we denote all the mode functions of the extra dimension as $f_{n}(z)$. The chirality of the fermions will be specified when it becomes necessary.
}

\section*{Acknowledgments}

The work of DVA is fully supported by his personal funds.
DVA and CYL thank Dimitri Schritt, Sebastian Horvath, and Tom Watson for their early contributions and extended discussions.
JMHS thanks CNPq for financial support through  grant number 303561/2018-1.
CYL thanks the generous hospitality offered by the Center of Theoretical Physics at Lanzhou University where part of this work was completed.
YXL was supported, in part, by the National Natural Science Foundation of China through Grants No. 11875151 and No. 12047501, and the 111 Project (Grant No. B20063).
SHP acknowledge financial support from Conselho Nacional de Desenvolvimento Cient\'ifico e Tecnol\'ogico (CNPq) (No. 303583/2018-5).


\end{document}